\documentclass{article}

\RequirePackage{etex}

\usepackage[utf8]{inputenc}
\usepackage[fleqn]{amsmath}
\usepackage{amsfonts, amssymb}

\usepackage[pdftex, colorlinks]{hyperref} 
 \hypersetup{
     colorlinks,
     linkcolor={red!50!blue},
     citecolor={blue!50!green},
     urlcolor={blue!80!black}
 }
\usepackage[numbers,sort&compress]{natbib} 

\usepackage[ntheorem]{empheq}
\usepackage[hyperref]{ntheorem}

\newtheorem{theorem}{Theorem}

\usepackage{extarrows} 
\usepackage{units} 
\usepackage{bm} 
\usepackage{bbold} 
\usepackage[subnum]{cases} 
\usepackage{tensor} 
\usepackage{soul} 
\usepackage{cancel} 
\usepackage[retainorgcmds]{IEEEtrantools} 
\IEEEeqnarraydefcolsep{0}{\leftmargini} 
\usepackage{graphicx}

\usepackage{array} 
\newcolumntype{C}[1]{>{\centering\let\newline\\\arraybackslash\hspace{0pt}}m{#1}}

\usepackage{color}
\usepackage{realboxes}
\usepackage{framed}
\usepackage{array} 
\usepackage{mathtools} 
\usepackage[version=4]{mhchem} 

\usepackage{enumerate} 
\usepackage{enumitem}

\usepackage{bookmark} 
\usepackage{float} 
\usepackage{rotating}

\usepackage{ytableau} 
\usepackage{youngtab}
\usepackage{adjustbox} 

\ytableausetup{mathmode, centertableaux}
\usepackage{xargs}
\usepackage{letltxmacro} 
\usepackage{xcolor}
\usepackage{tikz}
\usepackage{tkz-graph} 
\usepackage{tkz-euclide}
\usetikzlibrary{matrix,fit,arrows,decorations}
\usepackage{tikz-cd}
\usetikzlibrary{cd}

\usepackage{pbox}
 
\usepackage{environ}
\NewEnviron{angmatrix}{%
\tikz[baseline=0pt, every node/.style={inner sep=0pt, outer sep=0pt}]
\draw[line width=0.7pt] 
  node [append after command={
    (\tikzlastnode.north west) --
    ($(\tikzlastnode.west) + 0.07*(\tikzlastnode.west)!1!90:(\tikzlastnode.north west)$) -- 
    (\tikzlastnode.south west)
    (\tikzlastnode.north east) --
    ($(\tikzlastnode.east) + 0.07*(\tikzlastnode.east)!1!270:(\tikzlastnode.north east)$) -- 
    (\tikzlastnode.south east)
  }] 
  {\(\begin{matrix} \BODY \end{matrix}\)};
}%
{}

\usepackage{environ}
\NewEnviron{Yangmatrix}{%
\tikz[baseline=0pt, every node/.style={inner sep=0pt, outer sep=0pt}]
\draw[line width=0.7pt] 
  node (Bracket) [append after command={
    (\tikzlastnode.north west) --
    ($(\tikzlastnode.west) + 0.07*(\tikzlastnode.west)!1!90:(\tikzlastnode.north west)$) -- 
    (\tikzlastnode.south west)
    (\tikzlastnode.north east) --
    ($(\tikzlastnode.east) + 0.07*(\tikzlastnode.east)!1!270:(\tikzlastnode.north east)$) -- 
    (\tikzlastnode.south east)
    node at ($(Bracket.south east)+(0.25,0.1)$) {$Y$}
  }] 
  {\(\begin{matrix} \BODY \end{matrix}\)};
}%
{}

\usepackage{environ}
\NewEnviron{YAvgmatrix}{%
\tikz[baseline=0pt, every node/.style={inner sep=0pt, outer sep=0pt}]
\draw[line width=0.7pt] 
  node (Bracket) [append after command={
    (\tikzlastnode.north west) --
    ($(\tikzlastnode.west) + 0.07*(\tikzlastnode.west)!1!90:(\tikzlastnode.north west)$) -- 
    (\tikzlastnode.south west)
    (\tikzlastnode.north east) --
    ($(\tikzlastnode.east) + 0.07*(\tikzlastnode.east)!1!270:(\tikzlastnode.north east)$) -- 
    (\tikzlastnode.south east)
    node[right=8pt of \tikzlastnode.east, anchor=west] {$(Y)$}
  }] 
  {\(\begin{matrix} \BODY \end{matrix}\)};
}%
{}

\usepackage[a4paper,left=2.0cm]{geometry}
\newcommand{\blockmatrix}[9]{
  \draw[draw=#4,fill=#5,every node/.style={inner sep=0,outer sep=0}] (0,0) rectangle( #1,#2);
  \ifthenelse{\equal{#6}{true}}
  {
    \draw[draw=#7,fill=#8] (0,#2) -- (#9,#2) -- ( #1,#9) -- ( #1,0) -- ( #1 - #9,0) -- (0,#2 -#9) -- cycle;
  }
  {}
  \draw ( #1/2, #2/2) node[inner sep=0,outer sep=0]{ #3};
}

\newcommand{\mblockmatrix}[4][none]{
  \begin{tikzpicture} 
  \ifthenelse{\equal{#1}{none}}
  {
    \blockmatrix{#2}{#3}{#4}{none}{none}{false}{none}{none}{0.0}
  }
  {
    \definecolor{fillcolor}{rgb}{#1}
    \blockmatrix{#2}{#3}{#4}{none}{fillcolor}{false}{none}{none}{0.0}
  }
  \end{tikzpicture}
}

\newcommand{\fblockmatrix}[4][none]{%
  \begin{tikzpicture}[outer sep=0,inner sep=0] 
  \ifthenelse{\equal{#1}{none}}{\blockmatrix{#2}{#3}{#4}{black}{none}{false}{none}{none}{0.0}}{\definecolor{fillcolor}{rgb}{#1}\blockmatrix{#2}{#3}{#4}{black}{fillcolor}{false}{none}{none}{0.0}}\end{tikzpicture}
}

\newcommand{\dblockmatrix}[4][none]{
  \begin{tikzpicture} 
  \ifthenelse{\equal{#1}{none}}
  {\blockmatrix{#2}{#3}{#4}{black}{none}{true}{black}{none}{0.35cm}}{\definecolor{fillcolor}{rgb}{#1}\blockmatrix{#2}{#3}{#4}{black}{none}{true}{black}{fillcolor}{0.35cm}}\end{tikzpicture}
}

\newcommand{\diagonalblockmatrix}[5][none]{
  \begin{tikzpicture} 

  \ifthenelse{\equal{#1}{none}}
  {
    \blockmatrix{#2}{#3}{#4}{black}{none}{true}{black}{none}{#5}
  }
  {
    \definecolor{fillcolor}{rgb}{#1}
    \blockmatrix{#2}{#3}{#4}{black}{none}{true}{black}{fillcolor}{#5}
  }

  \end{tikzpicture}
}

\renewenvironment{abstract}{%
\centering\begin{minipage}{.95\textwidth}
\sffamily{\bf Abstract:}}
{
\end{minipage}\vskip 3em}
\makeatletter
\renewcommand\@maketitle{%
\hfill
\begin{minipage}{\textwidth}
\vskip 2em
\let\footnote\thanks 
{\LARGE \bf \@title \par }
\vskip 1.5em
{\large \@author \par}
\end{minipage}
\vskip 3em \par
}
\makeatother
\usepackage{authblk}

\newcommand{\bra}[1]{\langle #1 \vert}
\newcommand{\ket}[1]{\vert #1 \rangle}
\newcommand{\InnerProd}[2]{\left\langle #1 \middle| #2 \right\rangle}

\newcommand{\Tr}[1]{\text{tr}\left(#1\right)}

\newcommand{\diagram}[2][{}]{\pbox{\textwidth}{\includegraphics[#1]{{#2}}}}
\newcommand{\FPic}[2][{}]{\hspace{-0.27mm}\pbox{\textwidth}{\includegraphics[#1]{{#2}}}\hspace{-0.27mm}}

\newcommand{\ybox}[1]{\begin{ytableau} #1 \end{ytableau}}

\newcommand{\SUN}{\mathsf{SU}(N)}
\newcommand{\SU}[1]{\mathsf{SU}(#1)}
\newcommand{\GLN}{\mathsf{GL}(N)}
\newcommand{\MixedPow}[2]{V^{\otimes
    #1}\otimes V^{* \otimes #2}}
\newcommand{\PairPow}[1]{\left(V \otimes V^*\right)^{\otimes #1}}
\newcommand{\Pow}[1]{V^{\otimes #1}}

\newcommand{\Lin}[1]{\mathrm{Lin}\left( #1 \right)}
\newcommand{\PI}[1]{\mathsf{PI}\left( #1 \right)}
\newcommand{\API}[1]{\mathsf{API}\left( #1 \right)}

\newcommand{\vin}{\rotatebox[origin=c]{-90}{$\in$}}

\DeclareMathOperator{\tr}{\text{tr}}

\def\smath#1{\text{\scalebox{.8}{$#1$}}}
\def\sfrac#1#2{\smath{\frac{#1}{#2}}}

\let\oldytableau\ytableau
\let\endoldytableau\endytableau
\renewenvironment{ytableau}{\begin{adjustbox}{scale=.78}\oldytableau}{\endoldytableau\end{adjustbox}}

\makeatletter
\newcommand{\vast}{\bBigg@{3}}
\newcommand{\Vast}{\bBigg@{4}}
\makeatother

\tikzstyle basiclabel=[draw=none,fill=none,shape=rectangle,inner sep=2pt,scale=.8]
\tikzstyle leftlabel=[basiclabel,anchor=east]
\tikzstyle rightlabel=[basiclabel,anchor=west]

\title{Compact construction algorithms for the singlets of
  $\mathsf{SU}(N)$ over mixed tensor product spaces} 
\author[1]{J. Alcock-Zeilinger}
\author[2]{H. Weigert}
\affil[1]{\small Eberhard Karls Universit{\"a}t T{\"u}bingen, Fachbereich
  Mathematik, Auf der Morgenstelle 10, 72076 T{\"u}bingen, Germany}
\affil[2]{\small University of Cape Town, Deptartment of Physics, Private Bag X3, Rondebosch 7701, South Africa}
\date{}

\begin{document}
\maketitle

\begin{abstract}
  The irreducible representations of $\SUN$ over a mixed
  quark-antiquark Fock space component $\MixedPow{m}{n}$, where
  $\text{dim}(V)=N$, have been studied for many years
  (see~\cite{Young:1928, Littlewood:1950, Weyl:1946,
    Cartan:1894structure, Keppeler:2012ih} and many more). In analogy
  to the case for the quark-only Fock space component $\Pow{m}$, there exist
  efficient tools to classify the irreducible representations of
  $\SUN$ over $\MixedPow{m}{n}$ using tableaux. Unlike the case of
  $\Pow{m}$, the only general algorithm known to us for constructing
  the associated projection operators onto irreducible multiplets
  involves translating $n(N-1)$ fundamental factors into $n$
  antifundamental factors using the Leibniz rule, which turns out to
  be computationally extremely inefficient.

  If one is interested only in singlets, this problem can be entirely
  avoided as is demonstrated below where we provide an efficient
  algorithm to construct the projection operatators onto the
  irreducible representations of dimension $1$ of the special unitary
  group $\SUN$ over a mixed Fock space component $\MixedPow{m}{n}$ that
  transparently gives access to $N$ dependence, and discuss the
  relative merits in comparison to an alternative algorithm briefly
  discussed in a different context by Keppeler and Sjödahl
  in~\cite{Keppeler:2012ih}.



\end{abstract}

\tableofcontents

\section{Introduction}

The theory of invariants (in mathematics-circles also often referred
to through the \emph{Schur-Weyl duality}, see e.g.~\cite{Fulton:2004})
is a powerful method of characterizing the irreducible representations
of the general linear group $\GLN$ on a finite dimensional vector
space $W$~\cite{Weyl:1928-1929, Weyl:1946} that is formed as a tensor
product of a finite number of factors containing an $N$-dimensional
vector space $V$ and its dual $V^*$ (see~\cite{Tung:1985na} for a
textbook treatment).
This method exploits the fact that the set of linear maps on $W$ that
commute with the action of the group $\GLN$ on $W$ in the most general
case are easily described in terms of a set of maps called primitive
invariants of $\GLN$ on $W$~\cite{Cvitanovic:2008zz}, and will
collectively be denoted as $\PI{\GLN,W}$ in this paper. We will write
$\API{\GLN,W}$ to denote the algebra of \emph{real} linear
combinations of the primitive invariants.

Almost concurrently with the formulation of the theory of invariants,
Young contrived a combinatorial method of classifying the irreducible
representations of $\GLN$ on $\Pow{m}$ and many of its subgroups, in
particular the special unitary group $\SUN$~\cite{Young:1928}. In this
method, one constructs an object called a \emph{Young tableau}, and
from it obtains the irreducible idempotents (also known as the
\emph{Young projection operators}) corresponding to the irreducible
representations of $\SUN$ on $\Pow{m}$.

In the 1930's, Littlewood and
Richardson~\cite{LittlewoodRichardson:1934} were able to generalize
Young's tableaux to \emph{Littlewood-Richardson (LR)
  tableaux},\footnote{These are not to be confused with
  Littlewood-Richardson \emph{skew}
  tableaux~\cite{Fulton:1997,Fulton:2004}, which are sometimes simply
  referred to as Littlewood-Richardson tableaux in the literature.}
which correspond to the irrducible representations of $\SUN$ on
\emph{general} product spaces $W$ that may include spaces derived from
$V$ (such as $V^*$ or also specifically $V^{(\text{adj})}$, the
carrier space of the adjoint representation, which is also the
traceless part of $V\otimes V^*$) in addition to
$V$~\cite{Littlewood:1976nv}. If one considers a product space
consisting of $V$ and $V^*$ only, for example $\MixedPow{m}{n}$, then
the projection operators corresponding to the irreducible
representations of $\SUN$ on $\MixedPow{m}{n}$ can be extracted from
the LR tableaux via the Leibniz formula for
determinants~\cite{Jeevanjee:2015}, as is exemplified in
appendix~\ref{sec:LR-Example}.

The resulting procedure is, however, extremely longwinded and thus
only of limited use in practical
applications. Appendix~\ref{sec:LR-Example} contains an example
exhibiting the tedium. If one needs to keep $N$ generic, the method
becomes impractical already for small spaces such as $V\otimes V^*$,
and thus, by extrapolation, nearly unmanagable for spaces beyond
$V\otimes V^*$.

In physics applications one is often only interested in the
\emph{singlet representations} of $\SUN$ on $\MixedPow{m}{n}$. The
associated singlet states $\ket{\phi^S}\in\MixedPow{m}{n}$ of $\SUN$
satisfy, for every $U\in\SUN$,\footnote{Many casual readers will be
  irritated by the appearance of complex instead of Hermitian
  conjugation in this expression. See
  eqns.~\eqref{eq:IndexLeg-Arrow-Transformation}
  and~\eqref{eq:VVstar-VstarV-action} for the reason.}
\begin{equation}
  \label{eq:global-singlet-states-Def}
  \mathbf{U}
  \ket{\phi^S}
  =
  \ket{\phi^S}
  \qquad \text{and} \qquad
  \bra{\phi^S}
  \mathbf{U}^{\dagger}
  =
  \bra{\phi^S}
  \ ,
  \qquad \text{where }
  \mathbf{U} 
  :=
  U^{\otimes m} 
  \otimes 
  U^{*\otimes n}
  \ . 
\end{equation}
The singlet states of the color group $\SU{N_c}$ play a special role in
quantum chormodynamics since only these correspond to observable
particle configurations due to \emph{confinement}. Somewhat more
generally, group integrals will be non-vanishing only on singlet
integrands, which can be projected out if all singlet states are
known.

With such applications in mind, the present paper provides a compact,
and efficiently programmable algorithm to construct the
$1$-dimensional (singlet) representations of $\SUN$ on
$\MixedPow{k}{k}$, which we call \emph{generic singlets}, since they
remain singlets irrespective of $N$, once $N$ is large enough in
relation to $k$ for a specific singlet representation $\lambda$ to
appear in the decomposition of $\MixedPow{k}{k}$ into
irreducibles. The algorithm provides easy access to the associated
threshold values $N_{\lambda,k}$. We argue that singlets on
$\MixedPow{m}{n}$, where $m \neq n$ are \emph{non-generic} or
\emph{transient} in the sense that they are one dimensional only for
specific values of $N$. We will comment on their role more directly in section~\ref{sec:mqnqb-no-new-multiplets}.

The idea behind the algorithm that delivers generic singlets is
simple: To stay with the QCD example, consider a Fock space component
containing an equal number of quarks and antiquarks,
$\MixedPow{k}{k}$. To construct a particular singlet state and the
associated singlet projector of $\SUN$ on this Fock space component,
one first selects an element $P \in \API{\SUN,\MixedPow{i}{j}}$, where
the only restriction is that $i+j=k$. In particular, one may select
$P$ from $\API{\SUN,\Pow{k}}$ as we will discuss in
section~\ref{sec:Singlets}. (This is the reason the threshold values
$N_{\lambda,k}$ can be readily determined.) One then writes $P$ in
birdtrack notation (see section~\ref{sec:Background}) and bends $P$ to
obtain a singlet state in $\MixedPow{k}{k}$. Since $\MixedPow{m}{n}$
inherits an inner product from $V$, this state can easily be
normalized. The associated projection operator is obtained if we
multiply this by its Hermitian conjugate, obtained in birdtracks by
reflecting across a vertical axis, followed by a reversal of all
arrows. The procedure is summarized as
\begin{equation}
  \label{eq:Singlet-algorithm-schematic}
  \FPic[scale=0.75]{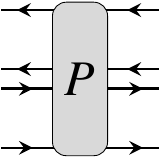}
  \xlongrightarrow{\text{bend \& normalize}} 
  \underbrace{ \; \alpha_P \;
    \FPic[scale=0.5]{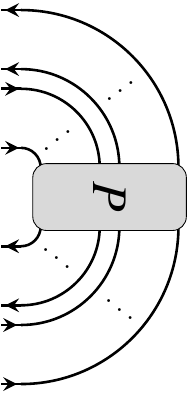} \;
  }_{\text{singlet state}}
  \xlongrightarrow{\text{mult. by conj.}} 
  \underbrace{ \;\alpha_P^2 \;
    \FPic[scale=0.5]{mqmqbSState-P-allLines}
    \hspace{2mm}
    \FPic[scale=0.5]{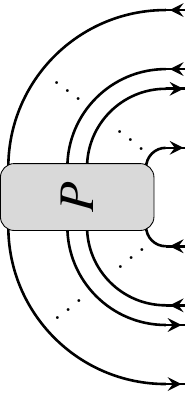} \;
  }_{\text{singlet projector}}
  \ ; 
\qquad
\frac{1}{\alpha_P^2} := \FPic[scale=0.5]{mqmqbSStateDag-P-allLines}\FPic[scale=0.5]{mqmqbSState-P-allLines}
\ .
\end{equation}
Converting a full set of basis elements $\{O_l\}$ of
$\API{\SUN,\MixedPow{i}{j}}$ in this manner leads to a maximal set of
linearly independent singlet basis states in $\MixedPow{k}{k}$ and their
associated projectors.

All inner products appearing in this context are induced by the inner
product on $V$: The inner product used in
eq.~\eqref{eq:Singlet-algorithm-schematic} features also as the inner
product on $\Lin{\MixedPow{m}{n}}$  through
\begin{equation}
  \label{eq:Sclar-Product-AB}
  \left\langle
    A
    \middle\vert
    B
    \right\rangle_{\Lin{\MixedPow{m}{n}}}
  :=
  \Tr{A^{\dagger} B}
  \qquad \forall\ A,B \in\Lin{\MixedPow{m}{n}}
  \ ,
\end{equation}
i.e. the inner product of the linear maps equals the inner product of
the states generated by the procedure.  Therefore any orthonormal
basis on $\Lin{\MixedPow{i}{j}}$ leads to an orthonormal set of
singlet states and projectors.

It is important to note that this procedure yields \emph{all} singlets on
$\MixedPow{k}{k}$, all of them are generic. Non-generic singlets are a
different matter, they appear \emph{only} in $\MixedPow{m}{n}$ where
$m\neq n$. At fixed $N$, they are relevant for
specific states in the spectrum of physical theories such as all
baryons in QCD. As soon as we consider varying $N$, for example to
implement a $1/N$ expansion, several options appear to implement the
process. All of them are related to the basic idea behind the
Littlewood-Richardson equivalence (as induced by the Leibniz formula
for determinants), which allows us to provide a canonical map that
takes these transient singlets onto generic ones at a given value of
$N$.

The main body of the paper aims to substantiate the claims made above:
Section~\ref{sec:Background} shortly recapitulates the use of the
birdtrack formalism, with special care devoted to the issues that
typically cause confusion in first time encounters with the
toolset. Section~\ref{sec:Singlets} provides the core statements on
generic and non-generic singlets.  Section~\ref{sec:Singlet-Examples}
demonstrates the tools at work for generic singlets by comparing the
MOLD-algorithm based construction with an alternative method while
comparing relative benefits and drawbacks for practical applications.

\section{Background: Theory of invariants \&
  birdtracks}\label{sec:Background}

This section serves to provide the theoretical and notational
background needed for the remainder of this paper: we present a short
overview of the theory of invariants in
section~\ref{sec:Intro-Invariants}, and an introduction to the
birdtrack formalism in section~\ref{sec:Intro-birdtracks}. For a more
comprehensive discussion of these topics, readers are referred to
references in the respective sections.

\subsection{Invariants of~\texorpdfstring{$\SUN$}{SU(N)}}\label{sec:Intro-Invariants}

This section presents a summary of the theory of invariants already
given in~\cite{Alcock-Zeilinger:2016bss, Alcock-Zeilinger:2016sxc,
  Alcock-Zeilinger:2016cva}, mainly to establish notation.

As mentioned earlier, the theory of invariants provides a method of
classifying the irreducible representations of $\GLN$ on a tensor product
space $W$. For this paper, our main interest is on the irreducible
representations of $\SUN\subset\GLN$ on the mixed product space
$\MixedPow{m}{n}$, and thus our treatment of the theory of invariants
will focus on this particular case. For a more general introduction to
the topic, readers are referred to standard textbooks such
as~\cite{Goodman:2009,Fulton:2004,Cvitanovic:2008zz}.

Consider first the case where $n=0$: we wish to classify the
irreducible representations of $\SUN$ on $\Pow{m}$. More precisely, we
consider the fundamental representation of $\SUN$ on a given vector
space $V$ with $\text{dim}(V) = N$, whose action will simply be
denoted by $v\mapsto U v$ for all $U\in\SUN$ and $v\in V$ (note that
we use the same symbol for the group element $U\in\SUN$ and its
fundamental representation on $V$). We then
explore the product representations of $\SUN$ on $\Pow{m}$ constructed
from its fundamental representation on $V$ as follows: Choosing a
basis $\{e_{(i)} |i = 1, \ldots,\text{dim}(V)\}$ such that
$v = v^i e_{(i)}$, the fundamental representation becomes
$v^i \mapsto \tensor{U}{^i_j} v^j$. This immediately induces a product
representation of $\SUN$ on $\Pow{m}$ if one uses this basis of $V$ to
induce a basis on $\Pow{m}$ so that a general element
$\bm{v}\in\Pow{m}$ takes the form
$\bm{v} = v^{i_1\ldots i_m} e_{(i_1)}\otimes\cdots\otimes e_{(i_m)}$:
\begin{align}
  \label{eq:SUN-Action}
  (\mathbf{U}\bm{v})^{i_1\ldots i_m} 
  :=
  \tensor{U}{^{i_1}_{j_1}} 
  \cdots 
  \tensor{U}{^{i_m}_{j_m}} 
  v^{j_1\ldots j_m} e_{(i_1)}\otimes\cdots\otimes e_{(i_m)}
\ .
\end{align} 
Since all the factors in $\Pow{m}$ are identical, the notion of
permuting the factors is a natural one and leads to a linear map on
$\Pow{m}$ according to
\begin{align}
  \label{eq:perm-facs-def}
  (\rho\bm{v})^{i_1\ldots i_m}
  :=
  v^{i_{\rho^{-1}(1)}\ldots i_{\rho^{-1}(m)}} e_{(i_1)}\otimes\cdots\otimes
  e_{(i_m)}
  \ ,
\end{align}
where $\rho$ is an element of the permutation group on $m$ objects, $S_m$.\footnote{Permuting the basis vectors instead yields
  $\rho^{-1}$: $v^{i_{\rho^{-1}(1)}\ldots i_{\rho^{-1}(m)}} \otimes \cdots
  \otimes e_{(i_m)} = v^{i_1\ldots i_m} e_{(i_{\rho(1)})} \otimes
  \cdots \otimes e_{(i_{\rho(m)})}$.} From 
definitions~\eqref{eq:SUN-Action} and~\eqref{eq:perm-facs-def} one
immediately infers that the product representation commutes with all
permutations on any $\bm{v}\in\Pow{m}$:
\begin{equation}
\label{eq:InvariantsIntro1}
  \mathbf{U} \circ \rho (\bm{v})  = \rho \circ \mathbf{U} (\bm{v})
  \ .
\end{equation}
In other words, any such permutation $\rho$ is an \emph{invariant} of
$\SUN$:
\begin{align}
  \label{eq:per-invariant}
  \mathbf{U}\circ\rho\circ \mathbf{U}^{\dagger} = \rho
  \quad \text{equivalently} \quad
  \mathbf{U}^{\dagger} \circ\rho\circ \mathbf{U} = \rho
  \ ,
\end{align}
where we used the fact that $\mathbf{U}^{-1}=\mathbf{U}^{\dagger}$ by
definition of the special unitary group. It can further be shown that
these permutations span the space of all linear invariants of $\SUN$
on $\Pow{m}$~\cite{Goodman:2009}. The permutations thus are the
\emph{primitive invariants} of $\SUN$ on $\Pow{m}$, and as sets,
\begin{equation}
  \label{eq:PI-is-Sm}
  S_m = \PI{\SUN,\Pow{m}}
  \ .
\end{equation}
The algebra of real linear invariants is then given by\footnote{The
  algebra of invariants as given here provides a representation of the
  group algebra $\mathbb R[S_m]$ and as such ``shares'' its
  multiplication table. Note, however, that its dimension will be
  smaller than $m!$ if $\dim(V) < N$. We will need to keep track of
  this information and thus we deal directly with $\API{\SUN,\Pow{m}}$
  instead of $\mathbb R[S_m]$.}
\begin{align}
  \label{eq:PI-def}
  \API{\SUN,\Pow{m}} 
  := 
  \Bigl\{ 
  \sum_{\sigma\in S_m} \alpha_\sigma \sigma \Big| 
  \alpha_\sigma\in \mathbb R, \sigma\in S_m \Bigr\}
  \subset
  \Lin{\Pow{m}} 
  \ .
\end{align}

As one considers the fundamental representation of $\SUN$ on a
vector space $V$, one may also consider the \emph{anti-fundamental}
representation of $\SUN$ on the dual space $V^*$. Again, the
irreducible representations of $\SUN$ on a mixed product space
$\MixedPow{m}{n}$ can be classified through the invariants living in
the algebra~\cite{Cvitanovic:2008zz,Tung:1985na}
\begin{equation}
  \label{eq:apimmp}
  \API{\SUN,\MixedPow{m}{n}} := \Bigl\{\sum_{\rho\in S_{m,n}} \alpha_\rho \rho \ \Big| 
  \ \alpha_\rho \in \mathbb R\Bigr\}
  \subset
  \Lin{\MixedPow{m}{n}}
  \ ,
\end{equation}
where $S_{m,n}$ (to be discussed in the following section, \emph{c.f.} eqns.~\eqref{eq:S3-to-S2plus1-map}) denotes the set of primitive invariants of $\SUN$ on
$\MixedPow{m}{n}$,\footnote{It should be noted that, unlike $S_m$,
  $S_{m,n}$ is \emph{not} a group; this is exemplified in
  eq.~\eqref{eq:q2qb1}, where four of the six elements of $S_{2,1}$ do
  not have an inverse.}
\begin{equation}
  \label{eq:PI-is-Smn}
S_{m,n} = \PI{\SUN,\MixedPow{m}{n}}
\ .
\end{equation}
We will explictly demonstrate below that its elements are in a
$1$-to-$1$ correspondence with the primitive invariants $S_{m+n}$. The
correspondence becomes exceptionally clear in the birdtrack formalism
we turn to next.

\subsection{The birdtrack formalism}\label{sec:Intro-birdtracks}

In the 1970's Penrose devised a graphical method of dealing with
invariants of Lie groups~\cite{Penrose:1971Mom,Penrose:1971Com}, which
was subsequently applied in a collaboration with
MacCallum~\cite{Penrose:1972ia}. This graphical method, now dubbed
the \emph{birdtrack formalism}, was modernized and further developed
by Cvitanovi{\'c}~\cite{Cvitanovic:2008zz} in recent years. The
immense benefit of this formalism is that it makes the actions of the
operators visually accessible and thus more intuitive.  For
illustration, we give as an example the permutations of $S_3$ written
both in their cycle notation (see~\cite{Tung:1985na} for a textbook
introduction) as well as birdtracks:
\begin{equation}
\label{eq:S3-Birdtracks}
\underbrace{\FPic{3ArrLeft}\FPic{3IdSN}\FPic{3ArrRight}}_{\mathrm{id}}\; , \quad \underbrace{\FPic{3ArrLeft}\FPic{3s12SN}\FPic{3ArrRight}}_{(12)}\; , \quad \underbrace{\FPic{3ArrLeft}\FPic{3s13SN}\FPic{3ArrRight}}_{(13)}\; , \quad \underbrace{\FPic{3ArrLeft}\FPic{3s23SN}\FPic{3ArrRight}}_{(23)}\; , \quad \underbrace{\FPic{3ArrLeft}\FPic{3s123SN}\FPic{3ArrRight}}_{(123)}\; , \quad \underbrace{\FPic{3ArrLeft}\FPic{3s132SN}\FPic{3ArrRight}}_{(132)}\; .
\end{equation}
The action of permutations on a tensor product can be naturally
defined as a reodering of factors, for example
\begin{subequations}
\begin{equation}
  \label{eq:v1v2v3-tensor-product}
  (123) \left(v_1\otimes v_2\otimes v_3\right) = v_3\otimes v_1\otimes v_2.
\end{equation}
In the birdtrack formalism, this equation is written as
\begin{equation}
  \label{eq:v1v2v3-tensor-product-birdtrack}
  \FPic{3ArrLeft}
  \FPic{3s123SN}
  \FPic{3ArrRight}
  \;
  \FPic{3v123Labels} 
  \; = \; 
  \FPic{3v312Labels} 
  \ ,
\end{equation}
\end{subequations}
where each factor in the product $v_1\otimes v_2\otimes v_3$ (written as
a tower $\FPic{3v123Labels}$) can be thought of as being moved along
the lines of
$\FPic{3ArrLeft}\FPic{3s123SN}\FPic{3ArrRight}$ in the direction of
the arrows. 
In fact, the arrow-direction on a particular index leg encodes its
transformation behaviour under the action of $\SUN$: We call a
Kronecker~$\delta$ a \emph{quark line} if we can interpret it as the
unit operator in $\mathrm{Lin}(V)$, and thus transforms under the
associated representation:
\begin{subequations}
  \label{eq:IndexLeg-Arrow-Transformation}
\begin{equation}
  \label{eq:IndexLeg-Arrow-Transformation-Quark}
  \text{quark line:} \qquad
  U \; 
  \FPic{1IdArr} \;
  U^\dagger\; 
  \; = \; 
  \FPic{1IdArr}
  \ \in \
  \mathrm{Lin}(V)
  \quad 
  \forall U \in \SUN
  \ .
\end{equation}
Similarly, an \emph{antiquark line} is a Kronecker~$\delta$ that acts as
the unit operator in $\mathrm{Lin}(V^*)$ and transforms accordingly as
\begin{equation}
  \label{eq:IndexLeg-Arrow-Transformation-AntiQuark}
  \text{antiquark line:} \qquad
   U^* \; 
  \FPic{1qbIdArr} \;
  U^t \; 
  \; = \; 
  \FPic{1qbIdArr}
  \ \in \
  \mathrm{Lin}(V^*)
  \quad 
  \forall U \in \SUN
  \ .
\end{equation}
\end{subequations}
Consistently, we interpret~\FPic{1q1qbTr12SStateArr}~as an element in
$V\otimes V^*$, and~\FPic{1qb1qTr12SStateArr}~as an element in
$V^*\otimes V$, transforming under the associated product
representations as invariants,
\begin{subequations}
  \label{eq:VVstar-VstarV-action}
\begin{align}
  \FPic{1q1qbUUComLabels} \;
  \FPic{1q1qbTr12SStateArr}
  & \; = \;
  \FPic{1q1qbTr12SStateArr}
  \ \in \ V\otimes V^*
  \\
  \FPic{1q1qbUComULabels} \;
  \FPic{1q1qbTr12SStateArr}
  & \; = \;
  \FPic{1q1qbTr12SStateArr}
  \ \in \ V^*\otimes V
  \ .
\end{align}
\end{subequations}
The appearance of $U^*$ instead of $U^\dagger$
in~\eqref{eq:VVstar-VstarV-action} is due to the fact that the $V^*$
factor it acts on is placed on its right.  This is simply how index
notation reflects that the anti-fundamental representation acts via a
left group action just like the fundamental one. The index positioning
is required to facilitate the cancellation of the group elements via
$U U^\dagger = \mathbb 1$.
Eq.~\eqref{eq:IndexLeg-Arrow-Transformation-AntiQuark} is a necessary
consequence of~\eqref{eq:VVstar-VstarV-action}
and~\eqref{eq:IndexLeg-Arrow-Transformation-Quark}.

The birdtrack formalism also offers an intuitive way of forming the
product of linear maps of this type by merely connecting the lines,
for example,
\begin{equation}
  \label{eq:S3-product-maps-Ex}
(12) \cdot (132)
=
\FPic{3ArrLeft}\FPic{3s12SN}\FPic{3ArrRight} \cdot
\FPic{3ArrLeft}\FPic{3s132SN}\FPic{3ArrRight}
=
\FPic{3ArrLeft}
\FPic{3s12SN}
\hspace{-0.2mm}
\FPic{3s132SN}
\FPic{3ArrRight}
=
\FPic{3ArrLeft}\FPic{3s13SN}\FPic{3ArrRight}
=
(13)
\ .
\end{equation}
The Hermitian conjugate of a
birdtrack can be formed by flipping the birdtrack about its
vertical axis and reversing the
arrows~\cite{Cvitanovic:2008zz,Alcock-Zeilinger:2016sxc}, e.g.
\begin{equation}
  \label{eq:Hermitian-conjugate-birdtrack-Ex}
\FPic{3ArrLeft}
\FPic{3s123SN}
\FPic{3ArrRight}
\;
\xrightarrow{\text{flip}}
\;
\reflectbox{%
\FPic{3ArrLeft}%
\FPic{3s123SN}%
\FPic{3ArrRight}%
}
\;
\xrightarrow{\text{reverse arrows}}
\;
\FPic{3ArrLeft}
\FPic{3s132SN}
\FPic{3ArrRight}
\; = \; 
\left(
\FPic{3ArrLeft}
\FPic{3s123SN}
\FPic{3ArrRight}
\right)^{\dagger}
\ .
\end{equation}
Particular linear combinations of permutations will be used throughout
this paper, such as symmetrizers and antisymmetrizers: A symmetrizer
$\bm{S}_{a_1 \ldots a_n}$ is defined as the sum over all permutations
of the set $\lbrace a_1 \ldots a_n\rbrace$, together with a prefactor
$\frac{1}{n!}$. Similarly, an antisymmetrizer
$\bm{A}_{a_1 \ldots a_n}$ differs from a symmetrizer only in the
prefactors of the terms that appear in the sum: each permutation in
the sum is \emph{weighted by the signature} of the permutation. In
birdtrack notation, a symmetrizer is denoted by an empty (white) box
drawn over the affected index lines, and an antisymmetrizer is
represented by a filled (black) box. For example,
\begin{subequations}
  \label{eq:Birdtrack-Sym-ASym-Definition}
  \begin{align}
    \bm{S}_{123} 
    \; & = \;
    \FPic{3ArrLeft}\FPic{3Sym123SN}\FPic{3ArrRight}
    \; := \;
    \frac{1}{3!}
    \left( \;
    \FPic{3ArrLeft}\FPic{3IdSN}\FPic{3ArrRight}
    \; + \; 
    \FPic{3ArrLeft}\FPic{3s12SN}\FPic{3ArrRight}
    \; + \; 
    \FPic{3ArrLeft}\FPic{3s13SN}\FPic{3ArrRight}
    \; + \; 
    \FPic{3ArrLeft}\FPic{3s23SN}\FPic{3ArrRight}
    \; + \; 
    \FPic{3ArrLeft}\FPic{3s123SN}\FPic{3ArrRight}
    \; + \; 
    \FPic{3ArrLeft}\FPic{3s132SN}\FPic{3ArrRight}
    \; \right)
    \label{eq:Birdtrack-Sym-Definition}
    \\
    \bm{A}_{123} 
    \; & = \;
    \FPic{3ArrLeft}\FPic{3ASym123SN}\FPic{3ArrRight}
    \; := \;
    \frac{1}{3!}
    \left( \;
    \FPic{3ArrLeft}\FPic{3IdSN}\FPic{3ArrRight}
    \; - \; 
    \FPic{3ArrLeft}\FPic{3s12SN}\FPic{3ArrRight}
    \; - \; 
    \FPic{3ArrLeft}\FPic{3s13SN}\FPic{3ArrRight}
    \; - \; 
    \FPic{3ArrLeft}\FPic{3s23SN}\FPic{3ArrRight}
    \; + \; 
    \FPic{3ArrLeft}\FPic{3s123SN}\FPic{3ArrRight}
    \; + \; 
    \FPic{3ArrLeft}\FPic{3s132SN}\FPic{3ArrRight}
    \; \right)
    \ .
    \label{eq:Birdtrack-ASym-Definition}
  \end{align}
\end{subequations}

The definition of $\SUN$ ($\text{dim}(V) = N$),
\begin{equation}
  \label{eq:SUNdef}
  \SUN := \{ U\in \Lin{V} | \langle U u| U v\rangle = \langle u|v \rangle \forall u,v\in V \land \det(U) = 1 \}
\ ,
\end{equation}
involves two explicit invariant algebraic structures: Kronecker deltas
that appear in the component expression of the inner product and the
$\varepsilon$-tensor that features in the Leibniz formula for
determinants. (See
also~\cite{Cvitanovic:2008zz,Tung:1985na,Weyl:1946,Fulton:2004} for more
background on its role in representation theory.) In birdtracks
\begin{equation}
  \label{eq:Kronecker-Delta-Levi-Civita-Birdtrack}
  \delta_{ij} \; = \;
  \FPic{1iLabels}\;\FPic{1IdArr}\;\FPic{1jLabels}
  \quad \text{and} \quad
  \frac{i^{\phi}}{\sqrt{N!}}\,\varepsilon_{a_1 \ldots a_N}
  \; = \;
\FPic[scale=0.75]{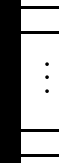}
\FPic[scale=0.75]{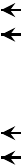}
\;
\FPic[scale=0.75]{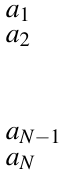}
\quad \text{or} \quad 
\frac{i^{-\phi}}{\sqrt{N!}}\,\varepsilon_{a_1 \ldots a_N} =
\FPic[scale=0.75]{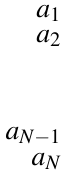}
\FPic[scale=0.75]{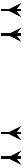}
\FPic[scale=0.75]{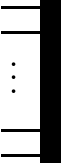} 
\quad \text{ where $\phi = \frac{N(N-1)}2$}
\ .
\end{equation}
While the Kronecker delta is evident in the birdtrack construction of
the primitive invariants above, the $\varepsilon$-tensor appears in a
more subtle manner, through the identity
(c.f.~\cite[eq.~(6.28)]{Cvitanovic:2008zz})
\begin{equation}
\label{eq:Epsilon-lengthN-ASym}
\FPic[scale=0.75]{NcArrLeft}
\FPic[scale=0.75]{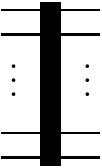}
\FPic[scale=0.75]{NcArrRight}
\;
  \FPic[scale=0.75]{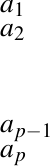}
\; \xlongequal{p=\text{dim}(V)=N} \; 
\FPic[scale=0.75]{NcArrLeft}
\FPic[scale=0.75]{Nc_EpsilonDag} 
\hspace{2mm}
\FPic[scale=0.75]{Nc_Epsilon}
\FPic[scale=0.75]{NcArrRight}
\;
\FPic[scale=0.75]{N_aLabelsRight}
\ ,
\end{equation}
which also motivates the convention used for the prefactor in the
relation between $\varepsilon$ and its
birdtack(s). (Eq.~\eqref{eq:conjClebschBirdtracks} identifies the two
versions as Hermitian conjugates of each other.)

\cite[Sec 6.2]{Cvitanovic:2008zz} provides a long list of
combinatorical identities (and their derivations) that are
essential in performing calculations involving antisymmetrizers of
length $p$ and
$\varepsilon$-tensors. The most relevant for the calculations in this
paper are the ``absorption identities''
\begin{equation}
  \label{eq:absorption-id}
  \FPic[scale=.75]{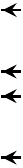}
  \FPic[scale=.75]{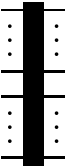}
  \FPic[scale=.75]{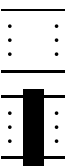}
  \FPic[scale=.75]{NckArr}
  \; = \; 
  \FPic[scale=.75]{NckArr}
  \FPic[scale=.75]{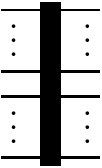}
  \FPic[scale=.75]{NckArr}
  \hspace{0.8cm}
  \text{so that (at $N=p$)}
  \hspace{0.8cm}
  \FPic[scale=.75]{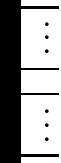}
  \FPic[scale=.75]{NckASymk-to-Nc-fatSN}
  \FPic[scale=.75]{NckArr}
  \; = \; 
  \FPic[scale=.75]{Nck_Epsilon}
  \FPic[scale=.75]{NckArr}
  \ ,
\end{equation}
and the partial trace  identities (where the trace is taken over the
top $p-k$ index lines and the remaining $k$ index lines are open)
\begin{equation}
  \label{eq:trace-id}
  \FPic[scale=.75]{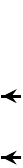}
  \FPic[scale=.75]{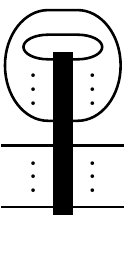}
  \FPic[scale=.75]{NckArrk-to-Nc}
  \; = \; 
  \frac{(N-k)! k!}{(N-p)! p!} \;
  \raisebox{0.5\height}{%
    \FPic[scale=.75]{NckArrk-to-Nc}%
    \FPic[scale=.75]{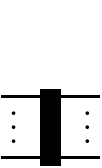}%
    \FPic[scale=.75]{NckArrk-to-Nc}%
  }
  \hspace{0.8cm}
  \text{and (at $N=p$)}
  \hspace{0.8cm}
  \FPic[scale=.75]{NckArrk-to-Nc}
  \FPic[scale=.75]{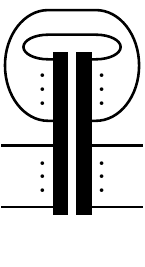}
  \FPic[scale=.75]{NckArrk-to-Nc}
  \; = \; 
  \frac{(N-k)! k!}{N!} \;
  \raisebox{0.5\height}{%
    \FPic[scale=.75]{NckArrk-to-Nc}%
    \FPic[scale=.75]{Nck-kASymk-to-Nc-fat}%
    \FPic[scale=.75]{NckArrk-to-Nc}%
  }
  \ .
\end{equation}


The birdtrack formalism also allows for an efficient way to
include antifundamental representations and the associated algebra of
invariants on the mixed space $\MixedPow{m}{n}$,
$\API{\SUN,\MixedPow{m}{n}}$. To do so, we start from $\Pow{(m+n)}$ and
replace factors of $V$, one by one, with a total of $n$ factors of
$V^*$ and, in parallel, modify the elements of
$S_{m+n} \subset\Lin{\Pow{(m+n)}}$ accordingly by swapping the left
and right endpoints of the associated level of its birdtrack.
An
example will give clarity: the primitive invariants
$S_3=\PI{\SUN,\Pow{3}}$ given in eq.~\eqref{eq:S3-Birdtracks} map onto
$S_{2,1}=\PI{\SUN,\MixedPow{2}{1}}$ as
\allowdisplaybreaks[0]\begin{subequations}
\label{eq:S3-to-S2plus1-map}
\begin{IEEEeqnarray}{0lCCCCCCCCCCCr}
  S_3:\hspace{1cm}
  &
  \FPic{3ArrLeft}
  \FPic{3IdSN}
  \FPic{3ArrRight}
  &
  \ , \quad 
  &
  \FPic{3ArrLeft}
  \FPic{3s12SN}
  \FPic{3ArrRight}
  &
  \ , \quad 
  &
  \FPic{3ArrLeft}
  \FPic{3s23SN}
  \FPic{3ArrRight}
  &
  \ , \quad 
  &
  \FPic{3ArrLeft}
  \FPic{3s13SN}
  \FPic{3ArrRight}
  &
  \ , \quad
  &
  \FPic{3ArrLeft}
  \FPic{3s123SN}
  \FPic{3ArrRight}
  &
  \ , \quad
  &
  \FPic{3ArrLeft}
  \FPic{3s132SN}
  \FPic{3ArrRight}
  & 
  \\
  &
  \FPic{StraightDownArrow}
  & & 
  \FPic{StraightDownArrow}
  & & 
  \FPic{StraightDownArrow}
  & & 
  \FPic{StraightDownArrow}
  & & 
  \FPic{StraightDownArrow}
  & & 
  \FPic{StraightDownArrow}
  &
  \nonumber \\
  S_{2,1}:
  &
  \FPic{2q1qbArr}
  \FPic{3IdSN}
  \FPic{2q1qbArr}
  &
  \ , \quad 
  &
  \FPic{2q1qbArr}
  \FPic{3s12SN}
  \FPic{2q1qbArr}
  &
  \ , \quad 
  &
  \FPic{2q1qbArr}
  \FPic{2q1qbTl23r23N}
  \FPic{2q1qbArr}
  &
  \ , \quad 
  &
  \FPic{2q1qbArr}
  \FPic{2q1qbTl13r13N}
  \FPic{2q1qbArr}
  &
  \ , \quad
  &
  \FPic{2q1qbArr}
  \FPic{2q1qbTl13r23N}
  \FPic{2q1qbArr}
  &
  \ , \quad
  &
  \FPic{2q1qbArr}
  \FPic{2q1qbTl23r13N}
  \FPic{2q1qbArr}
  & 
  \label{eq:q2qb1}
\end{IEEEeqnarray}
\end{subequations}\allowdisplaybreaks[1]%
in a direct $1$-to-$1$ correspondence. 

From the multiplication rule of birdtracks (as was exemplified
in~\eqref{eq:S3-product-maps-Ex}) it immediately follows that
$S_{2,1}$ (unlike $S_3$) is \emph{not} a group, as only the first two
elements in~\eqref{eq:q2qb1} have an inverse.

Generalizing the graphical procedure of swapping quarks into
antiquarks to arbitrary $m$ and $n$ yields
\begin{equation}
  \label{eq:apimmp-2}
  \API{\SUN,\MixedPow{m}{n}} := \Bigl\{\sum_{\rho\in S_{m,n}} \alpha_\rho \rho \ \Big| 
  \ \alpha_\rho \in \mathbb R\Bigr\}
  \subset
  \Lin{\MixedPow{m}{n}}
  \ .
\end{equation}
The multiplication table follows directly from the multiplication
rules of birdtracks (eq.~\eqref{eq:S3-product-maps-Ex}) and differs
significantly from that of $\API{\SUN,\Pow{(m+n)}}$.

Despite the bijection between the sets of primitive invariants
$\PI{\SUN,\Pow{m}} $ and $\PI{\SUN,\MixedPow{k}{l}}$, if $k+l=m$,
exemplified in eq.~\eqref{eq:S3-to-S2plus1-map}, the structures we can
associate with these two sets are radically different:
\begin{enumerate}
\item The
multplication table for elements in $\API{\SUN,\Pow{m}} $ is identical
to that of the permutation group $S_m$ and thus makes no reference to
$N$ or $\SUN$ for that matter: the product of any two elements
directly yields a specific element in the set of primitive invariants
$\PI{\SUN,\Pow{m}}$ and allows us to assign an associative
multiplication with the set itself (which furnishes even a
representation of $S_m$).  This is not the case for two elements of
$S_{m,n} =\PI{\SUN,\MixedPow{m}{n}}$. An arbitrary product of two
elements in $\PI{\SUN,\MixedPow{m}{n}}$ is not simply another element
in this set: Instead the result generically ends up in
$\API{\SUN,\MixedPow{m}{n}}$, with nontrivial $N$-dependent prefactors
appearing automatically. For example
\begin{equation}
  \label{eq:n-dep-mult-table}
  \FPic{2q1qbArr}
  \FPic{2q1qbTl13r23N}
  \FPic{2q1qbArr}
  \FPic{2q1qbTl23r13N}
  \FPic{2q1qbArr}
  = N   \FPic{2q1qbArr}
  \FPic{2q1qbTl13r13N}
  \FPic{2q1qbArr} \in \API{\SUN,\MixedPow{m}{n}}
  \ .
\end{equation}
\item While we can think of the group algebra $\mathbb R(S_m)$ in a way that
does not involve representations as primitive invariants on a vector
space $\Pow{m}$, we do not know of an equivalent structure for
$S_{m,n}$, all we have is $\API{\SUN,\MixedPow{m}{n}}$.

\item While $\mathbb R(S_m)$ has a fixed dimension
  ($\dim(\mathbb R(S_m) = |S_m| = m!$), its representations on
  $\API{\SUN,\Pow{m}}$ reach this dimension only if $m \ge N$. Below
  that threshold not all of the $\PI{\SUN,\Pow{m}}$ act as linearly
  independent maps on $\Pow{m}$ so that the dimension of
  $\API{\SUN,\Pow{m}}$ is smaller than $m!$. In this sense we say that
\begin{equation}
  \label{eq:maxdim}
  \text{max}(\text{dim}(\API{\SUN,\Pow{m}})) = m! 
\end{equation}
An analogous situation arises for $\API{\SUN,\MixedPow{m}{n}}$, which
also reaches is maximal dimension with $m \ge N$.

\end{enumerate}

\subsection{Orthogonal bases for \texorpdfstring{$\API{\SUN,\MixedPow{m}{n}}$}{APIij} via Clebsch Gordan coefficients}

\begin{sloppypar}
  We denote a general Clebsch-Gordan operator that implements the
  projection and basis change from a product of irreducible
  representations labelled $q_1,\ldots,q_m,\bar{q}_1,\ldots,\bar{q}_n$
  (with states labelled by
  $k_1,\ldots,k_m,\bar{k}_1,\ldots,\bar{k}_n$) into an irreducible
  representation labelled by $\lambda$ (where $\lambda$ stands in for
  an irreducible representation corresponding to a
  Littlewood-Richardson tableau, its states labelled by
  $\kappa$)~\cite{Tung:1985na}, by
  $C_{\lambda\kappa;q_{1\ldots m} k_{1\ldots m} \bar{q}_{1\ldots n}
    \bar{k}_{1\ldots n}}$,
\end{sloppypar}
\begin{align}
  \label{eq:Clebsch1}
  C_{\lambda\kappa;q_{1\ldots m} k_{1\ldots m} \bar{q}_{1\ldots n} \bar{k}_{1\ldots n}} 
  & =
    \scalebox{0.9}{$%
  \ket{\lambda,\kappa}
  \overbrace{
    \langle
    \lambda,\kappa
    \vert 
     q_1, k_1
    \rangle
    \ldots
    \ket{q_m, k_m}
    \ket{\bar{q}_1,\bar{k}_1}
    \ldots
    \ket{\bar{q}_n,\bar{k}_n}
    }^{%
    \mathfrak{C}_{\lambda\kappa;q_{1\ldots m} k_{1\ldots m} \bar{q}_{1\ldots n} \bar{k}_{1\ldots n}}}%
    \bra{q_1, k_1} 
    \ldots
    \bra{q_m, k_m}
    \bra{\bar{q}_1,\bar{k}_1}
    \ldots
    \bra{\bar{q}_n,\bar{k}_n} 
    $}
    \nonumber \\
  & =: \; 
  \scalebox{0.9}{$\ket{\lambda,\kappa}$} \;
  \FPic{mqnqbClebschKappaRepLabels} \;
  \FPic{mqnqbClebsch-Lambda}
  \FPic{mqnqbClebschqkRightLabels} \;
  \FPic{mqnqbClebschBraqkLabels}
\ .
\end{align}
The part marked by the overbrace,
\begin{equation}
  \label{eq:CGcoeff}
 \mathfrak{C}_{\lambda\kappa;q_{1\ldots m} k_{1\ldots m}
  \bar{q}_{1\ldots n} \bar{k}_{1\ldots n}} 
  = 
  \langle\lambda,\kappa\vert q_1,k_1\rangle\cdots |q_m,k_m\rangle |\Bar q_1,\Bar k_1\rangle\cdots |\Bar q_n,\Bar k_n\rangle 
\ ,
\end{equation}
is the usual Clebsch-Gordan coefficient, and the \emph{labelled}
diagram in the second line is its birdtrack
representation~\cite{Cvitanovic:2008zz,Alcock-Zeilinger:2016cva}.  The
full operator is obtained by summing over all the states and represented by an unlabelled diagram
\begin{equation}
  \label{eq:clebsch-simplify}
  C_{\lambda,m,n} :=
  \sum_{\kappa} \sum_{k_i, \bar{k}_i} 
  C_{\lambda\kappa;q_{1\ldots m} k_{1\ldots m} \bar{q}_{1\ldots n}
    \bar{k}_{1\ldots n}}
 = \;
  \FPic[scale=0.75]{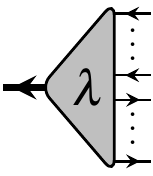}
  \ .
\end{equation}
It should be thought of as a linear map
\begin{equation}
  \label{eq:clebsch-map}
  \FPic[scale=0.75]{mqnqbClebsch-Lambda} : 
  \MixedPow{m}{n} 
  \to
  V^{(\lambda)}
  \ ,
\end{equation}
where $V^{(\lambda)}\subset V^{\otimes m}\otimes V^{* \otimes n} $
denotes the irreducible subspace associated with the representation
$\lambda$.

A familiar set of Clebsch-Gordan coefficients 
is given by the generator coefficients $[t^a]_{ik}$, which is
graphically denoted by a vertex between a solid (quark) and a dotted
(gluon) line~\cite{Cvitanovic:2008zz}, the arrow points from the right to the left matrix index on the generator:
\begin{equation}
\label{eq:tzBirdtracks}
  \frac{1}{\sqrt{2}} [t^a]_{ik} :=
  \FPic{1q1qbtaLabel}
\qquad \text{and} \qquad
  \frac{1}{\sqrt{2}} [t^b]_{lj} :=
  \FPic{1q1qbtbLabelHC}
\ .
\end{equation}
To graphically distinguish adjoint from fundamental lines, the latter are drawn as
a dotted lines.  The direction of the arrow removes any ambiguity about
the order of factors in the interpretation of the associated
Clebsch-Gordan operators $\FPic{1q1qbtz}$ and $\FPic{1q1qbtzHC}$ as
linear maps
\begin{equation}
  \label{eq:vvstartova}
  \FPic{1q1qbtz} : V\otimes V^*\to V^{(\text{adj})} \subset V\otimes V^*
\qquad \text{and} \qquad
  \FPic{1q1qbtzHC}:   V^{(\text{adj})} \to V\otimes V^* 
\ .
\end{equation}
  Another familiar set of Clebsch-Gordan coefficients are
$d^{abc}$ and $f^{abc}$, which, in birdtrack notation, are depicted by
an empty (white) circle respectively a filled (black) circle over
the gluon lines,
\begin{equation}
\label{eq:Fabc-Dabc-Birdtracks}
  d^{abc} := 
  \FPic{3gDabc2Labels}
  \qquad \text{and} \qquad
  i f^{abc} := 
  \FPic{3gFabc2Labels}
  \ .
\end{equation}
The associated Clebsch Gordan operators can be interpreted as linear
maps $V^{(\text{adj})}\otimes V^{(\text{adj})} \to V^{(\text{adj})}$.

It should be noted that for $N<3$, all $d^{abc}=0$. In particular,
we say that the coefficients $d^{abc}$ and with it the full associated
operator vanishes \emph{dimensionally}, since the dimension of the
vector space $\text{dim}(V)=N$ is too small to accommodate it; we give
a more comprehensive discussion on the conditions needed to avoid
dimensionally null operators after
eqns.~\eqref{eq:Clebsch-Tproperties}.

Likewise the $\varepsilon$-tensor with $N$ legs becomes a linear map from $\Pow{N}$ onto a singlet
\begin{equation}
  \label{eq:epsilon-Clebsch}
  \FPic[scale=0.75]{Nc_Epsilon}
\FPic[scale=0.75]{NcArrRight} : \Pow{N} \to \mathbb C
\ ,
\end{equation}
a one dimensional irreducible representation.

In the birdtrack spirit, the Hermitian conjugate of a Clebsch-Gordan
operator is given by (\emph{c.f.} eq.~\eqref{eq:Hermitian-conjugate-birdtrack-Ex})
\begin{equation}
  \label{eq:ClebschDag}
  C_{\lambda,m,n}^{\dagger} := \;
  \scalebox{0.75}{%
    \FPic{mqnqbClebschDag-Lambda}%
  }
\ .
\end{equation}
The birdtrack expression for $C_{\lambda,m,n}^{\dagger}$ is obtained
from that of $C_{\lambda,m,n}$ by reflecting it at a vertical axis
followed by a reversal of all arrows.  To illustrate this with our
earlier examples we observe that the prescription instructs us to set
\begin{equation}
\label{eq:conjClebschBirdtracks}
  \left[\FPic{1q1qbtz}\right]^\dagger
=
  \FPic{1q1qbtzHC}
\hspace{1cm}
  \left[\reflectbox{\FPic{3gFabc}}\right]^\dagger = \FPic{3gFabc}
\hspace{1cm}
\left[\reflectbox{\FPic{3gDabc}}\right]^\dagger   =  \FPic{3gDabc}
\hspace{1cm}
\left[ \;
\FPic[scale=0.75]{Nc_Epsilon}
\FPic[scale=0.75]{NcArrRight}
\; \right]^\dagger
=
\FPic[scale=0.75]{NcArrLeft}
\FPic[scale=0.75]{Nc_EpsilonDag} 
\end{equation}
This faithfully encodes hermiticity of the generators $t^{a\dagger} = t^a$,
both in the fundamental and adjoint representations (where
$[\Tilde t^a]_{i j} = i f^{i a j}$), the symmetry of the $d^{abc}$,
and, in the last expression, we obtain an interpretation for
the definitions for the two $\varepsilon$ birdtracks from
eq.~\eqref{eq:Kronecker-Delta-Levi-Civita-Birdtrack}.

By its very nature as a linear map $\MixedPow{m}{n} \to V^{(\lambda)}$
onto an irreducible image, the Clebsch-Gordan operator translates a
product representation into its irreducible sub-block labelled by
$\lambda$, i.e.
\begin{align}
  \label{eq:transl-to-lambda}
  \FPic[scale=0.75]{mqnqbClebsch-Lambda} \;
  \FPic[scale=0.75]{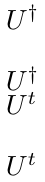}
  \; = \;
  U_{(\lambda)}
  \FPic[scale=0.75]{mqnqbClebsch-Lambda}
  \qquad \text{and} \qquad
  \FPic[scale=0.75]{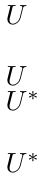}
  \FPic[scale=0.75]{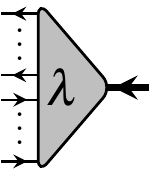}
  \; = \;
  \FPic[scale=0.75]{mqnqbClebschDag-Lambda} \;
  U_{(\lambda)}^{\dagger}
\end{align}
for all $U^{\dagger},U\in\SUN$ (or, more pendantically, in its (anti-)
fundamental representation) and $U_{(\lambda)}$ in the representation
$\lambda$ of $\SUN$. Furthermore, by definition, these new states are
chosen to be orthonormal,
\begin{equation}
  \label{eq:ClebschOn}
  \InnerProd{\lambda, \kappa}{\lambda', \kappa'}
  \; = \;
  \scalebox{0.75}{%
\FPic{mqnqbClebschKappaRepLabels}\;%
\FPic{mqnqbClebsch-orthogonal-LambdaLambdap}\;%
\FPic{mqnqbClebschKappapRepLabels}%
} 
\; = \;
\delta_{\lambda,\lambda'} \delta_{\kappa,\kappa'}
\ .
\end{equation}
Eq.~\eqref{eq:transl-to-lambda} guarantees that this statement remains
invariant under the group action.

Let $\lambda$ and $\lambda'$ denote two equivalent irreducible
representations of $\SUN$~\cite{Tung:1985na}, i.e. representations for
which there exists an isomorphism $\mathcal{S}_{\lambda\lambda'}$,
such that
$U_{(\lambda)} = \mathcal{S}_{\lambda\lambda'} U_{(\lambda')}
\mathcal{S}_{\lambda\lambda'}^{-1}$
for all $U\in\SUN$, as linear maps acting on their respective
domains. (If $\lambda =\lambda'$, then $\mathcal{S}_{\lambda\lambda'}=\mathcal{S}_{\lambda\lambda}$
becomes the identity map.) This, of course, implies that there exists
a pair of bases for $\lambda$ and $\lambda'$ such that the matrix
representations of $U_{(\lambda)}$ and $U_{(\lambda')}$ become
identical. Since the bases we work with are by definition orthonomal,
one may, on the level of matrix representations, interpret
$\mathcal{S}_{\lambda\lambda'}$ as a \emph{unitary} change of basis
matrix that synchronizes the basis choice for $\lambda$ and $\lambda'$
in this sense.  $\mathcal{S}_{\lambda\lambda'}$
can then be
used to construct linear maps on $\MixedPow{m}{n}$ from the
Clebsch-Gordan operators as follows:

\paragraph{Projection operators:} If $\lambda=\lambda'$, we define the
projection operators $P_{\lambda}$ through
\begin{subequations}
\label{eq:Clebsch-ProjOps-TransOps}
\begin{equation}
  \label{eq:Clebsch-ProjOps}
P_\lambda \; := \;
\sum\limits_\kappa \ket{\lambda,\kappa}\bra{\lambda,\kappa}
\; = \; 
C_{\lambda,m,n}^{\dagger} \mathcal{S}_{\lambda\lambda} C_{\lambda,m,n} 
\; = \; 
\scalebox{0.75}{%
\FPic{mqnqbClebschProjOps-Lambda}%
}
\ ,
\end{equation}
and general theory assures us that these yield projectors on all
irreducible subspaces contained in
$\MixedPow{m}{n}$~\cite{Tung:1985na} and accordingly provide a
decomposition of unity, $\sum\limits_\lambda P_\lambda = \mathbb 1$,
on $\MixedPow{m}{n}$. Clearly, the $P_\lambda$ are elements of the
algebra $\API{\SUN,\MixedPow{m}{n}}$ due to
eq.~\eqref{eq:transl-to-lambda}.

\paragraph{Transition operators:} 
If $\lambda\neq\lambda'$ (but $\lambda$ and $\lambda'$ are
equivalent), then the operator $T_{\lambda\lambda'} $ defined as
\begin{equation}
  \label{eq:Clebsch-TransOps}
  T_{\lambda\lambda'} 
  \; := \;
  \sum\limits_\kappa \ket{\lambda,\kappa}\bra{\lambda',\kappa}
  \; = \;
  C_{\lambda,m,n}^{\dagger} \mathcal{S}_{\lambda\lambda'} C_{\lambda',m,n} 
\; = \; 
  \scalebox{0.75}{%
  \FPic{mqnqbClebschTransOps-Lambda}%
}
\end{equation}
\end{subequations}
is called the \emph{transition
  operator}~\cite{Alcock-Zeilinger:2016cva} between $P_{\lambda}$ and
$P_{\lambda'}$. 

The combined set of operators~\eqref{eq:Clebsch-ProjOps-TransOps},
called the \emph{projector basis}\label{def:projector-basis} (this
name is justified below), satisfies
\begin{equation}
  \label{eq:Clebsch-Proj-orth}
  P_{\lambda} P_{\lambda'} = \delta_{\lambda\lambda'}
  P_{\lambda}
\end{equation}
and
  \begin{subequations}
    \label{eq:Clebsch-Tproperties}
    \begin{align}
      \label{eq:Clebsch-TP}
      & 
      T_{\lambda \lambda'}P_{\lambda'} 
      = T_{\lambda\lambda'} = P_{\lambda}T_{\lambda \lambda'} 
      \\
      \label{eq:Clebsch-Tunitary}
      & T_{\lambda\lambda'}^\dagger = T_{\lambda'\lambda} \\
      \label{eq:Clebsch-TTinv}
     &  T_{\lambda\lambda'} T_{\lambda'\lambda} = P_{\lambda}
       \ ;
    \end{align}
  \end{subequations}
  these properties are a consequence of the orthogonality of
  Clebsch-Gordan operators eq.~\eqref{eq:ClebschOn} (or, equivalently,
  Schur's Lemma~\cite{Schur:1901}). Since the
  operators~\eqref{eq:Clebsch-ProjOps-TransOps} are elements of
  $\API{\SUN,\MixedPow{m}{n}}$, they can be decomposed into linear
  combinations of the primitive invariants in $S_{m,n}$. Therefore,
  there are at most
  $\text{max}(\text{dim}(\API{\SUN,\MixedPow{m}{n}})) = |S_{m,n}| =
  (m+n)!$ of them (\emph{c.f.} eq.~\eqref{eq:maxdim}).

  If $\text{dim}(V)=N<m+n$, not all elements of $S_{m,n}$ are linearly
  independent (as maps on $\MixedPow{m}{n}$). This is reflected by the
  fact that some Clebsch-Gordan operators are represented by algebraic
  expressions that are strictly zero if $N$ is smaller than some
  threshold value $N_{\lambda,\MixedPow{m}{n}}$. The most familiar
  example for this are probably the $d^{a b c}$ coefficients, which
  vanish for $N < 3$.

  Since we are interested in keeping $N$ a parameter we include such
  Clebsch Gordan operators in our list so that the formal set of
  operators in $\API{\SUN,\MixedPow{m}{n}}$ we construct from them
  will always contain $(m+n)!$ operators, just like our generating set
  for $S_{m,n}$. Abusing nomeclature somewhat, we call this set the
  extended projector basis. The abuse of course lies in the fact that
  some operators in the combined
  set~\eqref{eq:Clebsch-ProjOps-TransOps} may act as the zero map if
  $N < N_{\lambda,\MixedPow{m}{n}}$ --- these operators are said to be
  \emph{dimensionally
    null}.\footnote{\emph{C.f.}~\cite{Alcock-Zeilinger:2016bss,
      Alcock-Zeilinger:2016cva} for a further discussion on
    dimensionally null operators.} It is important to remember that,
  for \emph{all} the operators~\eqref{eq:Clebsch-ProjOps-TransOps} to
  be dimensionally nonzero, we require $\text{dim}(V)=N\geq m+n$. In
  section~\ref{sec:Singlet-Examples} we will revisit the subject of
  dimensionally null operators in light of the singlet projectors to
  be constructed in
  section~\ref{sec:mqmqb-bending-basis-elements}. There, we will
  suggest a particular basis for the singlets that makes the
  identification of dimensionally null operators especially simple,
  \emph{c.f.}  Theorem~\ref{thm:practical-singlet-construction}.

  Unlike the projection operators $P_\lambda$ constructed
  in~\eqref{eq:Clebsch-ProjOps}, the transition operators
  $T_{\lambda\lambda'}$ in~\eqref{eq:Clebsch-TransOps} are clearly not
  Hermitian. However, they inherit a notion of unitarity (if properly
  restricted) from the underlying $\mathcal S_{\lambda\lambda'}$,
  since, $T_{\lambda\lambda'}$ reduces to
  $\mathcal S_{\lambda\lambda'}$ if restricted onto the target spaces
  of the representations $\lambda$ and $\lambda'$. This is conveniently reflected in
  the notation: It follows immediately from their definition in terms
  of the Clebsch-Gordan states~\eqref{eq:Clebsch-TransOps} that
\begin{align}
  \label{eq:T-not-herm}
  (T_{\lambda'\lambda})^\dagger = T_{\lambda\lambda'}
  \ .
\end{align}
By definition, these operators simply embed the equivalence
isomorphisms $\mathcal{S}_{\lambda\lambda'}$ into the algebra of
invariants $\API{\SUN,\MixedPow{m}{n}}$.

\begin{sloppypar}
  By their very definition, the dimensionally-non-null Clebsch-Gordan
  coefficients give a \emph{complete set of
    states} translating the product representation into the
  representation $\lambda$ (see, for
  example,~\cite[Thm.~3.12]{Tung:1985na}). Furthermore, the projector
  basis gives all nonzero combinations of Clebsch-Gordan operators of
  the form
  $C_{\lambda,m,n}^{\dagger}\mathcal{S}_{\lambda\lambda'}
  C_{\lambda',m,n}:\MixedPow{m}{n}\to\MixedPow{m}{n}$.  Since these
  are necessarily invariants of $\SUN$,
\begin{quote}
  the projection and transition operators exhaust the algebra of
  invariants and thus constitute a basis for
  $\API{\SUN,\MixedPow{m}{n}}$,
\end{quote}
justifying the name ``projector basis''. Unlike the generating sets
for $S_{m,n}$ and $S_{m+n}$, which are in $1$-to-$1$ correspondence so
that $\vert S_{m,n}\vert=\vert S_{m+n}\vert=(m+n)!$, the projector
basis exposes the dimensionally null operators: The set of
dimensionally null Clebsch-Gordan operators lead to elements in the
list of operators~\eqref{eq:Clebsch-ProjOps-TransOps} that are equally
dimensionally null, but can be uniquely related to linear combinations
of the generating sets $S_{m,n}$. The statement that these linear
combinations vanish as linear maps on $\MixedPow{m}{n}$ identifies the
full set of dimensionally null linear combinations.

Since the generating sets of the algebras $\API{\SUN,\MixedPow{m}{n}}$
and $\API{\SUN,\Pow{(m+n)}}$, $S_{m,n}$ and $S_{m+n}$ respectively,
are in $1$-to-$1$ correspondence (\emph{c.f.}
eq.~\eqref{eq:S3-to-S2plus1-map}), it follows that for $N\ge n+m$
\begin{equation}
  \text{dim}(\API{\SUN,\MixedPow{m}{n}}) = \vert S_{m,n}\vert = \vert
S_{m+n}\vert=(m+n)!
\end{equation}
(\emph{c.f.}~\cite{Alcock-Zeilinger:2016cva} for
the quark-only counterpart).
\end{sloppypar}

The observant reader will have noticed that all conclusions other than
the discussion of dimensionally null operators could have been drawn
from Schur's Lemma~\cite{Schur:1901}
(see~\cite{Sagan:2000,KosmannSchwarzbach:2000} and other standard
textbooks) --- this is, for example, done
in~\cite{Alcock-Zeilinger:2018}. However, in order to fully
justify the graphical singlet construction algorithm given in
eq.~\eqref{eq:Singlet-algorithm-schematic}, the explicit formulation
through Clebsch-Gordan operators (in the birdtrack formalism, as given
here) seems more intuitive than the abstract results derived from
Schur's Lemma, as will become clear in
section~\ref{sec:mqmqb-bending-basis-elements}.

\section{Singlets}\label{sec:Singlets}

We will now present a construction algorithm for the singlet
projection operators of $\SUN$ on a mixed quark-antiquark Fock space component
$\MixedPow{m}{n}$. By definition, a singlet space in $\MixedPow{m}{n}$
is an irreducible subspace on which the product representation acts
trivially (i.e. it acts as the identity map). The latter condition
implements the physical idea of an uncharged state, in QCD this refers
to global color neutrality: it states that the $U_{(\lambda)}$ in
\begin{equation}
  \label{eq:singlet}
   \FPic[scale=0.75]{mqnqbClebschUUComLabels}
  \FPic[scale=0.75]{mqnqbClebschDag-Lambda}
  \; = \;
  \FPic[scale=0.75]{mqnqbClebschDag-Lambda} \;
  U_{(\lambda)}^\dagger
\end{equation}
is simply the unit matrix. Irreducibility then requires the dimension
of the subspace of the representation $\lambda$ to be one. As a
consequence we may omit the state label leg on the right altogether --
we are confronted with a representation that consists of a single
\emph{invariant state}:
\begin{equation}
  \label{eq:invariant-state}
  \FPic[scale=0.75]{mqnqbClebschUUComLabels}
\FPic[scale=0.75]{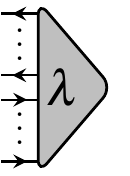}
\; = \;
\FPic[scale=0.75]{mqnqbClebschSinglet-Lambda}
\ .
\end{equation}
The states $\FPic{1q1qbTr12SStateArr} \in V\otimes V^*$, and
$\FPic{1qb1qTr12SStateArr} \in V^*\otimes V$ provide elementary
examples.

With this notation, singlet projection operators always split: For a singlet, there exists a birdtrack
representation which factorizes into disconnected left and right hand
sides in the form
\begin{equation}
\label{eq:Clebsch-Singlet-mqnbb}
\FPic[scale=0.75]{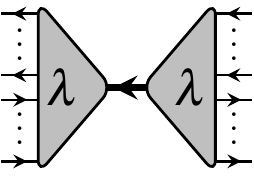}%
\quad \xrightarrow{\text{singlet rep.}} \quad
\FPic[scale=0.75]{mqnqbClebschSinglet-Lambda}%
\FPic[scale=0.75]{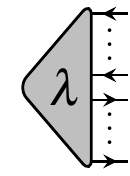}%
\ ,
\end{equation}
the two factors being conjugates of each other.

As with any irreducible representation in $\MixedPow{m}{n}$, the
projectors onto singlet representations may be dimensionally zero, but
even once they turn on, we need to distinguish two ``types'' of
singlets in $\MixedPow{m}{n}$:
\begin{itemize}
\item \emph{Generic singlets}, which turn on at some threshold value
  of $N$ and remain singlets for all larger values of $N$, and
\item \emph{Non-generic singlets}, which turn on at some threshold
  value of $N$ and turn into higher dimensional irreducible
  representations as we increase $N$ further. These singlets are in a
  sense transient phenomena.
\end{itemize}

The prototypical case of a non-generic singlet appears as we vary $N$
for the totally antisymmetric irreducible representation in $\Pow{p}$:
Its projector is dimensionally zero for $N < p$, and
switches on at $N=p$. The dimension of its associated subspace
\begin{equation}
  \label{eq:VpASdef}
  \Pow{p}\bigr\vert_{\text{AS}} :=  
    \FPic[scale=0.75]{NcArrLeft}
    \FPic[scale=0.75]{NcASym1-to-Nc-fat}
    \FPic[scale=0.75]{NcArrRight}
    \Pow{p}
    \ ,
\end{equation}
immediately follows from~\eqref{eq:trace-id} with $k=p$:
\begin{equation}
  \label{eq:dimAS}
  \text{dim}(\Pow{p}\bigr\vert_{\text{AS}})
  =
  \text{tr}\left(
    \FPic[scale=0.75]{NcArrLeft}
    \FPic[scale=0.75]{NcASym1-to-Nc-fat}
    \FPic[scale=0.75]{NcArrRight}
    \;
    \FPic[scale=0.75]{NcLabels-ap-Right}
  \right)
  =
  \binom{N}{p}
  =
  \frac{N!}{p!(N-p)!}
  =
  \frac{N(N-1)(N-2)\cdots (N-p+1)}{p!}
  \ .
\end{equation}
This is equal to zero for all $N < p$ where the operator is
dimensionally zero, equal to one for $p = N$ and strictly larger than
one for $N > p$. The irreducible multiplet is a singlet only for
$p = N$, and thus not generic in the the sense described
earlier. Correspondingly, the associated projector splits, precisely and only
at $N=p$, as already stated in eq.~\eqref{eq:Epsilon-lengthN-ASym}
\begin{equation}
\label{eq:Epsilon-lengthN-ASym-2}
\FPic[scale=0.75]{NcArrLeft}
\FPic[scale=0.75]{NcASym1-to-Nc-fat}
\FPic[scale=0.75]{NcArrRight}
\;
  \FPic[scale=0.75]{NcLabels-ap-Right}
\; \xlongequal{p=\text{dim}(V)=N} \; 
\FPic[scale=0.75]{NcArrLeft}
\FPic[scale=0.75]{Nc_EpsilonDag} 
\hspace{2mm}
\FPic[scale=0.75]{Nc_Epsilon}
\FPic[scale=0.75]{NcArrRight}
\;
\FPic[scale=0.75]{N_aLabelsRight}
\ .
\end{equation}

As it turns out, the singlets in $\MixedPow{m}{n}$ appear as
\emph{generic singlets} if $n=m$,
(section~\ref{sec:mqmqb-bending-basis-elements}). If $m\neq n$ the
only singlets that can appear are \emph{non-generic singlets}
(section~\ref{sec:mqnqb-no-new-multiplets}). In fact, the splitting
relation~\eqref{eq:Epsilon-lengthN-ASym} provides a ``canonical''
isomorphism between all singlets on a mixed product space
$\MixedPow{m}{n}$ (where $m\neq n$) into singlets on
$\MixedPow{\alpha}{\alpha}$ for some natural number $\alpha$, see
Theorem~\ref{thm:NcDepSingletEquivalence}.

We first present the general construction algorithm for generic
singlet states of $\SUN$ on $\MixedPow{m}{m}$, which was already
alluded to in eq.~\eqref{eq:Singlet-algorithm-schematic}. The
treatment we present below is a direct generalization of that given
in~\cite{Alcock-Zeilinger:2016cva} for the Clebsch-Gordan operators on
$\Pow{m}$. For a more comprehensive discussion of the simpler case
using different methods, readers are referred to~\cite[in
German]{Clebsch1866theorie} or~\cite{Tung:1985na} for a more modern
treatment.

\subsection{Singlets for an equal number of ``quarks and antiquarks'': bending basis elements}\label{sec:mqmqb-bending-basis-elements}

Our goal is to identify a complete set of linearly independent
invariant states in $\MixedPow{k}{k}$, i.e. states in
$\MixedPow{m}{m}$ that satisfy~\eqref{eq:invariant-state}.

The fact that the the full set of operators in $\API{\SUN,\Pow{k}}$ is
invariant and the logic that connects singlet states $V\otimes V^*$
with invariant maps in $\Lin{V}$
(c.f. eqns.~\eqref{eq:IndexLeg-Arrow-Transformation}
and~\eqref{eq:VVstar-VstarV-action}) readily allows us to interpret
$\API{\SUN,\Pow{k}}$ as the subspace of singlet states in $\MixedPow{k}{k}$.

Moreover, since
\begin{equation}
  \label{eq:singlets-VmV*n-isomorphic spaces}
  \MixedPow{m}{n}\otimes\MixedPow{n}{m}
  \cong
  \MixedPow{(m+n)}{(m+n)}
\end{equation}
by a simple reordering isomorphism, one can start the process from any
$\API{\SUN,\MixedPow{m}{n}}$ as long as the numbers add up, i.e. as
long as $m+n=k$. To emphasize that the distribution of quark and
antiquark legs is irrelevant we will frame our statements in terms of
$\API{\SUN,\MixedPow{m}{n}}$ in this section.

In birdtracks, the process is a simple graphical reshaping of the
diagrammatic representations 
that was
alluded to in eq.~\eqref{eq:Singlet-algorithm-schematic}: Starting from
an operator $T_{\lambda\lambda'}$\footnote{We allow for $\lambda$ and
  $\lambda'$ to be equal, in which case
  $T_{\lambda\lambda'}\xrightarrow{\lambda=\lambda'}P_{\lambda}$.} and
labelling the fundamental lines as $q,p$ and the antifundamental lines
as $\bar{q},\bar{p}$ for clarity, we obtain the following state
\begin{equation}
  \label{eq:Clebsch-mplusn-StingletState}
\scalebox{0.75}{%
\FPic{mqnqbClebschpLabels}
\FPic{mqnqbClebschDag-Lambda} 
\FPic{mqnqbClebsch-Lambdap}
\FPic{mqnqbClebschqLabels}%
} 
\quad \xlongrightarrow{\text{reshape}} \quad 
\scalebox{0.75}{%
\FPic{mqnqbClebschSLabels}%
}
\;
\scalebox{0.75}{%
\FPic{mqnqbClebschSState-Lambda}%
}
\; =: 
\ket{\mathfrak{m}_{\lambda\lambda'}}
\ .
\end{equation}
\begin{sloppypar}
  Due to the reshaping process the quark lines $q_1\ldots q_m$ have
  become antiquark lines, and similarly the antiquark lines
  $\bar{q}_1\ldots\bar{q}_n$ have become quark lines in that their
  transformation behaviour changed (\emph{c.f.}
  eqns.~\eqref{eq:IndexLeg-Arrow-Transformation}): The index lines
  $q_1\ldots q_m$ transformed as quark lines in the operators
  $T_{\lambda\lambda'}$ but now transform as antiquark lines after the
  reshaping procedure, and similarly for the index lines labelled
  $\bar{q}_1\ldots\bar{q}_n$. Hence, the
  states~\eqref{eq:Clebsch-mplusn-StingletState} are elements in the
  space $\MixedPow{m}{n}\otimes\MixedPow{n}{m}$ which is isomorphic to
  $\MixedPow{(m+n)}{(m+n)}$ --- due to
  eq.~\eqref{eq:singlets-VmV*n-isomorphic spaces} we will often merely
  say that the states~\eqref{eq:Clebsch-mplusn-StingletState} are
  elements of $\MixedPow{(m+n)}{(m+n)}$. Lastly, by the completeness
  of Clebsch-Gordan operators~\cite{Tung:1985na}, the
  construction~\eqref{eq:Clebsch-mplusn-StingletState} exhausts all
  possible linearly independent singlet states of $\SUN$ on
  $\MixedPow{(m+n)}{(m+n)}$, and thus spans the space of singlet
  states.
\end{sloppypar}%

The singlet states~\eqref{eq:Clebsch-mplusn-StingletState} can be used
to construct the singlet projection operator
$P^{S}_{\lambda\lambda'}$ (which lies in the algebra of invariants
$\API{\SUN,\MixedPow{(m+n)}{(m+n)}}$) as
\begin{subequations}
  \label{eq:Clebsch-Singlet-ProjOps}
\begin{equation}
  \label{eq:Clebsch-Singlet-ProjOps-beta-undefined}
  P^{S}_{\lambda\lambda'} \; := 
 \ket{\mathfrak{m}_{\lambda\lambda'}}\bra{\mathfrak{m}_{\lambda\lambda'}}
  \; = \; 
  \beta_{\lambda\lambda'} \cdot \;
  \FPic[scale=0.5]{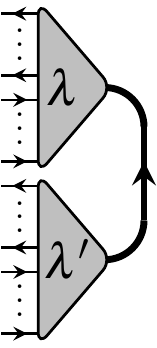}%
  \;\; 
  \FPic[scale=0.5]{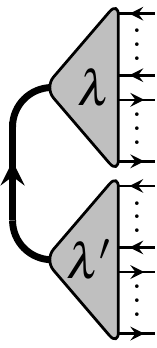}%
  \ ,
\end{equation}
where $\beta_{\lambda\lambda'}$ is a constant uniquely defined by
requiring $P^{S}_{\lambda\lambda'}\cdot
P^{S}_{\lambda\lambda'}=P^{S}_{\lambda\lambda'}$,
\begin{equation}
  \label{eq:Clebsch-Singlet-ProjOps-beta}
\beta_{\lambda\lambda'} := \; 
\left(
\scalebox{0.5}{%
\FPic{mqnqbClebschSStateDag-Lambda}%
\FPic{mqnqbClebschSState-Lambda}%
}
\right)^{-1}
\; = \;
\left[
 \Tr{\scalebox{0.5}{%
\FPic{mqnqbClebschSState-Lambda}%
}
 \;\; 
\scalebox{0.5}{%
\FPic{mqnqbClebschSStateDag-Lambda}%
}}
\right]^{-1}
\ ,
\end{equation}
\end{subequations}
unless the state $\ket{\mathfrak{m}_{\lambda\lambda'}}=0$ in which
case we define $\beta_{\lambda\lambda'}:=0$ (recall that
$\ket{\mathfrak{m}_{\lambda\lambda'}}=0$ can only occur if $\text{dim}(V)=N<m+n$). The operators~\eqref{eq:Clebsch-Singlet-ProjOps-beta-undefined} are
\emph{singlet projectors} satisfying
\begin{equation}
  \label{eq:singlet-invariant-U}
\mathbf{U} \cdot P^{S}_{\lambda\lambda'}
\; = \;
P^{S}_{\lambda\lambda'}
\; = \;
P^{S}_{\lambda\lambda'} \cdot \mathbf{U}^{\dagger}
\ ,
\qquad
\mathbf{U}:=
\underbrace{U\otimes\ldots\otimes U}_{\text{$m$ times}}
\otimes
\underbrace{U^{\dagger}\otimes\ldots\otimes U^{\dagger}}_{\text{$n$ times}}
\otimes
\underbrace{U\otimes\ldots\otimes U}_{\text{$n$ times}}
\otimes
\underbrace{U^{\dagger}\otimes\ldots\otimes U^{\dagger}}_{\text{$m$ times}}
\ ;
\end{equation}
this is an immediate consequence of
eq.~\eqref{eq:Clebsch-Singlet-mqnbb}. Thus, unless
$\ket{\mathfrak{m}_{\lambda\lambda'}}=0$ (in which case
$P^{S}_{\lambda\lambda'}$ projects onto a dimensionally null
representation), eq.~\eqref{eq:Clebsch-Singlet-ProjOps-beta} ensures
that $\text{dim}(P^{S}_{\lambda\lambda'})=1$.\footnote{We emphasize
    that $\beta_{\lambda\lambda'}$ is defined such that
    $P^{S}_{\lambda\lambda'}\cdot
    P^{S}_{\lambda\lambda'}=P^{S}_{\lambda\lambda'}$
    is true, \emph{not} to ensure $\Tr{P^{S}_{\lambda\lambda'}}=1$:
    while it is always possible to force the trace of a projection
    operator $O$ to $1$ by fixing its normalization constant, this
    constant may not be the correct one rendering $O$ idempotent.}


The projection operators $P^S_{\lambda\lambda'}$ on
$\MixedPow{(m+n)}{(m+n)}$ are clearly orthonormal from
eq.~\eqref{eq:ClebschOn}. Furthermore, we note that we have not fixed
$\text{dim}(V)=N$ to a particular value in our considerations so far,
but have rather kept it as a parameter. Thus, the
projector~\eqref{eq:Clebsch-Singlet-ProjOps} is either a
dimensionally null projector (which can only occur for $N<m+n$) or a
singlet projector for all $N\geq m+n$, inspiring us to call
$P^S_{\lambda\lambda'}$ a \emph{generic singlet}.

Lastly, we notice that the singlet projection
operators~\eqref{eq:Clebsch-Singlet-ProjOps} all correspond to
\emph{equivalent} irreducible representations of $\SUN$, since we can
explicitly construct the transition operators between them: Consider two singlet
projection operators 
\begin{equation}
  \label{eq:TwoClebsch-Singlet-ProjOps-lambda-xi}
  P^{S}_{\lambda\lambda'} = \beta_{\lambda\lambda'} \cdot
 \ket{\mathfrak{m}_{\lambda\lambda'}}\bra{\mathfrak{m}_{\lambda\lambda'}}
\qquad
\text{and}
\qquad
 P^{S}_{\xi\xi'} = \beta_{\xi\xi'} \cdot
 \ket{\mathfrak{m}_{\xi\xi'}}\bra{\mathfrak{m}_{\xi\xi'}}
\ ,
\end{equation}
where $\beta_{\lambda\lambda'}$ and $\beta_{\xi\xi'}$ are defined
according to eq.~\eqref{eq:Clebsch-Singlet-ProjOps-beta}. The object
$T^S_{\lambda\lambda',\xi\xi'}$ defined as
\begin{equation}
  \label{eq:Clebsch-Singlet-TransOps}
T^S_{\lambda\lambda',\xi\xi'}
:=
\sqrt{\beta_{\lambda\lambda'}\beta_{\xi\xi'}} \cdot 
\ket{\mathfrak{m}_{\lambda\lambda'}}\bra{\mathfrak{m}_{\xi\xi'}}
=
\sqrt{\beta_{\lambda\lambda'}\beta_{\xi\xi'}} \cdot \;
\scalebox{0.5}{%
\FPic{mqnqbClebschSState-Lambda}%
}
 \;\; 
\scalebox{0.5}{%
\FPic{mqnqbClebschSStateDag-Xi}%
}
\end{equation}
is the unique transition operator between $P^S_{\lambda\lambda'}$ and
$P^S_{\xi\xi'}$ as it satisfies the defining properties of transition
operators (\emph{c.f.} eqns.~\eqref{eq:Clebsch-Tproperties})
  \begin{subequations}
    \label{eq:Clebsch-Singlet-Tproperties}
    \begin{align}
      \label{eq:Clebsch-Singlet-TP}
      & 
      T^S_{\lambda\lambda',\xi\xi'} \cdot P^S_{\xi\xi'} 
      = T^S_{\lambda\lambda',\xi\xi'} = P^S_{\lambda\lambda'} \cdot T^S_{\lambda\lambda',\xi\xi'} 
      \\
      \label{eq:Clebsch-Singlet-Tunitary}
      & \left(T^S_{\lambda\lambda',\xi\xi'}\right)^\dagger = T^S_{\xi\xi',\lambda\lambda'} \\
      \label{eq:Clebsch-Singlet-TTinv}
      & T^S_{\lambda\lambda',\xi\xi'} \cdot T^S_{\xi\xi',\lambda\lambda'} = P^S_{\lambda \lambda'}
       \ ;
    \end{align}
  \end{subequations}
  this is again an immediate consequence of
  eq.~\eqref{eq:ClebschOn}. Note that, if one of the representations
  corresponding to $P^S_{\lambda\lambda'}$ and $P^S_{\xi\xi'}$ is
  dimensionally null, then $T^S_{\lambda\lambda',\xi\xi'}=0$ (as one
  of the states $\ket{\mathfrak{m}_{\lambda\lambda'}}$ or
  $\bra{\mathfrak{m}_{\xi\xi'}}$ vanishes). Let us summarize:

\begin{theorem}[Generic singlets \& singlet count]\label{thm:N-independent-Singlets}
  Consider the irreducible representations of $\SUN$ on a product
  space $\MixedPow{k}{k}$. If $N=\mathrm{dim}(V)\geq k$, there exist
  exactly $k!$ singlet states $\ket{\mathfrak{m}_{\lambda\lambda'}}$
  satisfying
  \begin{equation}
    \label{eq:SingletStates-no-color-rotation}
    \mathbf{U} \ket{\mathfrak{m}_{\lambda\lambda'}} = \ket{\mathfrak{m}_{\lambda\lambda'}}
    \quad {\text{and}} \quad
    \bra{\mathfrak{m}_{\lambda\lambda'}} \mathbf{U}^{\dagger} = \bra{\mathfrak{m}_{\lambda\lambda'}}
    \ ,
  \end{equation}
  where $\mathbf{U}$ is a tensor product defined as
  $\mathbf{U}:=U^{\otimes k}\otimes\left(U^{\dagger}\right)^{\otimes
    k}$,
  and $U\in\SUN$ is arbitrary. These singlet states are obtained from
  reshaping the basis elements of $\API{\SUN,\MixedPow{m}{n}}$ as
  described in eq.~\eqref{eq:Clebsch-mplusn-StingletState}, where
  $m,n$ are any non-negative integers satisfying $m+n=k$. The
  projection operators $P^S_{\lambda\lambda'}$ onto the singlet representations
  of $\SUN$ on $\MixedPow{k}{k}$ are given by
  \begin{equation}
    \label{eq:Singlet-ProjOps-Theorem}
    P^S_{\lambda\lambda'} := 
    \frac{
      \ket{\mathfrak{m}_{\lambda\lambda'}}\bra{\mathfrak{m}_{\lambda\lambda'}}
    }{
      \InnerProd{\mathfrak{m}_{\lambda\lambda'}}{\mathfrak{m}_{\lambda\lambda'}}
    }
  \end{equation}
  If $N=\mathrm{dim}(V)<k$, then several states
  $\ket{\mathfrak{m}_{\lambda\lambda'}}$ become zero, causing the
  associated projection operator $P^S_{\lambda\lambda'}$ to
  correspond to a (dimensionally) null representation of $\SUN$.

  Furthermore, all $1$-dimensionel irreducible representations of
  $\SUN$ on the product space $\MixedPow{k}{k}$ are equivalent; the
  transition operator $T^S_{\lambda\lambda',\xi\xi'}$ between the
  singlet projectors 
  $P_{\lambda\lambda'}^S$ and $P_{\xi\xi'}^S$ is given by
\begin{equation}
  \label{eq:TransitionOpsSinglet}
  T^S_{\lambda\lambda',\xi\xi'} :=
  \frac{
    \ket{\mathfrak{m}_{\lambda\lambda'}}\bra{\mathfrak{m}_{\xi\xi'}}
  }{
    \sqrt{
      \InnerProd{\mathfrak{m}_{\lambda\lambda'}}{\mathfrak{m}_{\lambda\lambda'}}
      \InnerProd{\mathfrak{m}_{\xi\xi'}}{\mathfrak{m}_{\xi\xi'}}
    }}
  \ .
\end{equation}
$T^S_{\lambda\lambda',\xi\xi'}$ is the zero map precisely when either
$P_{\lambda\lambda'}^S$ or $P_{\xi\xi'}^S$ (or both) are dimensionally
zero.

The singlet projection and transition operators form a subalgebra of
$\API{\SUN,\MixedPow{k}{k}}$ called the singlet subalgebra.
\end{theorem}

Two comments are in order:
\begin{enumerate}
\item Reshaping the projector basis (\emph{c.f.}
  page~\pageref{def:projector-basis}) generates orthogonal singlet
  projectors $P^S_{\lambda \lambda'}$, as will be discussed in
  section~\ref{sec:Singlet-Examples}. This is not automatic:
  Reshaping, for example, the primitive invariants $S_k$ spanning
  $\API{\SUN,\Pow{k}}$ does not lead to orthogonal operators.
\item In the case where several singlet projectors and all associated
  transition operators vanish dimensionally (this can only occur for
  $N<k$), a well chosen basis causes individual singlet operators to
  vanish, rather than establishing complicated constraint equations
  between the singlet projection operators. This will also be
  discussed in more detail in section~\ref{sec:Singlet-Examples}.
\end{enumerate}

\subsection{Singlets for an arbitrary number of quarks and antiquarks:
Littlewood-Richardson rule for singlets in the birdtrack formalism}
\label{sec:mqnqb-no-new-multiplets}

The singlets of $\SUN$ on $\MixedPow{k}{k}$ (where $k:=m+n$) discussed
thus far contain Kronecker~$\delta$s only since they can \emph{all} be
constructed from the (basis) elements of $\API{\SUN,\Pow{k}}$. The
second invariant of $\SUN$, the $\varepsilon$-tensor, does not play an
explicit role in this construction at all. For $\GLN$, where this
invariant is absent, this is not surprising, since each
fundamental index has to be contracted with an antifundamental index
to obtain a singlet state~\cite{Howe:1989}, but for $\SUN$ this is a
nontrivial result: all singlet states in $\MixedPow{k}{k}$ that can be
written in terms of the $\varepsilon$-tensor can be \emph{recast} in
terms of Kronecker $\delta$s entirely. If an $\varepsilon$-tensor appears in a
singlet expression with equal numbers of quarks and antiquarks, it must
appear in a pair combination that allows us to use
eq.~\eqref{eq:Epsilon-lengthN-ASym-2} to trade it for a (reshaped)
antisymmetrizer.

The role of the $\varepsilon$ invariant is more subtle: it allows singlet
representations of $\SUN$ over product spaces in which the number of
fundamental and antifundamental factors is distinct. Such singlets
contain $\varepsilon$-tensors $\varepsilon_{a_1 a_2 \ldots a_N}$ in addition to
Kronecker~$\delta$s and here they cannot be eliminated in favor of
Kronecker~$\delta$s. These singlets are non-generic or transient in the sense
that they are singlets only if $N$ is at the threshold value
$N_{\lambda,m,n}$ where this representation first appears. Moreover,
for this specific value of $N$, the $\varepsilon$-tensors provide a
canonical isomorphism onto an associated generic singlet.

However, as will be shown, no new information is produced
when including the second invariant into the projectors. This does not
come as a surprise: Due to the Leibniz identity~\cite{Jeevanjee:2015},
\begin{equation}
  \label{eq:Leibniz-Nminus1q-into-qb-Us}
  \tensor{\varepsilon}{_{b_1 b_2 \ldots b_{N}}}
  U^{\dagger}_{b_{N} a_{N}} =
  \tensor{\varepsilon}{_{a_1 a_2 \ldots a_{N}}}
  {U}_{b_1 a_1}
  {U}_{b_2 a_2} \ldots
  {U}_{b_{(N-1)} a_{(N-1)}}
  \ ,
\end{equation}
it is possible to translate $N-1$ fundamental indices into an
antifundamental one (\emph{c.f.}
appendix~\ref{sec:Singlets-Projectors-LRtableaux-Leibniz} for more
details).  Therefore, $\varepsilon_{a_1 a_2 \ldots a_N}$ can be
understood as a Clebsch-Gordan operator translating $N-1$ fundamental
index legs into an antifundamental leg,
\begin{equation}
  \label{eq:Epsilon-Nm1qb-to-1q-Clebsch}
  \FPic[scale=.75]{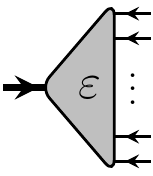}
  :=
  \FPic{nqClebsch-Epsilon}
  : V^{\otimes(N-1)} \to V^*
\end{equation}
(\emph{c.f.} eq.~\eqref{eq:Kronecker-Delta-Levi-Civita-Birdtrack} for
the graphical notation of $\varepsilon_{a_1 a_2 \ldots a_N}$).
Restricting~\eqref{eq:Epsilon-Nm1qb-to-1q-Clebsch} onto the
antisymmetrized subspace of $V^{\otimes(N-1)}$ produces an
isomorphism:
\begin{equation}
  \label{eq:Epsilon-Nm1qb-to-1q-Clebsch-restr}
   \FPic[scale=.75]{nqClebschb-Varepsilon}
  : V^{\otimes(N-1)}\bigr\vert_{\text{AS}} \xrightarrow{\cong} V^*
\hspace{1cm}\text{ since }\hspace{1cm}
  \text{dim}(\Pow{N-1}\bigr\vert_{\text{AS}})
  =
  \binom{N}{N-1} = N = \text{dim}(V^*)
  \ ,
\end{equation}
with its Hermitian conjugate acting as its inverse. I.e., in physics
parlance, we have the orthogonality and completeness relations
\begin{equation}
  \label{eq:eps-inverses}
  \FPic[scale=.75]{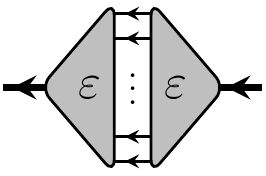}  
  \; = \;
  \mathbb 1_{V^*}
  \hspace{1cm}\text{ and } \hspace{1cm}
  \FPic[scale=.75]{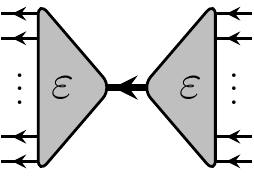} 
  \; = \;
  \mathbb 1_{\Pow{N-1}\bigr\vert_{\text{AS}}}
\end{equation}
respectively.
Therefore we have
\begin{equation}
\label{eq:Epsilon-lengthN-ASym2}
\underbrace{
\FPic[scale=0.75]{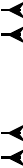}
\FPic[scale=0.75]{NcASym1-to-Nc-fat}
\FPic[scale=0.75]{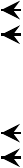}
\;
\FPic[scale=0.75]{N_aLabelsRight}
\; = 
\; 
\FPic[scale=0.75]{NcBigArrLeft}
\FPic[scale=0.75]{Nc_EpsilonDag} 
\hspace{2mm}
\FPic[scale=0.75]{Nc_Epsilon}
\FPic[scale=0.75]{NcBigArrRight}
\;
\FPic[scale=0.75]{N_aLabelsRight}
\; = \;
\FPic[scale=0.75]{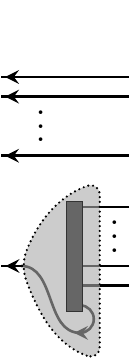}
\FPic[scale=0.75]{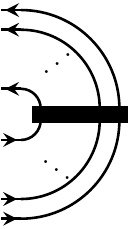}
\hspace{2mm}
\FPic[scale=0.75]{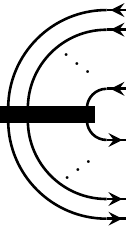} 
\FPic[scale=0.75]{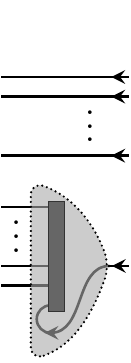}
\;
\FPic[scale=0.75]{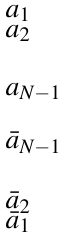}
}_{\in \API{\SUN,\Pow{N}}}
\quad \cong \quad
\underbrace{
  \vphantom{\FPic[scale=0.75]{mqmqbIdq-ClebschDagEpsilonqb}}
\FPic[scale=0.75]{mqmqbSState-TotASym}
\hspace{2mm}
\FPic[scale=0.75]{mqmqbSStateDag-TotASym} \;
\FPic[scale=0.75]{mqmqbTotASymState-aLabels-Right}
}_{\in\API{\SUN,\MixedPow{(N-1)}{(N-1)}}}
\end{equation}
Here $\cong$ indicates that this procecure induces a specific
isomorphism between two very specific singlets in $V^N$ and
$\MixedPow{(N-1)}{(N-1)}$.

The underlying ideas are of course well known: This
naturally arises from applying the Littlewood-Richardson method (see
appendix~\ref{sec:LR-Example}) to the specific case of singlets. Practical
calculations on larger product spaces suffer from the same algebraic
complexity exemplified in appendix~\ref{sec:LR-Example}.

Due to their relevance in physics applications, these ideas have been
visited and revisited many times, in particular in connection with the
large $N_c$ expansion of QCD. We will briefly comment on these
applications from our perspective in
section~\ref{sec:Baryon-Equiv-2q2qbASym} before we cast the
Littlewood-Richardson equivalence for singlets in the birdtrack formalism in
section~\ref{sec:Singlets-VmV*n-special-case}.



\subsubsection{Example: ``Baryon color'' singlet projectors at fixed
  and general \texorpdfstring{$N$}{N}}\label{sec:Baryon-Equiv-2q2qbASym}

The isomorphism of a given pair of singlet spaces spelled out in
eq.~\eqref{eq:Epsilon-lengthN-ASym2} is special in that it connects
singlets that appear as subspaces of $V^N$ and
$\MixedPow{(N-1)}{(N-1)}$ respectively. This is of relevance in many
physics applications since these spaces are typically tied to particle
content of wave functions or correlators that help identify the larger
spaces into which the individual singlets are embedded.

\textbf{QCD Wilson line correllators at $N_c=3$:}

We now consider $N = N_c = 3$ to be the number of colors in QCD. The
singlet states and singlet projectors then refer to global color
singlets. Due to confinement all asymptotic states of the theory must
live in the color singlet subspace of the theory, but the microscopic
particle content is by no means given simply in terms of a fixed
number of quarks, antiquarks, and gluons. Still, perturbation theory
and factorization arguments often isolate certain Fock space
components with fixed numbers of legs that can be readily interpreted
in terms of birdtrack diagrams. For example, observables accessed in
high energy collider experiments in the context of the Regge-Gribov
limit are described in terms of Wilson line correlators in a
factorzation approach that has been dubbed the Color Glass Condensate
framework. For the present purpose the Wilson lines can be thought of
as group valued fields $U : \mathbb R^2_\perp \to
\mathsf{SU}(N_c)$
(where $\mathbb R^2_\perp$ refers to the spatial directions orthogonal to the
collider axis), that, in a given Fock-space component whose color
space is $\MixedPow{m}{m}$, induce transitions between all available
global color singlet states.

For $m=1$ we have precisely one such state and one such correlator
that probes the Wilson line field $U_{\bm x}$ at $2m = 2$ positions:
\begin{equation}
  \label{eq:dipole}
  \langle
  \FPic{1q1qbTr12SStateDagArr}\,
  \FPic{1q1qbUxUDagytLabels}
\FPic{1q1qbTr12SStateArr} \rangle =  \langle \frac{\text{tr}( U_{\bm x} U^\dagger_{\bm y})}{N_c} \rangle
\end{equation}
The average indicated by $\langle\ \rangle$ is over soft gluon fields
and depends on kinematics of the observable under consideration via
the JIMWLK equation~\cite{JalilianMarian:1997jx,JalilianMarian:1996xn,Jalilian-Marian1998a,Iancu:2000hn,Ferreiro:2001qy}, but this is outside the scope of the present
discussion. What we are interested in here is solely the role played by
the singlet states.

For $m=2$ there are two states and thus $2\times 2$ correlators, that
probe the field at altogether $2m = 4$ positions:
\begin{equation}
  \label{eq:quadrupole-correlators}
 \left\langle \begin{pmatrix}
  \FPic{2q2qbSym12SStateDag}
  \FPic{2q2qbArr}\FPic{2q2qbUUDagt-x12-y12-Labels}
  \FPic{2q2qbArr}\FPic{2q2qbSym12SState}
  &
   \FPic{2q2qbSym12SStateDag}
  \FPic{2q2qbArr}\FPic{2q2qbUUDagt-x12-y12-Labels}
  \FPic{2q2qbArr}\FPic{2q2qbASym12SState}
  \\
  \FPic{2q2qbASym12SStateDag}
  \FPic{2q2qbArr}\FPic{2q2qbUUDagt-x12-y12-Labels}
  \FPic{2q2qbArr}\FPic{2q2qbSym12SState}
  & \FPic{2q2qbASym12SStateDag}
  \FPic{2q2qbArr}\FPic{2q2qbUUDagt-x12-y12-Labels}
  \FPic{2q2qbArr}\FPic{2q2qbASym12SState}
\end{pmatrix}
\right\rangle
\end{equation}
The explicit component expressions for the operator entries
\begin{equation}
  \label{eq:AAS-components}
  \begin{aligned}[t]
  [A]^{i j} = & \
  \frac{\tr(U_{\bm x_1} U^\dagger_{\bm y_2}) \tr(U_{\bm x_2} U^\dagger_{\bm y_1})}4
  + (-1)^{j-1}
  \frac{\tr(U_{\bm x_1} U^\dagger_{\bm y_1} U_{\bm x_2} U^\dagger_{\bm y_2})}4
  + (-1)^{i-1}
  \frac{\tr(U_{\bm x_1} U^\dagger_{\bm y_2} U_{\bm x_2} U^\dagger_{\bm y_1})}4
  \\ & \
  + (-1)^{i+j}
  \frac{\tr(U_{\bm x_1} U^\dagger_{\bm y_1}) \tr(U_{\bm x_2} U^\dagger_{\bm y_2})}4  
  \end{aligned}
\end{equation}
make it easy to verify that the matrix becomes diagonal in both the
limits $\bm x_1\to \bm x_2$ and $\bm y_1\to\bm y_2$.

If we probe a baryon, the leading Fock space component is formed from
$N_c = 3$ fundamental quarks. For this contribution, there is only one
possible singlet state and associated correlator, namely
\begin{equation}
  \label{eq:baryon3}
\FPic{3EpsilonDagN}
\FPic{3ArrRight}
\FPic{3U-x123-Labels}
  \FPic{3ArrLeft}
\FPic{3EpsilonN}
   = \varepsilon^{i_1 i_2 i_3}  [U_{\bm x_1}]_{i_1 j_1} [U_{\bm x_2}]_{i_2 j_2} [U_{\bm x_3}]_{i_3 j_3} \varepsilon^{j_1 j_2 j_3}
\end{equation}

The singlet to singlet relation~\eqref{eq:Epsilon-lengthN-ASym2} at $N_c = 3$,
\begin{equation}
\label{eq:ASym123-equivalent-ASym1q2qASym1qb2qb-text}
\FPic{3ArrLeft}
\FPic{3ASym123SN}
\FPic{3ArrRight}
\; = \; 
\FPic{3ArrLeft}
\FPic{3EpsilonN}\;\;
\FPic{3EpsilonDagN}
\FPic{3ArrRight}
=
\FPic{2q2qb3EpsilonDag2TopL}
\FPic{2q2qbArrL}
\FPic{2q2qbASym12SStateL}\;\; 
\FPic{2q2qbASym12SStateDagL}
\FPic{2q2qbArrL}
\FPic{2q2qb3Epsilon2TopL}
\quad 
\cong
\quad
\FPic{2q2qbArrL}
\FPic{2q2qbASym12SStateL}\;\; 
\FPic{2q2qbASym12SStateDagL}
\FPic{2q2qbArrL}
\quad
\text{for}
\quad
\text{dim}(V)=N=3
\ ,
\end{equation}
then provides a relationship between the ``baryon'' correlator
\eqref{eq:baryon3} and the $\bm y_1, \bm y_2 \to \bm x_3$ coincidence
limit of the bottom right entry in
eq.~\eqref{eq:quadrupole-correlators}, namely
\begin{equation}
  \label{eq:q2qb2baryonlimit}
  \left.\FPic{2q2qbASym12SStateDagL}
  \FPic{2q2qbArrL}
  \FPic{2q2qbUUDagt-x12-y12-LabelsL}
\FPic{2q2qbArrL}
\FPic{2q2qbASym12SStateL}
\right\vert_{\bm y_1,\bm y_2 = \bm x_3}
=
\FPic{2q2qbASym12SStateDagL}
\FPic{2q2qbArrL}
\FPic{2q2qb3Epsilon2TopL}
\FPic{2q2qbEpsilonU-x123-LabelsL}
\FPic{2q2qb3EpsilonDag2TopL}
\FPic{2q2qbArrL}
\FPic{2q2qbASym12SStateL}
=
\FPic{3EpsilonDagN}
\FPic{3ArrRight}
\FPic{3smallU-x123-Labels}
  \FPic{3ArrLeft}
\FPic{3EpsilonN}
\end{equation}
as already noted in~\cite{Marquet:2010cf}.

It is in this sense that correlators associated with transient
singlets, for any value of $N$, appear as coincidence limits of more
complicated correlators ``inside'' higher Fock space components of the
theory and, in this sense, are not ``new'' entities.

\textbf{QCD in the large $N_c$ limit:}

The situation becomes more involved if we consider taking the large
$N_c$ limit as suggested by
'tHooft~\cite{tHooft:1973alw,tHooft:1974pnl} as an alternative
approximation scheme for QCD. He argued that taking $N_c$ large (but
ultimately finite) while keeping the 'tHooft coupling
$g^2_{\text{'tHooft}} := g^2_{\text{QCD}} N_c$ constant leads to a
meaningful approximation scheme for QCD in (fractional) powers of
$1/N_c$, starting from some well defined leading order results.

In the traditional framework, quarks are taken to transform under the
fundamental representation of $\mathsf{SU}(N_c)$. The leading
Fock space component of mesons resides in the singlet subspace of
$V\otimes V^*$ and appears as $\text{tr}(q\otimes \Bar q) = q^i \Bar q^i$
while the leading Fock space contribution to baryons at $N_c=3$ is
formed according to
\begin{equation}
  \label{eq:baryon-singlet}
  \varepsilon^{i_1 i_2 i_3} q^{i_1} q^{i_2} q^{i_3}
\end{equation}
and relies on
$1 = \det(U) = \varepsilon^{i_1 i_2 i_3} U^{1 i_1}U^{2 i_2}U^{3 i_3}$ for
global color invariance.

To take the large $N_c$ limit one may leave the meson color structure
unchanged but the ``smallest'' fully antisymmetric product state of
fundamental quarks must contain $N_c$ of them to obtain an invariant:
\begin{equation}
  \label{eq:baryon-singlet-2}
  \varepsilon^{i_1 i_2 \ldots i_{N_c}} q^{i_1} q^{i_2} \cdots q^{i_{N_c}}
  \ .
\end{equation}
In birdtrack notation
\begin{equation}
  \label{eq:QCDFstates}
  N_c = 3: \text{mesons in } \FPic{1q1qbTr12SStateArr}
  \hspace{.5cm}
  \text{baryons in }  \FPic{3ArrLeft}\FPic{3EpsilonN}
  \hspace{2cm}
  N_c > 3: \text{mesons in } \FPic{1q1qbTr12SStateArr}
  \hspace{.5cm}
  \text{baryons in }\FPic[scale=0.75]{NcBigArrLeft}
\FPic[scale=0.75]{Nc_EpsilonDag} 
\end{equation}
As a result, and by way of a nontrivial
argument~\cite{tHooft:1973alw,tHooft:1974pnl,Witten:1979}, meson masses
scale like $N_c^0$; baryons become heavy, their masses scale like
$N_c^1$ (the number of elementary fields in the product state) and the
$m$-meson couplings scale like $N_c^{1-m/2}$.

Later Witten argued~\cite{Witten:1979, Witten:1983tw} that it is
natural to identify baryons with toplogical solitons, such as
Skyrmions~\cite{Skyrme1961,Skyrme1962} by showing that masses of
mesons and baryons as well as scattering amplitudes scale with
$1/\sqrt{N_c}$ the same way as the soliton model quantities scale with
the meson-meson coupling $g$.

The observation that this large $N_c$ extrapolation is not unique 
is based on the $\varepsilon$-isomorphism under scrutiny in this section,
\begin{align}
  \label{eq:epsilon-iso-fund=ASqbar2}
  \diagram[scale=.75]{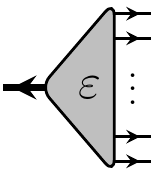}
  : V^{*\otimes(N-1)}\vert_{\text{AS}} \xrightarrow{\cong} V
  \ .
\end{align}
(Compare the arrow directions with its counterpart in
eq.~\eqref{eq:Epsilon-Nm1qb-to-1q-Clebsch-restr})

This implies that at $N_c=3$ we cannot distinguish a theory that has
quarks in the fundamental representation, $q^i \in V$, from a theory
that has its quarks in a two index antisymmetric subspace,
$\Bar Q^{[i j]} \in V^{*\otimes 2}$. (The square brackets indicate
antisymmetry in the indices of the tensor.) At $n=3$ (i.e. the number
of particles $n$ is three), the map that
translates between them is simply
\begin{align}
  \label{eq:QbartoQ}
  q^k = \frac12\varepsilon^{k l m} \Bar Q^{[l m]}
  \hspace{1cm}\text{ and its inverse }\hspace{1cm}
  \Bar Q^{[i j]} = \frac12\varepsilon^{i j k} q^k
  \ ,
\end{align}
with constistent behavior under the group action. We can indeed
interpret the $\varepsilon$-tensors as a basis for the 3-dimensional
space of antisymmetric matrices $\{T^k, k=1,2,3\}$ with
\begin{equation}
  \label{eq:taAS}
  [T^k]^{i j} := - \frac{1}2 \varepsilon^{i k j}
  \ ,
\end{equation}
normalized and mutually orthogonal under the inner product
\begin{equation}
  \label{eq:inner-product-Nc=3}
  \langle\ ,\ \rangle :V^{*\otimes 2}\times V^{*\otimes 2} \to \mathbb C,
 \hspace{1cm} \text{ defined by } \langle A, B\rangle := 2\text{tr}(A^\dagger B)
  \text{ for all $A,B\in V^{*\otimes 2}$,}
\end{equation}
such that
\begin{equation}
  \label{eq:Qbartoq}
  \Bar Q^{[i j]} = [T^k]^{i j} \langle T^k, Q \rangle
  \hspace{1cm}\text{ with }\hspace{1cm}
  \langle T^k, Q \rangle = q^k
  \ .
\end{equation}
We can extrapolate away from $N_c=3$ by noting that the antisymmetric
part of $V^{*\otimes 2}$ transforms as an $\frac{N_c(N_c-1)}2$
dimensional irreducible representation of $\mathsf{SU}(N_c)$,
irrespective of the exact value of $N_c$. Choosing an orthonormal basis
$\{T^k \in\mathbb R^{N_c\times N_c}| (T^k)^t = -T^k, \text{tr}((T^k)^t T^l) = \delta^{k l}, k=1,\ldots,\frac{N_c(N_c-1)}2\}$ of antisymmetric real
$N_c\times N_c$ matrices under the associated inner product
(generalizing the $N_c=3$ example above), we find that we can retain
eq.~\eqref{eq:Qbartoq} and use $\langle T^k, Q \rangle$ to
\emph{define} $q^k$ at any $N_c$. The associated representation
matrices are simply
\begin{equation}
  \label{eq:rep-Asymm}
  [U_{\text{AS}}]^{a b}  = 2\text{tr}\Bigl( (T^a)^\dagger (U^\dagger)^t T^b U^\dagger\Bigr)
\end{equation}
with the $U$ in the fundamental representation of
$\mathsf{SU}(N_c)$. Automatically
\begin{equation}
  \label{eq:detUAS=1}
  \det(U_{\text{AS}}) = 1
  \ ,
\end{equation}
so that the ``smallest''  totally antisymmetric singlet formed as a
product state of $\Bar Q^{[i j]}$ fields is given by
\begin{equation}
  \label{eq:Qbar-AS-singlet}
  \varepsilon^{a_1\ldots a_{d_{\text{AS}}}}
  \langle T^{a_1} ,\Bar Q\rangle
  \cdots
  \langle T^{a_{d_{\text{AS}}}} ,\Bar Q\rangle
  \hspace{1cm} \text{ where }\hspace{1cm} d_{\text{AS}} := \frac{N_c(N_c-1)}{2}
\end{equation}
(c.f.~\cite{Bolognesi:2006ws} for a technically different but equivalent
perspective).

It has been pointed out in~\cite{Armoni:2003gp} that the soliton mass
in this limit scales like $N_c^2$. This turns out to be in agreement
with the corresponding QCD extrapolation behavior. Like in the
traditional extrapolation, the scaling is foreshadowed by the particle
content scaling as
$\text{dim}(V^{*\otimes 2}|_{\text{AS}}) = \frac{N_c(N_c-1)}2 =
N_c^2\frac{1}2(1-\mathcal O(\frac1{N_c}))$ and proven
in~\cite{Cherman:2006iy,Cherman2006}. $m$-meson couplings scale like
$N_c^{2-m}$ and lead to a somewhat different phenomenology than the
traditional approach.

\subsubsection{Singlets on~\texorpdfstring{$\MixedPow{m}{n}$}{VmV*n}
from singlets
on~\texorpdfstring{$\MixedPow{k}{k}$}{VkV*k}}\label{sec:Singlets-VmV*n-special-case}

Both of the examples presented in
section~\ref{sec:Baryon-Equiv-2q2qbASym} speak of the importance of the
$\varepsilon$-induced isomorphisms and we want to close this section by
briefly sketching the general case.

A general singlet on $\MixedPow{k}{k}$ may contain \emph{several
}$\varepsilon$-tensors of length $N$. The remaining index legs (not
entering any $\varepsilon$-tensor) must be contained in a generic
subsinglet (consisting of Kronecker $\delta$s only) in order for the
overall operator to be color neutral. A schematic drawing of such a
general singlet projection operator of $\SUN$ on $\MixedPow{m}{n}$ (up
to the appropriate normalization constant) is
given in Figure~\ref{fig:GeneralSinglet} --- note that we have changed
notation slightly: the subscripts of the singlet projector $P^S$ from
now on refers to the number of fundamental and antifundamental factors
of the product space onto which the singlet acts (compare this with
eq.~\eqref{eq:Singlet-ProjOps-Theorem}), as this notation will be more
convenient for the present section.

\begin{figure}[H]
\centering
{\Large $P^S_{[m,n]} \; \rightarrow \quad $}
\scalebox{0.7}{\FPic{GeneralSinglet}}
\caption{This figure depicts a general singlet projector $P^S_{[m,n]}$ (up
to the appropriate normalization constant) on the
  space $\MixedPow{m}{n}$, where the top $m$ index lines
  ${\color{red!60!black}q_1} \ldots {\color{red!60!black}q_m}$
  (counted top to bottom) are in the fundamental representation, and
  the bottom $n$ index lines
  ${\color{red!60!black}\bar{q}_1} \ldots
  {\color{red!60!black}\bar{q}_n}$
  (counted bottom to top) are in the antifundamental
  representation. This singlet contains $(a+b)$ $\varepsilon$-tensors:
  $a$ of them over fundamental lines, and $b$ over antifundamental
  lines. The remaining $k=m-aN$ fundamental lines and $k=n-bN$
  antifundamental lines together form a generic subsinglet as is indicated by the (blue)
  shaded box.}
\label{fig:GeneralSinglet}
\end{figure}

\begin{figure}[H]
\centering
\scalebox{0.57}{%
\FPic{GeneralSinglet-BraceLeft}%
\FPic{GeneralSinglet-no-Labels}%
\; \FPic{GeneralSinglet-LabelsRight}%
}
{\Large $\quad \longrightarrow \quad$}
\FPic[scale=0.57]{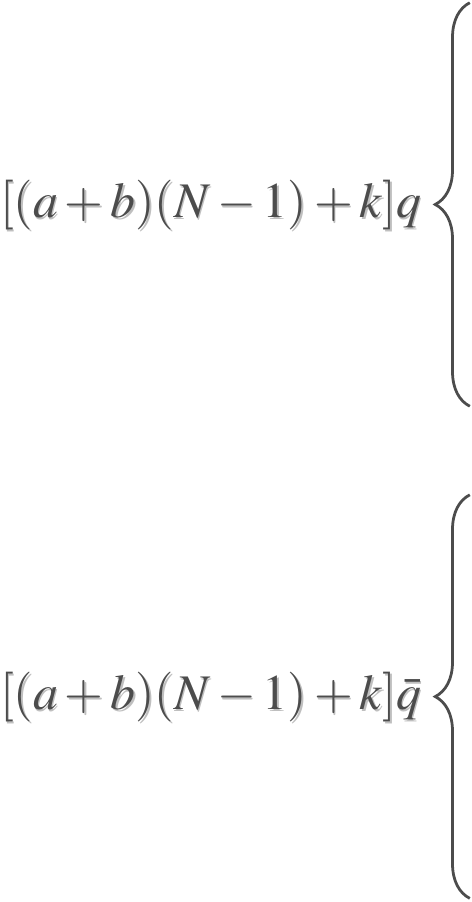}%
\FPic[scale=0.57]{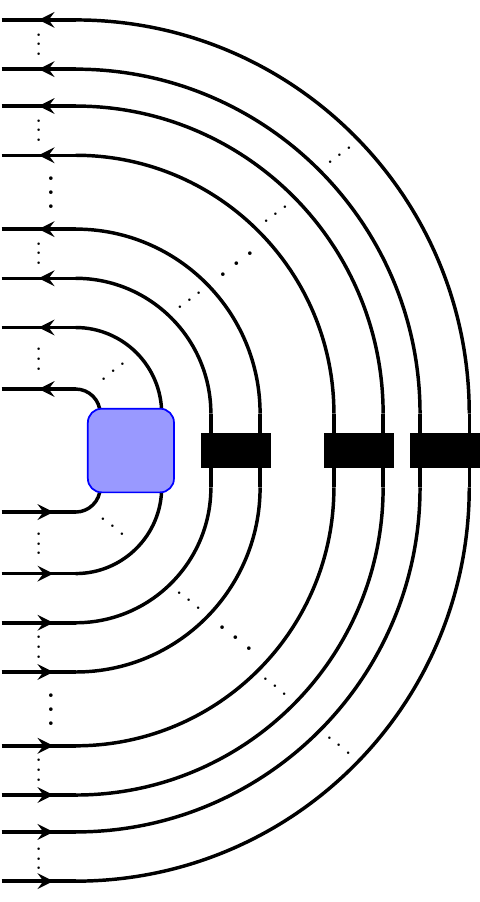}%
\hspace{0.27cm}
\reflectbox{\FPic[scale=0.57]{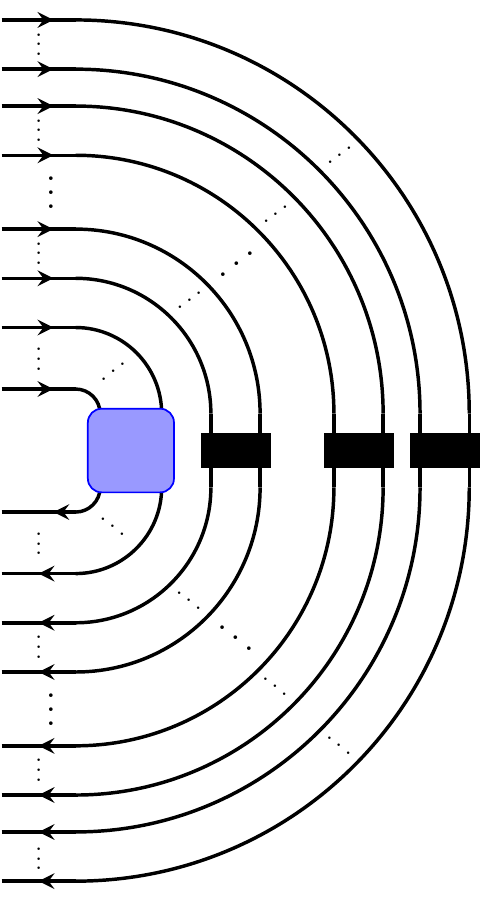}}%
\; \FPic[scale=0.57]{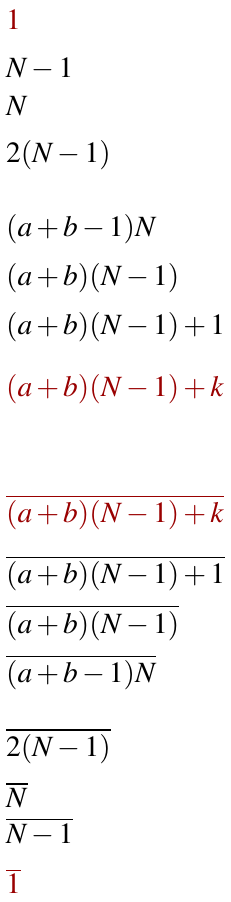}%
\caption{The operator of Figure~\ref{fig:GeneralSinglet} is
  transformed into a singlet projector consisting of Kronecker
  $\delta$s only (no $\varepsilon$-tensors): Each $\varepsilon$-tensor
  of length $N$ in the singlet in Figure~\ref{fig:GeneralSinglet} has
  been transformed into an antisymmetric subsinglet on $N-1$
  fundamental and antifundamental legs. The generic subsinglet of
  Figure~\ref{fig:GeneralSinglet} (blue shaded box) remains unchanged. (In
  this graphic, we have numbered each fundamental leg and each
  antifundamental leg (marked by the overbar) to keep track of their
  number.)}
\label{fig:GeneralSinglet-N-Indep}
\end{figure}

Each of the $\varepsilon$-tensors appearing in
Figure~\ref{fig:GeneralSinglet} can be related to an antisymmetric
(sub)singlet on $N-1$ fundamental and antifundamental legs,
analogously to the example in the previous
section~\ref{sec:Baryon-Equiv-2q2qbASym}. Thus, the singlet
$P^S_{[m,n]}$ of Figure~\ref{fig:GeneralSinglet} can be shown to be
isomorphic to a singlet on $\MixedPow{\alpha}{\alpha}$ for
$\alpha:=(a+b)(N-1)+k$, as depicted in Figure~\ref{fig:GeneralSinglet-N-Indep}.

Let us summarize:

\begin{theorem}[Non-generic singlets and Leibniz induced
  equivalences]\label{thm:NcDepSingletEquivalence}
  Let $N$ be a particular natural number and let $P^S_{[m,n]}$ be a singlet
  projection operator of $\SUN$ on $\MixedPow{m}{n}$. $P^S_{[m,n]}$
  has to fulfill the following conditions:
\begin{itemize}
\item $P_{m,n}^S$ contains exactly $(a+b)$ antisymmetrizers of length
  $N$ ($a$ of which are antisymmetrizers over fundamental legs, and $b$
  over antifundamental legs) such that
  \begin{equation}
    \label{eq:Ndep-Singlet-require-abk}
    m-aN = n-bN =:k
  \end{equation}
  for some natural number $k$. Note that $N$, $m$ and $n$ do not uniquely
  determine $a$, $b$ and $k$ through
  eq.~\eqref{eq:Ndep-Singlet-require-abk}, allowing for \emph{several}
  non-generic (transient) singlet projectors on $\MixedPow{m}{n}$.
\item The remaining $k$ fundamental and $k$ antifundamental legs not
  contained in any antisymmetrizer of length $N$ (equivalent to a
  product of $\varepsilon$-tensors of length $N$) are
  joined into a generic subsinglet $P^{S}_{m,n,(k)}$.
\end{itemize}
Then, there exists a generic singlet projection
operator $P_{[\alpha,\alpha]}^S$ in
$\API{\SUN,\MixedPow{\alpha}{\alpha}}$ with
\begin{equation}
  \label{eq:Ndep-Singlet-require-alpha}
\alpha := (a+b)(N-1)+k
\end{equation}
that is isomorphic to $P^S_{[m,n]}$ for the chosen value of
$N$
\begin{equation}
  \label{eq:Ndep-Singlet-equiv-to-Nindep-Singlet}
  \left. P_{[\alpha,\alpha]}^S \right\vert_{N} \cong P^S_{[m,n]} \ .
\end{equation}
In particular, $P_{[\alpha,\alpha]}^S$ will have the following
subsinglet structure:
\begin{itemize}
\item $k$ of its fundamental and antifundamental legs will constitute
  the subsinglet $P^{S}_{m,n,(k)}$,
\item the remaining legs constitute $(a+b)$ totally antisymmetric generic
  subsinglets, each containing exactly $N-1$ fundamental and
  antifundamental legs.
\end{itemize}
\end{theorem}

As mentioned in Theorem~\ref{thm:NcDepSingletEquivalence}, eq.~\eqref{eq:Ndep-Singlet-require-abk}
does not uniquely determine the integers $a,b$ and $k$ from $m,n$ and
$N$. An immediate consequence of this is that two different singlets
$Q_1^S$ and $Q_2^S$ of $\SUN$ over $\MixedPow{m}{n}$ may be equivalent
to singlets over $\MixedPow{\alpha_1}{\alpha_1}$ and
$\MixedPow{\alpha_2}{\alpha_2}$ respectively, where
$\alpha_1\neq\alpha_2$. Nonetheless, even though the product spaces
(Fock space components) are different,
$\MixedPow{\alpha_1}{\alpha_1}\neq\MixedPow{\alpha_2}{\alpha_2}$,
$Q_1^S$ and $Q_2^S$ are equivalent to each other.

\section{Practical construction of singlet
  projectors}\label{sec:Singlet-Examples}

Eq.~\eqref{eq:Clebsch-mplusn-StingletState} and
Theorem~\ref{thm:N-independent-Singlets} discuss the construction of
singlet states and singlet projection and transition operators of
$\SUN$ on $\MixedPow{k}{k}$ via bending products of Clebsch-Gordan
operators on $\MixedPow{m}{n}$ with
$m+n=k$. Theorem~\ref{thm:NcDepSingletEquivalence} (in particular
eq.~\eqref{eq:Ndep-Singlet-equiv-to-Nindep-Singlet}) ensures us that
this construction encompasses \emph{all} singlet projectors of $\SUN$,
also those on product spaces in which the number of fundamental
factors $V$ is different to the number of antifundamental factors
$V^*$. The singlet states on
$\MixedPow{m}{n}\otimes\MixedPow{n}{m}\cong\MixedPow{(m+n)}{(m+n)}$
constructed in eq.~\eqref{eq:Clebsch-mplusn-StingletState} allow for a
reordering of its index lines in order to obtain singlet states on
$\MixedPow{(m+n)}{(m+n)}$,
\begin{equation}
  \FPic[scale=0.75]{mqnqbClebschSState-Lambda}
  \quad \xlongrightarrow[\cong]{\text{reorder}} \quad
  \FPic[scale=0.75]{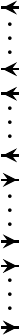}
  \FPic[scale=0.75]{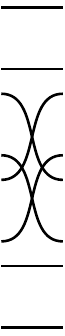}
  \FPic[scale=0.75]{mqnqbClebschSState-Lambda}
  \quad = \quad
  \FPic{mqmqbSState-O-allLines-ClebschSize}  
  \ ,
\end{equation}
where $O$ is an operator on $\Pow{(m+n)}=:\Pow{k}$. Rather than taking
$O$ to be a (product of) Clebsch-Gordan operators on $\Pow{k}$, we
will use the MOLD (Measure Of Lexical Disorder) projection and
transition operators
operators~\cite{Alcock-Zeilinger:2016bss,Alcock-Zeilinger:2016sxc} to
obtain the desired singlets: In~\cite{Alcock-Zeilinger:2016sxc}, we
gave a construction algorithm for compact, Hermitian versions of the
standard Young projection operators (which correspond to the
irreducible representations of $\SUN$ on $\Pow{k}$), called MOLD
operators. In~\cite{Alcock-Zeilinger:2016cva}, we expanded on this
topic and constructed compact transition operators between MOLD
projectors corresponding to equivalent irreducible
representations. The set containing all MOLD projection and transition
operators of $\SUN$ on $\Pow{k}$ is denoted by $\mathfrak{S}_k$. As
was shown in~\cite{Alcock-Zeilinger:2016cva}, $\mathfrak{S}_k$ spans
the algebra of invariants $\API{\SUN,\Pow{k}}$, and all of its
elements are mutually orthogonal under the scalar
product~\eqref{eq:Sclar-Product-AB} ($\InnerProd{A}{B}:=\Tr{A^{\dagger}B}$). In that, they satisfy the same
properties as the Clebsch-Gordan projection and transition operators
(\emph{c.f.} eqns.~\eqref{eq:Clebsch-ProjOps-TransOps}) used in
section~\ref{sec:mqmqb-bending-basis-elements}. 

Furthermore, the MOLD projection and transition operators have the
distinct advantage of making the identification of dimensionally null
operators very easy: As
was explained in~\cite{Alcock-Zeilinger:2016cva}, the elements of $\mathfrak{S}_k$ vanish \emph{individually} for some $N=N^*$,
if they contain an antisymmetrizer of length $>N^*$. Clearly, this can
only happen if $k>N$ as, for $k\leq N$, all operators in $\mathfrak{S}_k$
are dimensionally nonzero. Thus, not only do
individual MOLD operators vanish as $N$ decreases (rather than
establishing complicated constraint equations between operators), but,
furthermore, the MOLD operators allow one to immediately identify
vanishing operators through a simple, visual criterion.

In summary:
\begin{theorem}[Practical construction of generic
  singlets]\label{thm:practical-singlet-construction}
  To construct the singlet states of $\SUN$ in $\MixedPow{k}{k}$, it
  is sufficient to bend certain elements of the algebra
  $\API{\SUN,\Pow{k}}$; these states can then be used to construct the
  singlet projection and transition operators of $\SUN$ on
  $\MixedPow{k}{k}$.
  
  In particular, the fact that the MOLD projection and transition
  operators in $\mathfrak{S}_k$  
  \begin{enumerate}
  \item are easy to construct
    (\emph{c.f.}~\cite{Alcock-Zeilinger:2016sxc,Alcock-Zeilinger:2016sxc})
  \item are mutually orthogonal under the scalar
    product~\eqref{eq:Sclar-Product-AB} ($\InnerProd{A}{B}:=\Tr{A^{\dagger}B}$)
  \item allow for an easy identification of dimensionally null
    operators
  \end{enumerate}
  makes them ideally suited for the construction of singlet states of
  $\SUN$ on $\MixedPow{k}{k}$. The singlet projectors resulting from
  bending the MOLD operators form an orthonormal basis for the singlet
  subalgebra of $\API{\SUN,\MixedPow{k}{k}}$ with respect to the
  scalar product~\eqref{eq:Sclar-Product-AB}.
\end{theorem}

\subsection{Example: Singlet projectors on \texorpdfstring{$\MixedPow{3}{3}$}{V3xV*3} from MOLD projectors}\label{sec:MOLD-basis}

As an example, let us construct all singlet projection and transition
operators of $\SUN$ on $\MixedPow{3}{3}$. To accomplish this, we
bend the elements of
$\mathfrak{S}_3$~\cite[eq.~(141)]{Alcock-Zeilinger:2016cva}
(\emph{c.f.} the birdtrack notation introduced in
eqns.~\eqref{eq:S3-Birdtracks} and~\eqref{eq:Birdtrack-Sym-ASym-Definition})
\begin{equation}
  \label{eq:frakS3}
\mathfrak{S}_3 = 
\left\lbrace
\FPic{3ArrLeft}
\FPic{3Sym123SN}
\FPic{3ArrRight} \; ,
\quad
\frac{4}{3} \cdot 
\FPic{3ArrLeft}
\FPic{3Sym12ASym23Sym12}
\FPic{3ArrRight} \; ,
\quad
\sqrt{\frac{4}{3}} \cdot 
\FPic{3ArrLeft}
\FPic{3Sym12s23ASym12}
\FPic{3ArrRight} \; ,
\quad
\sqrt{\frac{4}{3}} \cdot 
\FPic{3ArrLeft}
\FPic{3ASym12s23Sym12}
\FPic{3ArrRight} \; ,
\quad
\frac{4}{3} \cdot 
\FPic{3ArrLeft}
\FPic{3ASym12Sym23ASym12}
\FPic{3ArrRight} \; ,
\quad
\FPic{3ArrLeft}
\FPic{3ASym123SN}
\FPic{3ArrRight}
\right\rbrace
\end{equation}
into singlet states
\begin{equation}
\label{eq:3q3qb-SingletStatesProjOps}
  \chi_1 \cdot \FPic{3q3qbArr}\FPic{3q3qbSym123SState} \ ,
\quad
  \chi_2 \cdot \FPic{3q3qbArr}\FPic{3q3qbSym12ASym23Sym12SState} \ ,
\quad
  \chi_2 \cdot \FPic{3q3qbArr}\FPic{3q3qbSym12s23ASym12SState} \ ,
\quad
  \chi_2 \cdot \FPic{3q3qbArr}\FPic{3q3qbASym12s23Sym12SState} \ ,
\quad
  \chi_2 \cdot \FPic{3q3qbArr}\FPic{3q3qbASym12Sym23ASym12SState} \ ,
\quad \text{and} \quad
  \chi_3\cdot \FPic{3q3qbArr}\FPic{3q3qbASym123SState} \ .
\end{equation}
The
normalization constants $\chi_i$ are given by
\begin{equation}
  \label{eq:ZetaNormConstants}
  \chi_1 
  = 
  \sfrac{6}{(N+2)(N+1)N} \ , 
  \quad 
  \chi_2 
  =
  \sfrac{3 \cdot \theta_{N>1}}{N (N^2-1)} 
  \quad
  \text{and} 
  \quad 
  \chi_3 
  =
  \sfrac{6 \cdot \theta_{N>2}}{(N-2)(N-1)N}
  \ ,
\end{equation}
where the function $\theta_{N>p}$, defined as
\begin{equation}
  \theta_{N>p} :=
  \begin{cases}
    1 & \text{if $N>p$} \\
    0 & \text{if $N\leq p$}
  \end{cases}
  \ ,
  \qquad
  p \in \mathbb{N}
  \ ,
\end{equation}
reminds us that the affected states
in~\eqref{eq:3q3qb-SingletStatesProjOps} are dimensionally zero for
values of $N$ that are smaller than the threshold $p$, which is simply
determined by the length of the longest antisymmetrizer in the
associated birdtrack.  (\emph{C.f.} the end of the present section for
a further discussion.)

Using the singlet states~\eqref{eq:3q3qb-SingletStatesProjOps} we can
construct the singlet projection and transition operators of $\SUN$
on $\MixedPow{3}{3}$ according to
Theorem~\ref{thm:N-independent-Singlets}. Arranging these operators
into a matrix (for visual clarity), which has the projection
operators on the diagonal and the transition operators on the
off-diagonal, we obtain
\begin{equation}
  \label{eq:3q3qb-ASymSingletTransOps}
\scalebox{0.8}{%
\raisebox{-0.5\height}{%
\includegraphics{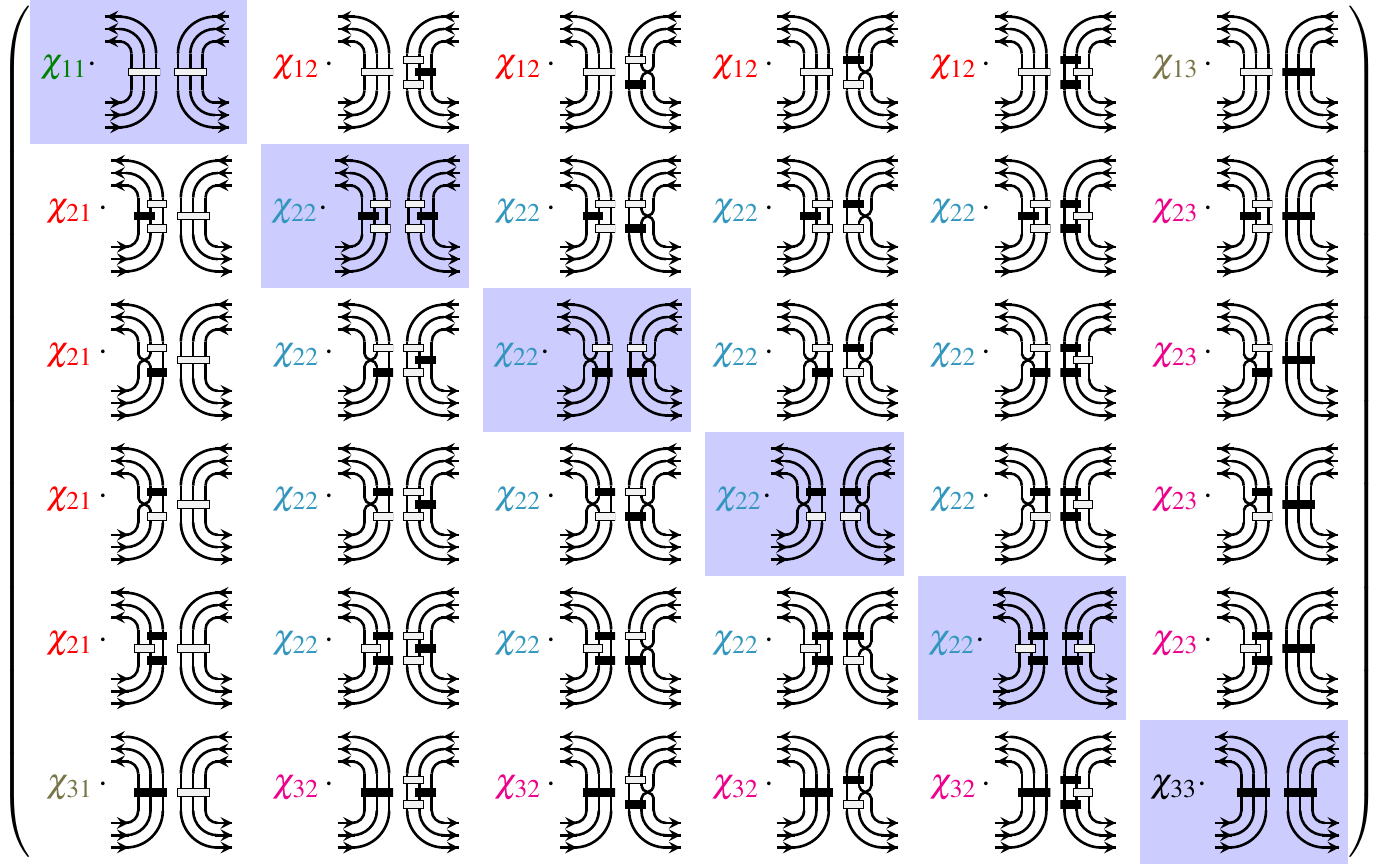}%
}%
}
\ ;
\end{equation}
the constants $\chi_{ij}$ are defined as
\begin{equation}
  \label{eq:ZetaijNormConstant}
  \chi_{ij} := \sqrt{\chi_i \cdot \chi_j}
\qquad \text{with $\chi_j$ given in
  eq.~\eqref{eq:ZetaNormConstants}}
\ .
\end{equation}

\subsection{The trace basis of singlet projectors}\label{sec:Trace-basis}

The basis of singlet states obtained from bending smaller operators
according to Theorem~\ref{thm:N-independent-Singlets} has the
advantage of being an efficient algorithm that is guaranteed to
produce all singlet states of $\SUN$ in
$\MixedPow{k}{k}$. Additionally, if the MOLD projection and transition
operators are used for the bending procedure, one immediately gains
access to which singlet representations vanish dimensionally. Lastly,
such a basis leads to significant simplification when implementing
(partial) coincidence limits between Wilson lines corresponding to
particles in the same representation, as will be discussed in a
follow-on paper~\cite{Weigert:2017}.

However, if one requires the adjoint representations contained within
the singlets to be explicit, a different construction is
mandatory. The algorithm we suggest is extremely efficient, exposes
the properties of coincidencs limits of quark with antiquark lines,
but at the price that a case by case post-processing is necessary to
expose dimensional zeros. Since the number of operators and the
threshold values $N_{\lambda,k}$ at which they occur follow uniquely
from the MOLD construction of the previous section, we can at least
use that as a requirements checklist for the post-processing steps.

To proceed, let us regroup the fundmental and antifundamental lines of
$\MixedPow{m}{m}$ pairwise,
\begin{IEEEeqnarray}{0rCCCl}
  & \MixedPow{m}{m} 
  & \cong & 
  \PairPow{m}
  \nonumber \\
  & \vin && \vin& \\
  & \FPic{mqmqbId}
  & \xlongrightarrow[\text{lines}]{\text{reorder}} &
  \FPic{mqmqbAltId} &
  \ . 
  \nonumber
\end{IEEEeqnarray}
Each pair $V\otimes V^*$ contains a singlet and an adjoint
representation as its irreducible components,
\begin{subequations}
  \label{eq:Fierz-Identity-Irreps1q1qb}
\begin{IEEEeqnarray}{0rCCCl}
  \delta_{\bar{p}\bar{q}} \delta_{pq}
  & \; = \; &
  \delta^{ab}
  [t^a]_{\bar{q}q} [t^b]_{\bar{p}p}
  &  \; + \; &
  \frac{1}{N}
  \delta_{\bar{q}q} \delta_{\bar{p}p}
    \label{eq:Fierz-Identity-Irreps1q1qb-Tensors}
    \\
  \FPic{1q1qbIdArr}
  & \; = \; &
  \underbrace{
    \FPic{1q1qbAdj12}
  }_{\text{adjoint}}
  & \; + \; &
  \underbrace{
    \frac{1}{N} \;
    \FPic{1q1qbTr12Arr}
  }_{\text{singlet}}
    \label{eq:Fierz-Identity-Irreps1q1qb-Birdtracks}
  \ ;
\end{IEEEeqnarray}
\end{subequations}
\eqref{eq:Fierz-Identity-Irreps1q1qb} is known as the Fierz
identity~\cite{Okun:1982ap,Cvitanovic:2008zz}. Singlet states that
make the adjoint and singlet components of the pairs $V\otimes V^*$ in
$\PairPow{m}$ explicit can be generated directly from the permutations
in $S_m$: 

\paragraph{Trace basis algorithm:}
\begin{enumerate}
  \item Write any $\rho\in S_m$ in  its disjoint 
cycle form, and also explicitly display the conventionally omitted $1$-cycles:
\begin{equation}
  \rho = \sigma_k \sigma_{k-1} \ldots \sigma_1
  \ ;
\end{equation}
\item Replace every cycle $\sigma$ (in the permutation $\rho$) of length $>1$ containing elements
  $i_j$, with the trace
  \begin{subequations}
  \begin{equation}
    \Tr{ 
      t^{a_{i_{\sigma(1)}}} 
      t^{a_{i_{\sigma(2)}}}
      \ldots
      t^{a_{i_{\sigma(\vert\sigma\vert)}}}
    }
    \ ,
  \end{equation}
  and multiply this trace with the tensor product
  \begin{equation}
    [t^{a_{i_1}}]_{\bar{q}_{i_1} q_{i_1}}
    \otimes
    [t^{a_{i_2}}]_{\bar{q}_{i_2} q_{i_2}}
    \otimes
    \cdots
    \otimes
    [t^{a_{i_{\vert\sigma\vert}}}]_{\bar{q}_{i_{\vert\sigma\vert}} q_{i_{\vert\sigma\vert}}}
    \ .
  \end{equation}
  \end{subequations}
  using a summation convention for all repeated indices $a_k$.
  \item Replace every $1$-cycle $(j)$, with the Kronecker delta
    $\delta_{\bar{q}_j q_j}$.
  \end{enumerate}
  The resulting object is a singlet state $\ket{\rho}\in\PairPow{m}$,
  presented in index notation. The $q_1,\ldots q_m$ refer
  to the $V$ factors, the $\Bar q_1,\ldots \Bar q_m$ refer to the
  $V^*$ factors in $\in\PairPow{m}$.  The procedure creates a unique
  state for each permutation $\rho$, since the disjoint cycle decomposition of $\rho$
  is unique.
  
  This algorithm automatically produces the correct amount of singlet
  states if $N \ge m$ where we are guaranteed that no dimensional
  zeros are present. Each of the $m!$ elements in $S_m$ leads to its
  own state, all of which are linearly independent. If $N < m$
  dimensional zeros will occur and the states generated by the
  trace basis algorithm will no longer be linearly independent. Unlike
  the MOLD states of section~\ref{sec:MOLD-basis} it is not individual
  states that turn off below threshold, instead we typically get
  genuine linear combinations involving several states that vanish in
  that situation. The most familiar example for this is the set of
  $d^{a b c}$ which appear as a linear combination of trace basis
  states at $m=3$ (see eq.~\eqref{eq:dandf}) and vanish unless
  $N \ge 3$. As long as we do not have a general deterministic
  algorithm to map out the dimensional zeros this remains an
  important drawback -- we need to post-process the states at each $m$
  to expose the dimensional zeros.

  If we restrict ourselves to permutations that do not contain one
  cycles (derangements) we arrive at singlet states in $V^{(\text{adj})\otimes m}$
  where $V^{(\text{adj})}$ is the traceless part of $V\otimes V^*$, i.e. the adjoint
  representation
  (c.f.~\eqref{eq:Fierz-Identity-Irreps1q1qb-Tensors}). This provides
  a convenient way to directly construct singlets in this subspace
  which has been suggested earlier by Keppeler and
  Sjödahl~\cite{Keppeler:2012ih}.

  Let us illuminate the trace basis algorithm through an example:
  Consider the permutations in $S_3$ written in their disjoint cycle
  structures, $\mathrm{id}=(1)(2)(3)$, $(12)(3)$, $(13)(2)$,
  $(23)(1)$, $(123)$ and $(132)$. Then, the (non-normalized) singlet
  states corresponding to these cycles are
\begin{subequations}
  \label{eq:TraceSinglets-3q3qb}
\begin{IEEEeqnarray}{0CCCCl}
  \mathrm{id} = (1)(2)(3)
  &  \; \longrightarrow \;  &
  \delta_{\bar{q}_1 q_1}
  \otimes
  \delta_{\bar{q}_2 q_2}
  \otimes
  \delta_{\bar{q}_3 q_3}
  & \; = \; &
  \FPic{3q3qbAltArr}
  \FPic{3q3qbTr12Tr34Tr56SState}
  \\[2mm]
  (12)(3)
  &  \; \longrightarrow \;  &
  \Tr{
    t^{a_1}
    t^{a_2}
  }
  [t^{a_1}]_{\bar{q}_1 q_1}
  \otimes
  [t^{a_2}]_{\bar{q}_2 q_2}
  \otimes
  \delta_{\bar{q}_3 q_3}
  & \; = \; &
  \FPic{3q3qbAltArr}
  \FPic{3q3qbTr56tExp1234SState}
  \; = \;
  \FPic{3q3qbAltArr}
  \FPic{3q3qbTr56t1234SState}
  \\[2mm]
  (13)(2)
  &  \; \longrightarrow \;  &
  \Tr{
    t^{a_1}
    t^{a_3}
  }
  [t^{a_1}]_{\bar{q}_1 q_1}
  \otimes
  \delta_{\bar{q}_2 q_2}
  \otimes
  [t^{a_3}]_{\bar{q}_3 q_3}
  & \; = \; &
  \FPic{3q3qbAltArr}
  \FPic{3q3qbTr34tExp1256SState}
  \; = \;
  \FPic{3q3qbAltArr}
  \FPic{3q3qbTr34t1256SState}
  \\[2mm]
  (23)(1)
  &  \; \longrightarrow \;  &
  \Tr{
    t^{a_2}
    t^{a_3}
  }
  \delta_{\bar{q}_1 q_1}
  \otimes
  [t^{a_2}]_{\bar{q}_2 q_2}
  \otimes
  [t^{a_3}]_{\bar{q}_3 q_3}
  & \; = \; &
  \FPic{3q3qbAltArr}
  \FPic{3q3qbTr12tExp3456SState}
  \; = \; 
  \FPic{3q3qbAltArr}
  \FPic{3q3qbTr12t3456SState}
  \\[2mm]
  (123)
  &  \; \longrightarrow \;  &
  \Tr{
    t^{a_1}
    t^{a_2}
    t^{a_3}
  }
  [t^{a_1}]_{\bar{q}_1 q_1}
  \otimes
  [t^{a_2}]_{\bar{q}_2 q_2}
  \otimes
  [t^{a_3}]_{\bar{q}_3 q_3}
  & \; = \; &
  \FPic{3q3qbAltArr}
  \FPic{3q3qbtExp123456SState}
  \label{eq:TraceSinglets-3q3qb-s123}
  \\[2mm]
  (132)
  &  \; \longrightarrow \;  &
  \Tr{
    t^{a_1}
    t^{a_3}
    t^{a_2}
  }
  [t^{a_1}]_{\bar{q}_1 q_1}
  \otimes
  [t^{a_2}]_{\bar{q}_2 q_2}
  \otimes
  [t^{a_3}]_{\bar{q}_3 q_3}
  & \; = \; &
  \FPic{3q3qbAltArr}
  \FPic{3q3qbtExp125634SState}
  \label{eq:TraceSinglets-3q3qb-s132}
  \ .
\end{IEEEeqnarray}
\end{subequations}
(In eqns.~\eqref{eq:TraceSinglets-3q3qb} we have used the fact that
the trace is cyclic, e.g. $\Tr{t^{a_3}t^{a_1}}=\Tr{t^{a_1}t^{a_3}}$
and $\Tr{t^{a_2}t^{a_3}t^{a_1}}=\Tr{t^{a_1}t^{a_2}t^{a_3}}$, etc..)

At $N > 2$ all the states in~\eqref{eq:TraceSinglets-3q3qb} are
linearly independent, but the singlet
states~\eqref{eq:TraceSinglets-3q3qb-s123}
and~\eqref{eq:TraceSinglets-3q3qb-s132} are not orthogonal to each
other,
\begin{equation}
  \FPic{3q3qbtExp123456SStateDag}
  \FPic{3q3qbAltArr}
  \FPic{3q3qbtExp125634SState}
  \; = \; 
  - \frac{N^2-1}{N}
  \; \neq \;
  0
  \ .
\end{equation}
The most useful mutually orthogonal linear combinations arise by
forming the symmetric and antisymmetric linear combinations
\begin{equation}
  \label{eq:dandf}
  \FPic{3q3qbAltArr}
  \FPic{3q3qbDabcSState}
  \; := \;
  \frac{1}{2}
  \left(
    \FPic{3q3qbAltArr}
    \FPic{3q3qbtExp123456SState}
    \; + \;
    \FPic{3q3qbAltArr}
    \FPic{3q3qbtExp125634SState}
  \right)
  \qquad \text{and} \qquad
  \FPic{3q3qbAltArr}
  \FPic{3q3qbFabcSState}
  \; := \;
  \frac{1}{2}
  \left(
    \FPic{3q3qbAltArr}
    \FPic{3q3qbtExp123456SState}
    \; - \;
    \FPic{3q3qbAltArr}
    \FPic{3q3qbtExp125634SState}
  \right)
  \ ,
\end{equation}
where the empty (white) circle and the filled in (black) circle over
the gluon lines correspond to the structure constants $d^{abc}$ and
$if^{abc}$ (\emph{c.f.} eq.~\eqref{eq:Fabc-Dabc-Birdtracks}).  In
addition to being orthogonal these combinations are adapted to
dimensional zeros: The symmetric combination, containing the
$d^{a b c}$ vanishes for $N < 3$ representing a dimensional zero while
the antisymmetric combination is nontrivial for all $N >
1$. Unfortunately, we do not have a general post-processing algorithm for arbitray $m$
to achieve such a mutually orthogonal result in a generic fashion.

One
obtains the following orthonormal basis of singlet states on
$\PairPow{3}$,
\begin{subequations}
  \label{eq:FierzBasis-3q3qbStates}
\begin{equation}
  \label{eq:3q3qbFierzSStates}
  \xi_1 \cdot
  \FPic{3q3qbAltArr}\FPic{3q3qbTr12Tr34Tr56SState} \ , 
  \quad 
  \xi_2\cdot
  \FPic{3q3qbAltArr}\FPic{3q3qbTr12t3456SState} \ ,
  \quad 
  \xi_2\cdot
  \FPic{3q3qbAltArr}\FPic{3q3qbTr56t1234SState} \ , 
  \quad 
  \xi_2\cdot
  \FPic{3q3qbAltArr}\FPic{3q3qbTr34t1256SState} \ ,
  \quad 
  \xi_3\cdot
  \FPic{3q3qbAltArr}\FPic{3q3qbFabcSState} 
  \quad \text{and} \quad
  \xi_4\cdot 
  \FPic{3q3qbAltArr}\FPic{3q3qbDabcSState}
\end{equation}
with normalization constants $\xi_i$ given by
\begin{equation}
  \label{eq:xiNormConstants-FierzBasis}
\xi_1 = \frac{1}{\sqrt{N^3}} 
\; , \quad
\xi_2 = \frac{\theta_{N>1}}{\sqrt{N (N^2-1)}}
\; , \quad
\xi_3 = \frac{\theta_{N>1}}{\sqrt{2 N (N^2-1)}}
\quad \text{and} \quad
\xi_4 = \sqrt{\frac{\theta_{N>2} \cdot N}{2 (N^2-4) (N^2-1)}}
\ .
\end{equation}
\end{subequations}
Besides being orthonormal, the basis~\eqref{eq:FierzBasis-3q3qbStates}
also gives immediate access to which singlet states become
dimensionally null as $N$ decreases: the structure
constant $d^{abc}$ vanishes for $N<3$, and every operator containing a
generator $t^a$ vanishes for $N<2$.

\section{Conclusion}\label{sec:Conclusion}

Singlet representations of $\SUN$ are of vital importance in many physics applications. The most prominent example is arguably QCD, where
confinement requires color-charged particles to combine into
color-neutral states. However, the general method to construct the
multiplets of $\SUN$ on $\MixedPow{m}{n}$ from the Leibniz formula for
determinants (\emph{c.f.} appendix~\ref{sec:LR-Example}) is
computationally costly and thus not useful in practice (as exemplified
in appendix~\ref{sec:LR-Example}).

In this paper, we gave an alternative, \emph{simple} construction
method for the singlet projection operators of $\SUN$ on
$\MixedPow{k}{k}$ (section~\ref{sec:mqmqb-bending-basis-elements}):
these singlets are obtained from \emph{bending} the basis elements of
the algebra of invariants of $\SUN$ on $\MixedPow{m}{n}$ with $m+n=k$
(\emph{c.f. }Theorem~\ref{thm:N-independent-Singlets}), and were
referred to as \emph{generic singlets}. We argued that the MOLD projection
and transition operators of $\SUN$ on $\Pow{k}$ are ideally suited for
this process. 
Theorem~\ref{thm:N-independent-Singlets} also gave a counting argument
predicting the number of singlet representation of $\SUN$ on
$\MixedPow{m}{m}$ to be maximally $m!$. If $N < m$ the number of
singlets is smaller than $m!$, and the MOLD operators/states give direct
access to which of the underlying states become dimensionally zero.

Singlets on $\MixedPow{m}{n}$ with $m\neq n$ are always non-generic,
they only appear at isolated values of $N$, right at the threshold
above which the irreducible representation ceases to be dimensionally
zero. We referred to these as \emph{transient singlets}. At that value
the Littlewood-Richardson correspondence, mediated by the Leibniz
formula for determinants, maps these canonically onto generic singlets
in some $\MixedPow{\alpha}{\alpha}$, typically with $\alpha \ll m$ or
$\alpha\ll n$ (but not both). As such, they are most efficiently
reconstructed from their Littlewood-Richardson partner as exemplified
in section~\ref{sec:mqnqb-no-new-multiplets}.


We used an explicit example to demonstrate the general singlet construction
algorithm: we constructed the singlets of $\SUN$ on $\MixedPow{3}{3}$
in section~\ref{sec:Singlet-Examples}. This exemplifies that the MOLD
projection and transition
operators~\cite{Alcock-Zeilinger:2016sxc,Alcock-Zeilinger:2016cva} are
particularly well suited for the bending procedure to generate
singlets, as they are easily constructed and encode important
information on dimensionally vanishing operators in a visually
explicit manner. We, furthermore, provided an efficient algorithm that
constructs the singlet states of $\MixedPow{k}{k}$ directly from the
permutations in the group $S_k$ (section~\ref{sec:Trace-basis}); we
referred to this as the \emph{trace basis algorithm}. While this
latter algorithm presents a starting point towards constructing a
basis of singlet states that makes the adjoint (in QCD parlance gluon)
components explicit, it does not, by itself, give rise to an
orthogonal basis, nor does it encode information on dimensionally
vanishing singlets as $N$ decreases (in the sense that not a
particular singlet vanishes, but rather a linear combination of basis
states). This warrants further research on the topic.

\appendix

\section{Irreducible representations of
  \texorpdfstring{$\SUN$}{SU(N)} on mixed product spaces: the textbook
  method}\label{sec:LR-Example}

Young's contributions to the representation theory of $\SUN$ on
$\Pow{m}$~\cite{Young:1928} allow for a simple construction algorithm
of (Hermitian) projection operators onto the irreducible
representations of $\SUN$~\cite{Keppeler:2013yla,Alcock-Zeilinger:2016sxc}. The
situation for the irreducible representations of $\SUN$ on a mixed
product space $\MixedPow{m}{n}$ is not as well developed, despite what a
casual glance at the literature might lead us to believe. The aim of
this appendix is to give a brief account of the existing (standard)
methods to construct the projection operators corresponding to the
irreducible representations of $\SUN$ on $\MixedPow{m}{n}$ from the
appropriate tableaux (Littlewood-Richardson tableaux). This will
illustrate that the standard methods are only adequate for
classification purposes, not for explicit calculations. While all
pieces of information given in this section are present in the
standard
literature~\cite{Tung:1985na,Sagan:2000,Fulton:1997,Jeevanjee:2015},
we are not aware of a text that describes the entire method from start
(constructing the tableaux) to finish (obtaining the projection
operators), and thus have chosen to give a full account here.

\subsection{The irreducible representations of
  \texorpdfstring{$\SUN$}{SU(N)} on
  \texorpdfstring{$\MixedPow{m}{n}$}{VmV*n} with standard methods:
  Littlewood-Richardson tableaux}\label{sec:Intro-LR-tableaux}

When constructing the (Hermitian) Young projection operators from
Young tableaux, one presupposes each factor $V$ in the product space
on which $\SUN$ acts to be in the fundamental representation, and
therefore represents it by a single box in the Young
tableau. Antifundamental factors $V^*$ can therefore not be
represented by a single box. However, as a result of the Leibniz
formula for determinants (\emph{c.f.}
eq.~\eqref{eq:Leibniz-Nminus1q-into-qb-Us} or~\cite{Jeevanjee:2015}
for a textbook treatment), an antifundamental component can be viewed
as an antisymmetric combination of $N-1$ fundamental
ones~\cite{Tung:1985na}, \ytableausetup{mathmode, boxsize=1.8em}
\allowdisplaybreaks[0]
\begin{IEEEeqnarray}{0rCCCCCl}
& \text{fundamental:} 
& \qquad 
& \text{antifundamental:} 
&& \nonumber \\
& \ybox{1}
&
& \begin{ytableau}
    1 \\
    2 \\
    \vdots \\
    \scriptstyle{N-1}
   \end{ytableau}
& \ ,
 \label{eq:tableaux-various-reps}
\end{IEEEeqnarray}
\allowdisplaybreaks where the numbers here help to keep track of the
amount of boxes, but are not necessarily the filling of the box in the
tableau sense.\footnote{\label{footnote:factors-over-hooks}Further
  motivation behind the claim~\eqref{eq:tableaux-various-reps} is
  given by the dimension of the representation corresponding to these
  tableaux, which can be calculated using the
  \emph{factors-over-hooks}
  formula~\cite{Cvitanovic:2008zz,Georgi:1999wka}: Using this formula,
  one finds that each of the tableaux in
  eq.~\eqref{eq:tableaux-various-reps} corresponds to an
  $N$-dimensional representation, namely the fundamental and the
  antifundamental representation, respectively.}

While Young's algorithm gives a prescription on how to add fundamental
factors $V$ to the product space (via adding individual boxes to the
Young tableau), a different method is needed to add antifundamental
factors $V^*$.  In the 1970's, Littlewood and Richardson (LR)
generalized Young's method to include factors of \emph{all}
representations~\cite{Littlewood:1950}. Since, in this paper, our
focus lies on adding factors $V^*$ to the product space on which we
consider $\SUN$, we quote a simplified version of the LR prescription,
also referred to as \emph{Pieri's
  formula}~\cite{Sagan:2000,Fulton:1997}. For the fully general
algorithm, readers are referred to Littlewood's
book~\cite{Littlewood:1950} or Sagan's book~\cite{Sagan:2000}, the
latter offering a more modern combinatorial view. Furthermore, Howe
and Lee~\cite{Howe:2011} provide a wonderfully intuitive proof of the
general LR rule using only classical invariant theory.

\begin{theorem}[adding antifundamental factors to the product space
  (LR rule, Pieri's formula)]
\label{thm:add-antiquark-LR}
Let $\Theta$ be a standard Young
tableau\footnote{See~\cite{Sagan:2000,Fulton:1997} or other 
  textbooks for a definition of \emph{standard Young tableaux}.} consisting of $m$
boxes (corresponding to an irreducible representation of $\SUN$ on
$\Pow{m}$), and let an antiquark be represented by the Young tableau
consisting of one column of length $N-1$ (\emph{c.f.}
eq.~\eqref{eq:tableaux-various-reps}),
\begin{equation}
  \label{eq:antiquark-tableau-BarPhi}
  \Bar{\Phi} =
\FPic[scale=0.8]{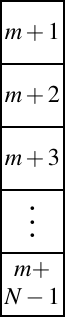}
\; =: \;
\FPic[scale=0.8]{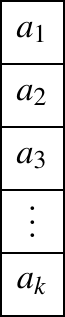}
\qquad
\text{with $a_{j+1}:=a_j+1$, and end points $a_1=m+1$ and $a_k=m+N-1$}
\ .
\end{equation}
Then, the product $\Theta\otimes\Bar{\Phi}$ yields the direct sum of all
tableaux that can be constructed as follows: Take the tableau $\Theta$
and add each box $\ybox{a_j}\in\Bar{\Phi}$ that preserves the
left-alignedness and top-alignedness of standard Young
tableaux. Additionally, we require that each box $\ybox{a_j}$ appears
in a row strictly above $\ybox{\scriptstyle a_{j+1}}$, and that the
resulting tableau has a maximum of $N$ rows. Evidently, all tableaux
in this sum are standard Young tableaux with $m+N-1$ boxes and
correspond to the irreducible representations of $\SUN$ on
$\Pow{m}\otimes V^*$.

We may iterate the above described procedure to form the tableaux
\begin{equation}
  \label{eq:n-antiquark-tableaux-BarPhi}
  \Theta \otimes \Bar{\Phi}_1 \otimes \cdots \otimes \Bar{\Phi}_n
  \ ,
\end{equation}
where each $\Bar{\Phi}_i$ is a tableau consisting of a single column
of length $N-1$. Let the set of all tableaux appearing in the
sum~\eqref{eq:n-antiquark-tableaux-BarPhi} be denoted by
$\left\lbrace
  \Theta\otimes\Bar{\Phi}_1\otimes\cdots\otimes\Bar{\Phi}_n\right\rbrace$,
each of which corresponds to an irreducible representation of $\SUN$
on $\MixedPow{m}{n}$. All tableaux constructed according to the
Littlewood-Richardson rule will be referred to as
Littlewood-Richardson tableaux in this paper.
\end{theorem}
\ytableausetup{mathmode, boxsize=normal}%

For the sake of brevity, we will assume that all Young tableaux
discussed in this section are standard (unless explicitly stated
otherwise!) and merely refer to them as Young tableaux, dropping the
adjective ``standard''.

The requirement that the resulting LR tableau has at most $N$ rows
ensures that the corresponding operator is not dimensionally zero,
since a column of size $>N$ would give rise to an antisymmetrizer of
more than $\text{dim}(V)=N$ objects (the steps involved in obtaining a
projection operators from an LR tableau are explained by means of an
example in section~\ref{sec:Singlets-Projectors-LRtableaux-Leibniz}).

As an example of the LR rule described in
Theorem~\ref{thm:add-antiquark-LR}, consider the Young tableau
\begin{equation}
  \Theta = 
  \begin{ytableau}
    1 & 3 \\
    2
  \end{ytableau}
  \ ,
\end{equation}
and let $N=4$ such that
\begin{equation}
  \Bar{\Phi} = 
  \begin{ytableau}
    *(blue!25) 4 \\
    *(blue!25) 5 \\
    *(blue!25) 6 
  \end{ytableau}
\ .
\end{equation}
Then, according to Theorem~\ref{thm:add-antiquark-LR}, the product
$\Theta\otimes\Bar{\Phi}$ yields
\begin{equation}
  \Theta\otimes\Bar{\Phi} = 
  \begin{ytableau}
    1 & 2 & *(blue!25) 4 \\
    3 & *(blue!25)  5 \\
    *(blue!25) 6 
  \end{ytableau}
\oplus \;
\begin{ytableau}
  1 & 2 & *(blue!25) 4 \\
  3 \\
  *(blue!25) 5 \\
  *(blue!25) 6 
\end{ytableau}
\oplus \;
\begin{ytableau}
  1 & 2 \\
  3 & *(blue!25) 4 \\
  *(blue!25) 5 \\
  *(blue!25) 6 
\end{ytableau}
\ ,
\end{equation}
where each tableau in the direct sum has $6=3+N-1$ boxes and corresponds to
an irreducible representation of $\SU4$ on $\Pow{3}\otimes V^*$.

\subsection{Projection operators from Littlewood-Richardson tableaux
  (using the Leibniz formula)}\label{sec:Singlets-Projectors-LRtableaux-Leibniz}

The LR rule (Theorem~\ref{thm:add-antiquark-LR}) allows us to
build up the tableaux corresponding to the irreducible
representations of $\SUN$ on $\MixedPow{m}{n}$. Let us now discuss
how to construct the corresponding projection operators:


Recall the Leibniz formula for determinants~\cite{Jeevanjee:2015},
which allows one to express a group element $U^{\dagger}\in\SUN$ in
the antifundamental representation as a product of $N-1$ factors of
the same group element $U$ in the fundamental representation,
\begin{subequations}
  \label{eq:AntiQuarkMethodEx-4and5}
\begin{equation}
  \label{eq:AntiQuarkMethodEx4}
  \tensor{\varepsilon}{_{b_1 b_2 \ldots b_{N}}}
  U^{\dagger}_{b_{N} a_{N}} =
  \tensor{\varepsilon}{_{a_1 a_2 \ldots a_{N}}}
  {U}_{b_1 a_1}
  {U}_{b_2 a_2} \ldots
  {U}_{b_{(N-1)} a_{(N-1)}}
\end{equation}
(where we used the symbol $U^{\dagger}$ for both the group element as
well as its antifundamental representation on $V^*$).  Even further,
one may write
\begin{equation}
  \label{eq:AntiQuarkMethodEx5}
  \tensor{\varepsilon}{_{b_1 b_2 \ldots b_{N}}}
  \underbrace{
    U^{\dagger}_{b_{N} a_{N}}
    \ldots
    U^{\dagger}_{b_{(N-j+1)} a_{(N-j+1)}}
  }_{N-j \text{ elements}}  
  = 
  \tensor{\varepsilon}{_{a_1 a_2 \ldots a_{N}}} 
  \underbrace{
    {U}_{b_1 a_1}
    {U}_{b_2 a_2}
    \ldots 
    {U}_{b_{(N-j)} a_{(N-j)}}
  }_{j \text{ elements}}
\ .
\end{equation}
\end{subequations}
In birdtrack notation (\emph{c.f.}
eq.~\eqref{eq:Kronecker-Delta-Levi-Civita-Birdtrack}), the Leibniz
formulae~\eqref{eq:AntiQuarkMethodEx-4and5} allow one
to ``bend'' one, respectively, $N-j$ legs of the $\varepsilon$-tensor
(\emph{c.f.} eq.~\eqref{eq:Epsilon-Nm1qb-to-1q-Clebsch}),
\begin{equation}
  \label{eq:Leibniz-UUDag1}
  \FPic[scale=0.75]{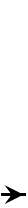}
  \FPic[scale=0.75]{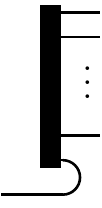}
  \FPic[scale=0.75]{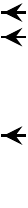}
  \qquad \text{and} \qquad
  \FPic[scale=0.75]{Nc-1-ArrLabels-top-q}
  \reflectbox{\FPic[scale=0.75]{Nc_Epsilon1qb}}
  \FPic[scale=0.75]{Nc-1-ArrLabels-bottom-qb}
  \ ,
  \qquad \text{respectively}
  \ , \qquad
  \FPic[scale=0.75]{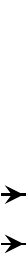}
  \FPic[scale=0.75]{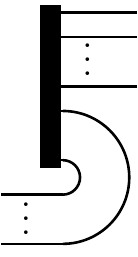}
  \FPic[scale=0.75]{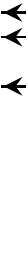}
  \qquad \text{and} \qquad
  \FPic[scale=0.75]{Nc-j-ArrLabels-top-q}
  \reflectbox{\FPic[scale=0.75]{Nc_Epsilonjqb}}
  \FPic[scale=0.75]{Nc-j-ArrLabels-bottom-qb}
  \ .
\end{equation}


We now illustrate how the Leibniz identity helps construct projection
operators from Littlewood-Richardson tableaux by means of an example:
The Fock space component containing a $q\bar{q}$-pair, $V\otimes V^*$,
decomposes into two irreducible representations of $\SUN$ (\emph{c.f.}
eq.~\eqref{eq:Fierz-Identity-Irreps1q1qb-Birdtracks}):
\begin{equation}
\label{eq:irrep-projectors-1q1qb}
  \text{the ``singlet":} \quad
  \frac{1}{N}\delta_{ik}\delta_{jl}
  \; = \;
  \frac{1}{N} \;
  \FPic{1q1qbTr12Arr}
  \hspace{1cm} \text{and the ``octet":} \quad
  [t^a]_{ik} [t^a]_{jl}
  \; = \;
  \FPic{1q1qbAdj12}
  \ ,
\end{equation}
where $[t^a]_{ik}$ is the generator of the group (\emph{c.f.} 
eq.~\eqref{eq:tzBirdtracks}).
The projection operators~\eqref{eq:irrep-projectors-1q1qb} can indeed
be recovered from the appropriate Littlewood-Richardson tableaux
corresponding to $1$ fundamental and $1$ antifundamental factor by
means of the Leibniz identity: Consider the branching tree of
LR tableaux for $1q+1\bar{q}$ constructed according to
Theorem~\ref{thm:add-antiquark-LR}:
\begin{equation}
  \label{eq:1q1qb-branching-tree}
\tikz[baseline=(current bounding box.west), yscale=1, xscale=1,
  shift={(0,0)},%
every node/.style={rectangle, inner sep = 4pt, outer sep = 1pt}]{
\node(T1)at(0,0){$\begin{ytableau} 1 \end{ytableau}$};
\node(T21)at(2,-1.5){$
  \begin{ytableau}
    1 & *(blue!25) 2 \\
    *(blue!25) 3 \\
    *(blue!25) \FPic{vdots-N} \\
    *(blue!25) N 
  \end{ytableau}$};
\node(T22)at(-2,-1.5){$
  \begin{ytableau}
    1 \\
    *(blue!25) 2 \\
    *(blue!25) 3 \\
    *(blue!25) \FPic{vdots-N} \\
    *(blue!25) N 
  \end{ytableau}$};
\draw[-{stealth}](T21)--(T1);
\draw[-{stealth}](T22)--(T1);
}
\ .
\end{equation}
The left tableau in the second level of the tree corresponds to a $1$-dimensional representation of
$\SUN$, while the right tableau corresponds to a $N^2-1$-dimensional 
representation of $\SUN$.\footnote{This can be checked using the
  \emph{factors-over-hooks}
  formula~\cite{Cvitanovic:2008zz,Georgi:1999wka}, \emph{c.f.}
  footnote~\ref{footnote:factors-over-hooks}.}

The MOLD algorithm~\cite{Alcock-Zeilinger:2016sxc} can be used to
construct Hermitian Young
projection operators on $\Pow{N}$ corresponding to the
tableaux~\eqref{eq:1q1qb-branching-tree},\footnote{The
  prefactors of the operators in
  eq.~\eqref{eq:1q1qb-branching-tree-Ops} arise from the MOLD
  algorithm: since both operators correspond to \emph{lexically
    ordered} tableaux (\emph{c.f.}~\cite{Alcock-Zeilinger:2016sxc}),
  the MOLD algorithm predicts that the Hermitian operators have the same
  normalization factor as their Young counterparts.}%
\ytableausetup{mathmode, boxsize=0.7em}
\begin{equation}
\label{eq:1q1qb-branching-tree-Ops}
P_{\scalebox{0.75}{\begin{ytableau}
    \scriptstyle 1 \\
    *(blue!25) \scriptstyle 2 \\
    *(blue!25) \scriptstyle 3 \\
    *(blue!25) \scriptstyle \FPic[scale=0.5]{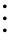} \\
    *(blue!25) \scriptstyle N \\
  \end{ytableau}}} = 
\scalebox{0.75}{%
\FPic{NcArrLeft}\FPic{NcASym1-to-Nc}\FPic{NcArrRight}%
}
\qquad \text{and} \qquad
  P_{\scalebox{0.75}{\begin{ytableau}
    \scriptstyle 1 & *(blue!25) \scriptstyle 2 \\
    *(blue!25) \scriptstyle 3 \\
    *(blue!25) \scriptstyle \FPic[scale=0.5]{vdots-N} \\
    *(blue!25) \scriptstyle N \\
  \end{ytableau}}} = 
\frac{2 (N-1)}{N} \cdot
\scalebox{0.75}{%
\FPic{NcArrLeft}\FPic{NcSym12ASym1-to-Nc-1Sym12}\FPic{NcArrRight}%
}
\ .
\end{equation}
\ytableausetup{mathmode, boxsize=normal}%
The Leibniz identity~\eqref{eq:Leibniz-UUDag1} allows us to transform
$N-1$ fundamental legs (corresponding to the antiquark) into $1$
antifundamental leg. This is done by acting an $\varepsilon$-tensor of
length $N$ (\emph{c.f.}  eqns.~\eqref{eq:AntiQuarkMethodEx4}
and~\eqref{eq:Leibniz-UUDag1}) on the bottom $N-1$ legs on either side
of the operator, for example,
\begin{equation}
\label{eq:1q1qb-Singlet-ops-from-Nc-via-Leibniz}
\FPic[scale=0.75]{NcArrLeft}%
\FPic[scale=0.75]{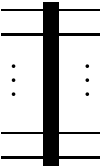}%
\FPic[scale=0.75]{NcArrRight}%
\quad \xrightarrow{\text{transform}} \quad
N \cdot
\FPic[scale=0.75]{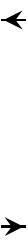}%
\FPic[scale=0.75]{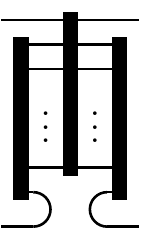}%
\FPic[scale=0.75]{1q1qb-from-Nc-Arr}%
\ ;
\end{equation}
notice that the additional factor $N$ arising from the action of
$\varepsilon_{i_2\ldots i_N i_{(N+1)}}$ is needed to ensure that the resulting
operator is idempotent.

After a significant computational effort (the details of which are
presented in the following section), one arrives at the desired
outcome:
\begin{subequations}
\label{eq:1q1qb-Epsilon1qb}
\begin{IEEEeqnarray}{0rCCCl}
\FPic[scale=0.75]{NcArrLeft}\FPic[scale=0.75]{NcASym1-to-Nc}\FPic[scale=0.75]{NcArrRight}%
\quad & \rightarrow & \quad
N \cdot
\FPic[scale=0.75]{1q1qb-from-Nc-Arr}%
\FPic[scale=0.75]{pASym1-to-pEpsilons2-to-p+1-LR}
\FPic[scale=0.75]{1q1qb-from-Nc-Arr}%
\; & = & \;\FPic{1q1qbTr12Arr}
\label{eq:1q1qb-Epsilon1qb-Singlet}
\\
\frac{2 (N-1)}{N} \cdot
\FPic[scale=0.75]{NcArrLeft}\FPic[scale=0.75]{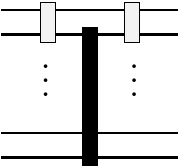}\FPic[scale=0.75]{NcArrRight}%
\quad & \rightarrow & \quad
2 (N-1)\cdot
\FPic[scale=0.75]{1q1qb-from-Nc-Arr}%
\FPic[scale=0.75]{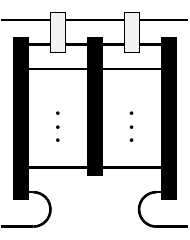}
\FPic[scale=0.75]{1q1qb-from-Nc-Arr}%
\; & = & \;
\FPic{1q1qbAdj12}
\label{eq:1q1qb-Epsilon1qb-Adjoint}
\ .
\end{IEEEeqnarray}
\end{subequations}
The immense computational expense involved in obtaining the last
equalities in eqns.~\eqref{eq:1q1qb-Epsilon1qb} makes this method
undesirable in practice, thus warranting a compact construction
method, as is given in the present paper for singlet operators.

The example presented here shows that the textbook method of talking
about irreducible representations of $\SUN$ on a mixed space
$\MixedPow{m}{n}$ (using LR tableaux) is rather more indirect than
that on a monotone space $\Pow{m}$ (using Young tableaux): The
primitive invariants of $\SUN$ on $\MixedPow{m}{n}$, and thus all
elements of $\API{\SUN,\MixedPow{m}{n}}$, are included in
$\API{\SUN,\Pow{[m+n(N-1)]}}$ in the sense of a subalgebra, as is
evident from the corresponding Littlewood-Richardson tableaux
(\emph{c.f.}  Theorem~\ref{thm:add-antiquark-LR}), which form a subset
of the Young tableaux of $m+n(N-1)$ boxes. This is due to the presence
of $\varepsilon_{i_1 \ldots i_N}$ as a second invariant in addition to
$\delta_{ij}$ (\emph{c.f.}  eq.~\eqref{eq:Kronecker-Delta-Levi-Civita-Birdtrack}). To
be a little more explicit: $\API{\SUN,\Pow{[m+n(N-1)]}}$ encompasses
representations on $\MixedPow{m}{n}$ via the LR rule, which exploits
the determinant condition for $\SUN$, i.e. the invariance of the
$\varepsilon$-tensor in $N$ dimensions, and so formally gives access
to all these representations as well. The drawback clearly is the size
of the algebra: $[m+n(N-1)]!\gg(m+n)!$ generically and keeping $N$ as
a parameter will not be a trivial task.

\subsection{Expanding on
  eqns.~\texorpdfstring{\eqref{eq:1q1qb-Epsilon1qb}}{eq1q1qbEpsilon1qb}:
  Simplifying the \texorpdfstring{$1q+1\bar{q}$}{1q+1qb}
  operators}\label{sec:Simpify-1q1qb-LR-ops}

This section provides all steps involved in simplifying the operators
in equations~\eqref{eq:1q1qb-Epsilon1qb}
using the birdtrack formalism.

\paragraph{The singlet
operator~\eqref{eq:1q1qb-Epsilon1qb-Singlet}:} 
Starting from the operator
\begin{equation}
  N \cdot 
  \FPic[scale=0.75]{1q1qb-from-Nc-Arr}%
  \FPic[scale=0.75]{pASym1-to-pEpsilons2-to-p+1-LR}
  \FPic[scale=0.75]{1q1qb-from-Nc-Arr}%
  \ ,
\end{equation}
we pull
the left $\varepsilon$-tensor to the right 
\begin{equation}
\label{eq:1q1qb-Simplify-Singlet-1}
N \cdot 
\FPic[scale=0.75]{1q1qb-from-Nc-Arr}%
\FPic[scale=0.75]{pASym1-to-pEpsilons2-to-p+1-LR}%
\FPic[scale=0.75]{1q1qb-from-Nc-Arr}%
\quad \xlongequal{\text{pull $\varepsilon$}} \quad
N \cdot 
\FPic[scale=0.75]{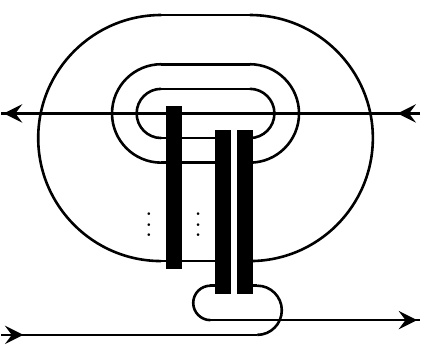}%
\ ,
\end{equation}
and then use identity~\eqref{eq:Epsilon-lengthN-ASym} to combine the
two $\varepsilon$-tensors of length $N$ into an antisymmetrizer,
\begin{equation}
\label{eq:1q1qb-Simplify-Singlet-1b}
N \cdot 
\FPic[scale=0.75]{pASym1-to-pEpsilons2-to-p+1-RR-ArrLabels}%
\quad \xlongequal{\text{eq.~\eqref{eq:Epsilon-lengthN-ASym}}} \quad
N \cdot 
\FPic[scale=0.75]{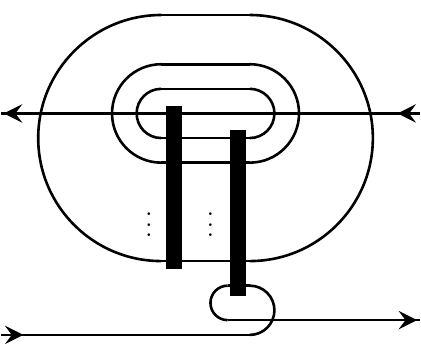}
\ .
\end{equation}
For simplicity, we will ``disentangle'' the $\bar{q}$-leg for the
remainder of this calculation and treat it as a quark leg,
\begin{equation}
  \label{eq:1q1qb-Simplify-Singlet-2}
N \cdot 
\FPic[scale=0.75]{pASym1-to-pASym2-to-qbTr2-to-p-ArrLabels}%
\quad \xrightarrow[\text{$\bar{q}$ line}]{\text{disentangle}} \quad
N \cdot 
\FPic[scale=0.75]{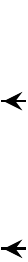}
\FPic[scale=0.75]{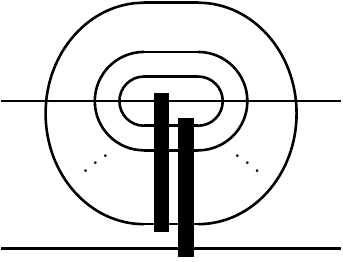}
\FPic[scale=0.75]{pqqArr}%
\ ,
\end{equation}
but we will reverse this at the end of the calculation (in
eq.~\eqref{eq:1q1qb-Simplify-Singlet-10}). Furthermore, we wish to
explicitly \emph{distinguish} the lengths of the two antisymmetrizers
in~\eqref{eq:1q1qb-Simplify-Singlet-2} for the sake of clarity in the
argument to follow. We thus say that the left antisymmetrizer
(originating from the LR tableau) has length $p$, while the right
antisymmetrizer (originating from the $\varepsilon$-tensors) has length
$N$. At the end of the calculation we will once again set $p=N$.

By identity~\cite[eq.~(6.19)]{Cvitanovic:2008zz}, the right
antisymmetrizer of length $N$ in~\eqref{eq:1q1qb-Simplify-Singlet-2}
can be decomposed as follows:
\begin{equation}
  \label{eq:1q1qb-Simplify-Singlet-3}
N \cdot 
\scalebox{0.75}{%
\FPic{pqqArr}\FPic{pASym1-to-pASym2-to-p+1Tr2-to-p}\FPic{pqqArr}%
} 
\; = \;
\scalebox{0.75}{%
\FPic{pqqArr}\FPic{pASym1-to-p-1ASym2-to-p-1Tr2-to-p-1}\FPic{pqqArr}%
}
- (N-1) \cdot
\scalebox{0.75}{%
\FPic{pqqArr}\FPic{pASym1-to-p-1ASym2-to-p-1sp-1pASym2-to-p-1Tr2-to-p-1}\FPic{pqqArr}%
}
\ .
\end{equation}
We may absorb the shorter antisymmetrizer(s) into the longer one in
each of the terms --- in the last term, we pull
the rightmost antisymmetrizer over the top to the left of the longer
antisymmetrizer and absorb it from the left. This yields
\begin{equation}
  \label{eq:1q1qb-Simplify-Singlet-4}
N \cdot
\scalebox{0.75}{%
\FPic{pqqArr}\FPic{pASym1-to-pASym2-to-p+1Tr2-to-p}\FPic{pqqArr}%
} 
\; = \;
\scalebox{0.75}{%
\FPic{pqqArr}\FPic{pASym1-to-p-1Tr2-to-p-1}\FPic{pqqArr}%
}
- (N-1) \cdot
\scalebox{0.75}{%
\FPic{pqqArr}\FPic{pASym2-to-p-1Tr2-to-p-1}\FPic{pqqArr}%
}
\ .
\end{equation}
We consider each term in eq.~\eqref{eq:1q1qb-Simplify-Singlet-4}
separately: The first term is just a free index line and a trace over
all but one of the indices of the antisymmetrizer of length $p$. Using
identity~\cite[eq.~(6.23)]{Cvitanovic:2008zz}, we can carry out this
trace, yielding
\begin{equation}
  \label{eq:1q1qb-Simplify-Singlet-5}
\scalebox{0.75}{%
\FPic{pqqArr}\FPic{pASym1-to-p-1Tr2-to-p-1}\FPic{pqqArr}%
}
\quad  = \;
\left[ 
\frac{(N-1)! / 
\left[ N-(p-1)-1 
\right]!}{p!} 
\right] \;
\FPic{2IdArr} 
\ .
\end{equation}
Setting $p=N$, we find
\begin{equation} 
  \label{eq:1q1qb-Simplify-Singlet-6}
\scalebox{0.75}{%
\FPic{pqqArr}\FPic{pASym1-to-p-1Tr2-to-p-1}\FPic{pqqArr}%
} \quad 
\; = \;
\left[ 
\frac{(N-1)! 0!}{N!} 
\right] \; 
\FPic{2IdArr} 
\; = \;
\frac{1}{N} \;
\FPic{2IdArr} 
\ .
\end{equation}

We now simplify the second term
in~\eqref{eq:1q1qb-Simplify-Singlet-4}. This term traces
all but two indices of the antisymmetrizer of length $p$. Again we
follow~\cite[eq.~(6.23)]{Cvitanovic:2008zz} to evaluate this trace,
\begin{equation}
  \label{eq:1q1qb-Simplify-Singlet-7}
(N-1) \cdot
\scalebox{0.75}{%
\FPic{pqqArr}\FPic{pASym2-to-p-1Tr2-to-p-1}\FPic{pqqArr}%
}
\quad =
(N-1) 
\left[ 
\frac{(N-2)! / \left[ N-(p-1)-1 \right]!}{p!/2!}
\right] \; 
\FPic{2ArrLeft}\FPic{2ASym12}\FPic{2ArrRight}
\ .
\end{equation}
Again setting $p=N$ we find
\begin{equation}
  \label{eq:1q1qb-Simplify-Singlet-8}
(N-1) \cdot
\scalebox{0.75}{%
\FPic{pqqArr}\FPic{pASym2-to-p-1Tr2-to-p-1}\FPic{pqqArr}%
} 
=
\frac{2}{N} \; 
\FPic{2ArrLeft}\FPic{2ASym12}\FPic{2ArrRight}
\ .
\end{equation}
Substituting~\eqref{eq:1q1qb-Simplify-Singlet-6}
and~\eqref{eq:1q1qb-Simplify-Singlet-8} back into~\eqref{eq:1q1qb-Simplify-Singlet-4} yields
\begin{equation}
  \label{eq:1q1qb-Simplify-Singlet-9}
N \cdot
\scalebox{0.75}{%
\FPic{pqqArr}\FPic{pASym1-to-pASym2-to-p+1Tr2-to-p}\FPic{pqqArr}%
}
\quad =
\frac{1}{N} \; 
\FPic{2IdArr} \; - 
\frac{2}{N} \; 
\FPic{2ArrLeft}\FPic{2ASym12}\FPic{2ArrRight}
=
\frac{1}{N} \; 
\FPic{2s12Arr} 
\ .
\end{equation}
Lastly, we have to transform the bottom index leg back into an
antiquark leg,
\begin{equation}
  \label{eq:1q1qb-Simplify-Singlet-10}
   \frac{1}{N} \; \FPic{2s12Arr} 
   \qquad \longrightarrow \qquad
   \frac{1}{N} \; \FPic{1q1qbTr12Arr}
\ .
\end{equation}
In summary, we found that
\begin{equation}
    \label{eq:1q1qb-Simplify-Singlet-Final}
\FPic[scale=0.75]{1q1qb-from-Nc-Arr}%
\FPic[scale=0.75]{pASym1-to-pEpsilons2-to-p+1-LR}%
\FPic[scale=0.75]{1q1qb-from-Nc-Arr}%
\; = \;
\frac{1}{N} \; \FPic{1q1qbTr12Arr}
\ ,
\end{equation}
as was claimed in eq.~\eqref{eq:1q1qb-Epsilon1qb-Singlet}.

\paragraph{The adjoint operator~\eqref{eq:1q1qb-Epsilon1qb-Adjoint}:}
Once again, we pull the left $\varepsilon$-tensor to the right and use
eq.~\eqref{eq:Epsilon-lengthN-ASym} to combine the two
$\varepsilon$-tensors into an antisymmetrizer of length
$N$. Furthermore, as we did in the calculation of the singlet
operator, we will treat the antiquark leg $\bar{q}$ as a quark leg,
transforming it back at the end of our calculation. We have:
\begin{equation}
  \label{eq:1q1qb-Simplify-Adjoint-1}
2 (N-1) \cdot
\FPic[scale=0.75]{1q1qb-from-Nc-Arr}
\FPic[scale=0.75]{pSym12ASym1-to-p-1Sym12Epsilons2-to-p+1-LR}
\FPic[scale=0.75]{1q1qb-from-Nc-Arr}
\quad \longrightarrow \quad
2 (N-1) \cdot
\FPic[scale=0.75]{pqqArr}
\FPic[scale=0.75]{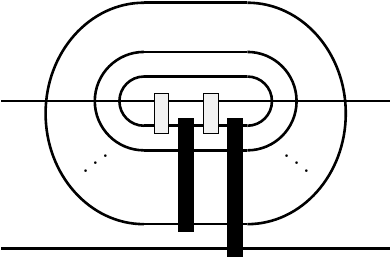}
\FPic[scale=0.75]{pqqArr}
\ .
\end{equation}
Using the cancellation rules derived
in~\cite{Alcock-Zeilinger:2016bss}, this operator may be simplified as 
\begin{equation}
  \label{eq:1q1qb-Simplify-Adjoint-2}
2 (N-1) \cdot
\scalebox{0.75}{%
\FPic{pqqArr}\FPic{pSym12ASym1-to-p-1Sym12ASym1-to-pTr2-to-p-1}\FPic{pqqArr}%
}
\; = \; 
2 (N-1) \cdot
\scalebox{0.75}{%
\FPic{pqqArr}\FPic{pSym12ASym1-to-pTr2-to-p-1}\FPic{pqqArr}%
}
\ ,
\end{equation}
where no additional constant is induced.\footnote{This is true since
  the cancelled part of the operator is a Young projection operator
  with normalization constant $2(N-1)$, as can be verified by direct
  calculation.} We begin the simplification process by decomposing the
symmetrizer in~\eqref{eq:1q1qb-Simplify-Adjoint-2} into its primitive
invariants (\emph{c.f.} eq.~\eqref{eq:Birdtrack-Sym-Definition}),
\begin{align}
2 (N-1) \cdot
\scalebox{0.75}{%
\FPic{pqqArr}\FPic{pSym12ASym1-to-pTr2-to-p-1}\FPic{pqqArr}%
} 
& = 
(N-1)
\left[ \;
\scalebox{0.75}{%
\FPic{pqqArr}\FPic{pASym2-to-pTr2-to-p-1}\FPic{pqqArr}%
}
+
\scalebox{0.75}{%
\FPic{pqqArr}\FPic{ps12ASym2-to-pTr2-to-p-1}\FPic{pqqArr}%
}
\; \right]
\nonumber \\
& \nonumber \\
& =
(N-1)
\left[ \;
\scalebox{0.75}{%
\FPic{pqqArr}\FPic{pASym2-to-pTr2-to-p-1}\FPic{pqqArr}%
}
+
\scalebox{0.75}{%
\FPic{pqqArr}\FPic{pASym1-to-pTr2-to-p-1}\FPic{pqqArr}%
}
\; \right]
\ ,
\label{eq:1q1qb-Simplify-Adjoint-3}
\end{align}
where we merely disentangled the index lines of the second term in~\eqref{eq:1q1qb-Simplify-Adjoint-3}. The first term
in eq.~\eqref{eq:1q1qb-Simplify-Adjoint-3} is an antisymmetrizer of
length $N$ with all but one of its legs traced. Thus,
using~\cite[eq.~(6.23)]{Cvitanovic:2008zz}, this term simplifies to
\begin{equation}
  \label{eq:1q1qb-Simplify-Adjoint-4}
(N-1) \cdot 
\scalebox{0.75}{%
\FPic{pqqArr}\FPic{pASym2-to-pTr2-to-p-1}\FPic{pqqArr}%
}
\quad =
(N-1)
\left[ 
\frac{(N-1)! / 
\left[ N-(N-1)-1 \right]!}{N!} 
\right] \; 
\FPic{2IdArr} 
=
\frac{(N-1)}{N} \;
\FPic{2IdArr}
\ .
\end{equation}
The second term in eq.~\eqref{eq:1q1qb-Simplify-Adjoint-3} is
an antisymmetrizer with all but two indices traced, yielding
\begin{equation}
  \label{eq:1q1qb-Simplify-Adjoint-5}
(N-1) \cdot
\scalebox{0.75}{%
\FPic{pqqArr}\FPic{pASym1-to-pTr2-to-p-1}\FPic{pqqArr}%
}
\quad =
(N-1)
\left[ 
\frac{(N-2)! / 
\left[N-(N-1)-1 \right]!}{N! / 2!} 
\right] \; 
\FPic{2ArrLeft}\FPic{2ASym12}\FPic{2ArrRight}
=
\frac{2}{N} \; 
\FPic{2ArrLeft}\FPic{2ASym12}\FPic{2ArrRight}
\ ,
\end{equation}
where we again used identity~\cite[eq.~(6.23)]{Cvitanovic:2008zz}.
Substituting expressions~\eqref{eq:1q1qb-Simplify-Adjoint-4}
and~\eqref{eq:1q1qb-Simplify-Adjoint-5} back
into~\eqref{eq:1q1qb-Simplify-Adjoint-3} yields
\begin{equation}
  \label{eq:1q1qb-Simplify-Adjoint-6}
2 (N-1) \cdot
\scalebox{0.75}{%
\FPic{pqqArr}\FPic{pSym12ASym1-to-pTr2-to-p-1}\FPic{pqqArr}%
} 
=
\frac{\left(N-1\right)}{N} \; 
\FPic{2IdArr} 
\; + \; 
\frac{2}{N} \;
\FPic{2ArrLeft}\FPic{2ASym12}\FPic{2ArrRight}  
\ .
\end{equation}
Decomposing the antisymmetrizer~$\FPic{2ArrLeft}\FPic{2ASym12SN}\FPic{2ArrRight}$~into its primitive invariants allows
for further simplification,
\begin{equation}
  \label{eq:1q1qb-Simplify-Adjoint-7}
2 (N-1) \cdot
\scalebox{0.75}{%
\FPic{pqqArr}\FPic{pSym12ASym1-to-pTr2-to-p-1}\FPic{pqqArr}%
} 
=
\frac{\left(N-1\right)}{N} \; 
\FPic{2IdArr} 
\; + \; 
\frac{1}{N} 
\left( 
\FPic{2IdArr} -
\FPic{2s12Arr} 
\right) 
=
\FPic{2IdArr} 
\; - \; 
\frac{1}{N} \;
\FPic{2s12Arr} 
\ .
\end{equation}
It remains to transform the bottom leg back into the antifundamental representation,
\begin{equation}
  \label{eq:1q1qb-Simplify-Adjoint-8}
\FPic{2IdArr} 
\; - \;
\frac{1}{N} \; \FPic{2s12Arr} 
\qquad \longrightarrow \qquad
\FPic{1q1qbIdArr} 
\; - \; 
\frac{1}{N} \;
\FPic{1q1qbTr12Arr} 
\; = \; 
\FPic{1q1qbAdj12}
\ ,
\end{equation}
where we used the Fierz identity~\cite{Cvitanovic:2008zz,Okun:1982ap}
(\emph{c.f.} eq.~\eqref{eq:Fierz-Identity-Irreps1q1qb})
\begin{equation}
  \label{eq:Fierz-Identity-Birdtrack}
  \FPic{1q1qbIdArr}
  \; = \; 
    \FPic{1q1qbAdj12}
  \; + \;
    \frac{1}{N} \;
    \FPic{1q1qbTr12Arr}
\end{equation}
to obtain the operator $\FPic{1q1qbAdj12}$ in the last step. Thus, we
found
\begin{equation}
  \label{eq:1q1qb-Simplify-Adjoint-Final}
2 (N-1) \cdot
\FPic[scale=0.75]{1q1qb-from-Nc-Arr}
\FPic[scale=0.75]{pSym12ASym1-to-p-1Sym12Epsilons2-to-p+1-LR}
\FPic[scale=0.75]{1q1qb-from-Nc-Arr}
\; = \;
\FPic{1q1qbAdj12}
\ ,
\end{equation}
confirming eq.~\eqref{eq:1q1qb-Epsilon1qb-Adjoint}.

\section{The equivalence between the \texorpdfstring{$3q$}{3q}-singlet
  and the totally antisymmetric \texorpdfstring{$2q+2\bar{q}$}{2q2qb}-singlet}\label{sec:3qSinglet-2q2qbSinglet-equivalent}

The equivalence between the operators
\begin{equation}
\label{eq:ASym123-equivalent-ASym1q2qASym1qb2qb-appendix}
\FPic{3ArrLeft}
\FPic{3ASym123SN}
\FPic{3ArrRight}
\quad 
\text{and} 
\quad
\FPic{2q2qbArrL}
\FPic{2q2qbASym12SStateL}\;\; 
\FPic{2q2qbASym12SStateDagL}
\FPic{2q2qbArrL}
\end{equation}
for
$\text{dim}(V)=N=3$~\cite{tHooft:1973alw,Witten:1979kh,Corrigan:1979xf}
is shown using the Leibniz
identity~\eqref{eq:Leibniz-Nminus1q-into-qb-Us}
(equivalently~\eqref{eq:Leibniz-UUDag1}): this identity translates
$N-1=2$ antifundamental index lines into a fundamental line through
the Levi-Civita symbols (i.e. $\varepsilon$-tensors of length $N=3$)
\begin{equation}
  \label{eq:2U1UDagLeibniz}
\FPic{3Epsilon1qbArrLabels-bottom-q}
\FPic{3Epsilon1qbN}
\FPic{3Epsilon1qbArrLabels-top-qb}
\qquad \text{and} \qquad
\FPic{3Epsilon1qbArrLabels-top-qb}
\FPic{3EpsilonDag1qbN} 
\FPic{3Epsilon1qbArrLabels-bottom-q}
\ .
\end{equation}
We act each such Levi-Civita tensor on the bottom two antifundamental
legs of the antisymmetric $2q+2\bar{q}$-singlet in
eq.~\eqref{eq:ASym123-equivalent-ASym1q2qASym1qb2qb-appendix} and then
absorb the antisymmetrizers into the Levi-Civita tensors,
\begin{equation}
\label{eq:2q2qb-absorb-ASym-into-Epsilon}
\FPic{2q2qbArrL}
\FPic{2q2qbASym12SStateL} \;\; \FPic{2q2qbASym12SStateDagL}
\FPic{2q2qbArrL}
  \quad \xrightarrow{\text{act $\varepsilon$}} \quad
\FPic{3EpsilonDag2qb-ArrLabels-proton-q}
\FPic{2q2qb3EpsilonDag2Top}
\FPic{2q2qbASym12SStateL-proton} \;\; \FPic{2q2qbASym12SStateDagL-proton}
\FPic{2q2qb3Epsilon2Top}
\FPic{3EpsilonDag2qb-ArrLabels-proton-q}
   \quad \xlongequal[\text{antisym.}]{\text{absorb}} \quad
\FPic{3EpsilonDag2qb-ArrLabels-proton-q}
\FPic{2q2qb3EpsilonDag2Top}
\FPic{2q2qbIdSStateL-proton} \;\; \FPic{2q2qbIdSStateDagL-proton}
\FPic{2q2qb3Epsilon2Top}
\FPic{3EpsilonDag2qb-ArrLabels-proton-q}
  \ .
\end{equation}
We now flip each antisymmetrizer about its vertical axis, keeping the
end points fixed,
\begin{equation}
\label{eq:3Epsilon-flip}
\FPic{3EpsilonDag2qb-ArrLabels-proton-q}
\FPic{2q2qb3EpsilonDag2Top}
\FPic{2q2qbIdSStateL-proton} \;\; \FPic{2q2qbIdSStateDagL-proton}
\FPic{2q2qb3Epsilon2Top}
\FPic{3EpsilonDag2qb-ArrLabels-proton-q}
  \quad \xlongequal{\text{flip}} \quad
(-1)^2 \;
\FPic{3EpsilonDag2qb-ArrLabels-proton-q}
\FPic{3Epsilon-proton} \;\;
\FPic{3EpsilonDag-proton}
\FPic{3EpsilonDag2qb-ArrLabels-proton-q}
\ ,
\end{equation}
where we had to absorb a transposition $(12)$ into each
$\varepsilon$-tensor in the process, inducing a factor of
$(-1)^2=1$.\footnote{Defining
  $\kappa_k$ to be the transposition between $k$ and $(N-k$), longer
  and thus more ``entangled'' $\varepsilon$-tensors will have a
  prefactor
  $\left[\text{sign}\left(\kappa_1 \kappa_2 \kappa_3 \ldots
    \right)\right]^2=1$.\label{footnote:absorbing-permutations-ASym-flip}}
It should be noted that through this flipping procedure each
$\varepsilon$-tensor in~\eqref{eq:3Epsilon-flip} will be accompanied
by a factor $i^{\pm\phi}$ (\emph{c.f.} eq.~\eqref{eq:Kronecker-Delta-Levi-Civita-Birdtrack})
with the \emph{wrong} sign in the exponent.
However, in the product~\eqref{eq:3Epsilon-flip}, each prefactor can
be \emph{reassigned} to the other $\varepsilon$-tensor, thus
remedying the incorrect sign. If this ``prefactor conundrum'' caused
by the flip in~\eqref{eq:3Epsilon-flip} seems undesirable to the
reader, we present a work-around in
section~\ref{sec:Untwisiting-epsilon-tensors}, which leaves the
prefactors untouched, but still yields the desired result.

It now remains to recombine the two Levi-Civita tensors
in~\eqref{eq:3Epsilon-flip} into the antisymmetrizer $\bm{A}_{123}$
according to eq.~\eqref{eq:Epsilon-lengthN-ASym} in order to obtain
the desired result,
\begin{equation}
  \label{eq:ASym123-N3-Epsilons}
\FPic{3EpsilonDag2qb-ArrLabels-proton-q}
\FPic{3Epsilon-proton} \;\;
\FPic{3EpsilonDag-proton}
\FPic{3EpsilonDag2qb-ArrLabels-proton-q}
\; = \;
\FPic{3ArrLeft}
\FPic{3Epsilon}\;\;
\FPic{3EpsilonDag}
\FPic{3ArrRight}
\; \xlongequal{\text{eq.~\eqref{eq:Epsilon-lengthN-ASym}}} \;
\FPic{3ArrLeft}
\FPic{3ASym123SN}
\FPic{3ArrRight}
\ .
\end{equation}

In this example, we chose to transform $N-1$ antifundamental lines into a
fundamental line by means of the Leibniz
identity~\eqref{eq:Leibniz-UUDag1}. Equally effectively, we could have
transformed $(N-j)$ antifundamental lines into $j$ fundamental ones
according to eq.~\eqref{eq:AntiQuarkMethodEx5}, for example, if $j=2$,
\begin{equation}
  \label{eq:1q1qb-to-Baryon-singlet}
\FPic{1q1qbTr12Arr}
\quad \xrightarrow{\text{act $\varepsilon$}} \quad
\FPic{3Epsilon2qb-ArrLabels-proton-q}
\FPic{3EpsilonDag2qb-1q1qbSState-proton}
\;\;
\FPic{3Epsilon2qbN-1q1qbSStateDag-proton}
\FPic{3Epsilon2qb-ArrLabels-proton-q}
\quad \xlongequal{\text{flip}} \quad
\FPic{3ArrLeft}
\FPic{3Epsilon}\;\;
\FPic{3EpsilonDag}
\FPic{3ArrRight}
\; = \; 
\FPic{3ArrLeft}
\FPic{3ASym123SN}
\FPic{3ArrRight}
\ .
\end{equation}
However, the first way of obtaining the baryon singlet projector will
be more useful when looking at Wilson line correlators and coincidence
limits, as is done in a future paper~\cite{Weigert:2017}.

\subsection{Untwisting \texorpdfstring{$\varepsilon$}{epsilon}-tensors
  without flipping the sign in the factor \texorpdfstring{$i^{\pm\phi}$}{iphi}}\label{sec:Untwisiting-epsilon-tensors}

Instead of the flip conducted in eq.~\eqref{eq:3Epsilon-flip}, we may
obtain the desired
equivalence~\eqref{eq:ASym123-equivalent-ASym1q2qASym1qb2qb-appendix}
in a way that does not cause havoc with any prefactors. Let us pick up
at eq.~\eqref{eq:2q2qb-absorb-ASym-into-Epsilon}: Keeping the end
points fixed, we may move the left
$\varepsilon^{\dagger}$ to the right of $\varepsilon$ \emph{without
  flipping it}; this yields a somewhat entangled operator,
\begin{equation}
\FPic{3EpsilonDag2qb-ArrLabels-proton-q}
\FPic{2q2qb3EpsilonDag2Top}
\FPic{2q2qbIdSStateL-proton} \;\; \FPic{2q2qbIdSStateDagL-proton}
\FPic{2q2qb3Epsilon2Top}
\FPic{3EpsilonDag2qb-ArrLabels-proton-q}
\; \xlongequal{\text{move $\varepsilon$-tensors}} \;
\FPic{3ArrTangledLabelsLeft-q}%
\FPic{3TwoEpsilonsTangled}%
\FPic{3ArrTangledLabelsRight-q}
\ .
\end{equation}
The two Levi-Civita tensors combine into
an antisymmetrizer of length $N=3$ according to eq.~\eqref{eq:Epsilon-lengthN-ASym},
\begin{equation}
\FPic{3ArrTangledLabelsLeft-q}%
\FPic{3TwoEpsilonsTangled}%
\FPic{3ArrTangledLabelsRight-q}
\; \xlongequal{\text{eq.~\eqref{eq:Epsilon-lengthN-ASym}}} \;
\FPic{3ArrTangledLabelsLeft-q}%
\FPic{3ASym123Tangled}%
\FPic{3ArrTangledLabelsRight-q}
\ .
\end{equation}
The antisymmetrizer in the middle may now be flipped to disentangle
the index lines; this does \emph{not} produce any phase factors, as the
antisymmetrizer is a real quantity,
\begin{equation}
  \label{eq:2q2qb-to-Proton-singlet-Final}
\FPic{3ArrTangledLabelsLeft-q}%
\FPic{3ASym123Tangled}%
\FPic{3ArrTangledLabelsRight-q} 
\; = \; 
\FPic{3ArrLeft}
\FPic{3s12N}
\FPic{3ASym123SN}
\FPic{3s12N}
\FPic{3ArrRight}
\; = \;
(-1)^2 \;
\FPic{3ArrLeft}
\FPic{3ASym123SN}
\FPic{3ArrRight}
\ ,
\end{equation}
where, in the disentanglement process, we absorb a transposition
$(12)$ on either side of the antisymmetrizer, inducing an additional
prefactor $(-1)^2=1$ in the last step (this is in analogy to the
prefactor $(-1)^2$ in eq.~\eqref{eq:3Epsilon-flip}, \emph{c.f.}
footnote~\ref{footnote:absorbing-permutations-ASym-flip}). Thus, we
once again arrive at the desired
result~\eqref{eq:ASym123-N3-Epsilons}.

 \bibliographystyle{utphys}
 \bibliography{PaperLibrary,BookLibrary,GroupTheory}

\end{document}